\DocumentMetadata{}
\documentclass[amcsmall,authorversion,nonacm]{acmart}
\usepackage[figuresright]{rotating}
\usepackage{adjustbox}
\usepackage{amsmath}
\usepackage{enumitem}
\usepackage{xspace}
\usepackage{tablefootnote}
\usepackage{multicol, multirow}
\newcommand{\eat}[1]{}

\AtBeginDocument{%
  \providecommand\BibTeX{{%
    \normalfont B\kern-0.5em{\scshape i\kern-0.25em b}\kern-0.8em\TeX}}}

\setcopyright{acmcopyright}
\copyrightyear{2018}
\acmYear{2018}
\acmDOI{XXXXXXX.XXXXXXX}

\acmJournal{CSUR}
\acmVolume{37}
\acmNumber{4}
\acmArticle{111}
\acmMonth{8}
\begin{document}

\title{Graph and Sequential Neural Networks in Session-based Recommendation: A Survey}

\author{Zihao Li}
\email{zihao.li@student.uts.edu.au}
\orcid{}
\affiliation{%
  \institution{University of Technology Sydney}
  \country{Australia}
}

\author{Chao Yang}
\email{chao.yang@student.uts.edu.au}
\affiliation{%
  \institution{University of Technology Sydney}
  \country{Australia}
}

\author{Yakun Chen}
\email{yakun.chen@student.uts.edu.au}
\affiliation{%
  \institution{University of Technology Sydney}
  \country{Australia}
}

\author{Xianzhi Wang}
\email{xianzhi.wang@uts.edu.au}
\affiliation{%
  \institution{University of Technology Sydney}
  \country{Australia}
}

\author{Hongxu Chen}
\email{hongxu.chen@uts.edu.au}
\affiliation{%
  \institution{University of Technology Sydney}
  \country{Australia}
}

\author{Guandong Xu}
\email{gdxu@eduhk.hk}
\affiliation{%
  \institution{The Education University of Hong Kong, Hong Kong Special Administrative Region of China and University of Technology Sydney}
  \country{Australia}
}

\author{Lina Yao}
\affiliation{%
  \institution{University of New South Wales}
  \country{Australia}
 }
\email{lina.yao@unsw.edu.au}

\author{Quan Z. Sheng}
\affiliation{%
  \institution{Macquarie University}
  \country{Australia}
 }
\email{michael.sheng@mq.edu.au}







\renewcommand{\shortauthors}{Li et al.}

\begin{abstract}
    Recent years have witnessed the remarkable success of recommendation systems (RSs) in alleviating the information overload problem. As a new paradigm of RSs, session-based recommendation (SR) specializes in users' short-term preference capture and aims to provide a more dynamic and timely recommendation based on the ongoing interacted actions. In this survey, we will give a comprehensive overview of the recent works on SR. First, we clarify the definitions of various SR tasks and introduce the characteristics of session-based recommendation against other recommendation tasks. Then, we summarize the existing methods in two categories: sequential neural network based methods and graph neural network (GNN) based methods. The standard frameworks and technical details are also introduced.
   Finally, we discuss the challenges of SR and new research directions in this area. 
\end{abstract}



\maketitle

\section{Introduction}

With the prosperity and prevalence of the Internet, a surge of companies inclined to provide their products or services in an E-commerce modal, \textit{e.g.}, Amazon, YouTube, Yelp, LinkedIn.
Although the abundance of products and information provides more probability for users to satisfy their various personality requirements, the information overload problem becomes more serious. 
As an effective and efficient information filtering technology, Recommendation Systems (RSs) are capable of mining user's \textit{Point-of-Interest} (\textit{POI})  based on her historical interaction records (\textit{e.g.}, click, watch, read, add cart, and purchase) and recommend interested items in an automatic fashion. Recently, RSs have evolved into a prominent solution and attracted widespread attention from academia and industry~\cite{wu2020graph,fang2020deep,wang2021survey}.

Overall, in the past decades, the conventional mainstream attempts for RSs can be divided into content-based models~\cite{balabanovic1997fab} 
, collaborative filtering-based (CF-based) models~\cite{sarwar2001item,wang2006unifying}
, and hybrid models (\textit{i.e.}, the combination of the content-based and CF-based two models)~\cite{lu2015content}.  
As for CF-based methods, they can be further subdivided into the memory-based models (\textit{e.g.}, item-based and user-based)~\cite{sarwar2001item,wang2006unifying} and model-based attempts (\textit{e.g.}, Naive Bayes, matrix factorization, latent factor model, and neural networks)~\cite{miyahara2000collaborative,koren2009matrix,koren2008factorization,ma2008sorec,shen2012learning,zhang2013combining,salakhutdinov2007restricted,wu2020graph,wang2021survey}. However, most of above methods are committed to utilizing users' whole historical information and capture the long-term, statistic preference for recommendation~\cite{saadat2022knowledge,elahi2024knowledge}. It may not be applicable for practical scenarios since (1) the user's preference will shift and evolve over time, and (2) those methods are difficult to capture the user's ongoing behaviors and realize a real-time recommendation. 
We believe modeling users' short-term preferences and intentions based on the current interaction behaviors are also critically important for dynamic and accurate recommendation.

\begin{figure}[t]
  \centering
  \includegraphics[width=0.9\linewidth]{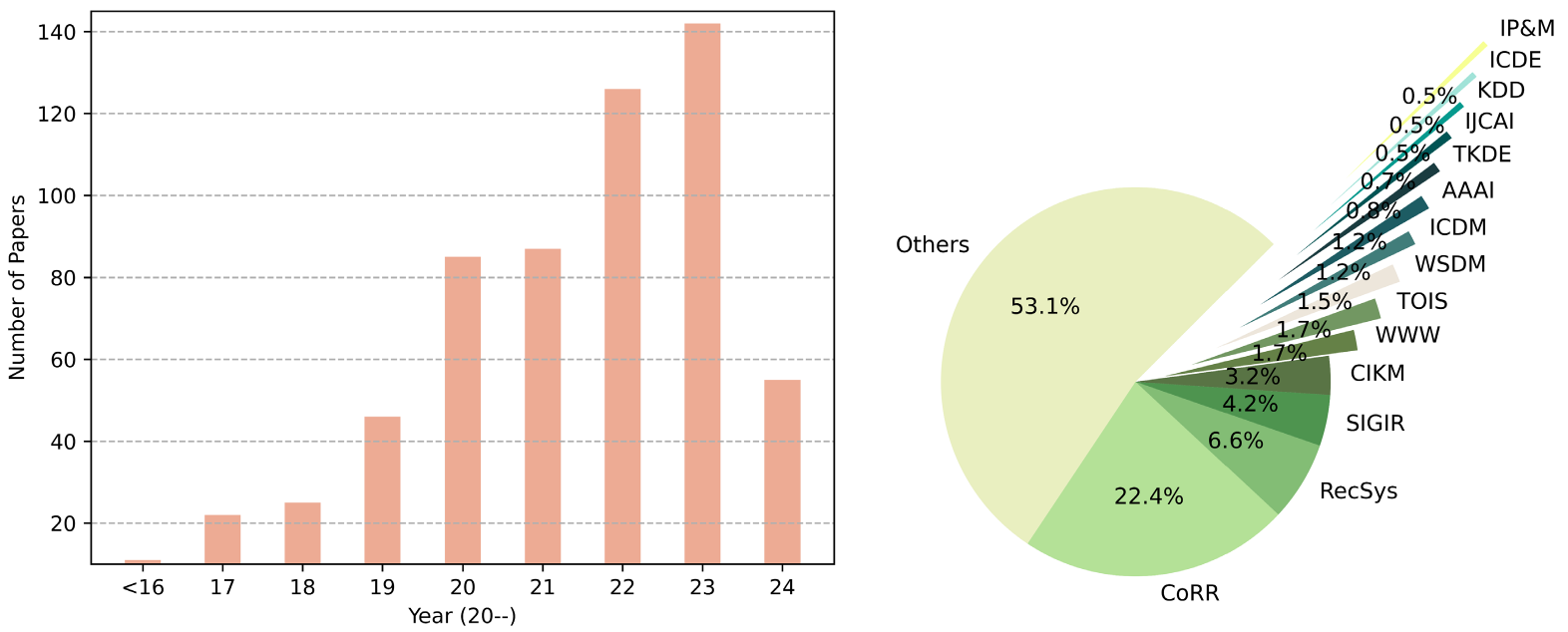}
  \caption{\textcolor{black}{The statistics of publications with regard to SR. "<16" means 2016 and before. The bar chart (left) displays the number of published papers each year, and the pie chart (right) illustrates the percentage of papers published in each top venue.}}
  \label{lab:papers}
\end{figure}

To bridge this gap, \textit{Session-based Recommendation (SR)}~\cite{wang2021survey} or \textit{Session-aware Recommendation}~\footnote{Quadrana \textit{et al.}~\cite{quadrana2017personalizing} point out for \textit{session-aware recommendation}, the user's all historical sessions are known, but in terms of \textit{session-based recommendation}, the users are anonymous and we only focus on the current session. However, in most of the works, we do not distinguish these two terminologies.} have emerged with increasing attention in recent years. 
Reviewing the development of SR, Modani \textit{et al.}~\cite{modani2002series} first proposed the concept of \textit{session} for a dynamic recommendation in 2002. 
On top of that, in 2005, Modani \textit{et al.}~\cite{modani2005framework} further devised a basic framework based on a bipartite graph, which is the pioneering work for the SR task. 
Since then, SR gradually become a hot topic in recommendation research avenues. \textcolor{black}{Figure~\ref{lab:papers} shows the number of papers (593 in total 
retrieved from DBLP Database\footnote{https://dblp.org/} by June 2024) published in top venues. 
We can find that the relevant work has increased significantly since the year 2019. In 2020 and 2021 the number of papers (85 and 87) is almost twice of 2019 (46). In 2022 and 2023, the number rises to 126 and 142 respectively. Besides, RecSys, SIGIR, CIKM, WWW, WSDM, TOIS and TKDE are the most popular target venues for SR.}


In general, the early works for SR focus more on popularity-based solutions~\cite{adomavicius2005toward}, and machine learning-based methods, including KNN~\cite{davidson2010youtube,ludewig2018evaluation,yang2014continuous}, Markov Chain~\cite{rendle2010factorizing,eirinaki2005web,rendle2010factorizing,zhang2007efficient} and matrix factorization~\cite{rendle2012bpr,cheng2013you,lian2013collaborative,shani2005mdp}.
Benefit from the powerful ability of feature extraction and representation, deep learning solutions, such as sequential neural networks~\cite{ren2019repeatnet,hidasi2015session,li2017neural,kang2018self,sun2019bert4rec,guo2020session} and graph neural networks (GNNs)~\cite{wu2019session,chen2021dual,xia2021self_1,zhang2020personalized,zhou2021temporal,chen2021session}, become ubiquitous and achieve promising results for SR in recent years. 
More concretely, the sequential neural networks, \textit{e.g.}, recurrent neural network (RNN), long short-term memory neural network (LSTM), and gated recurrent unit neural network (GRU), model the input session as a sequence and capture the order dependency among items for recommendation. In contrast, GNNs are required to predefine a graph structure based on sessions first, then items' correlations will be modeled via the information propagation and aggregation for recommendation. 

Although there are some surveys related to our work on recommendation systems~\cite{quadrana2018sequence,zhang2019deep,chen2020bias,fang2020deep,wu2020graph,wang2021survey}, they either put the lens on sequential recommendation~\cite{quadrana2018sequence,fang2020deep} or discuss the application of GNNs in recommendation systems~\cite{wu2020graph,chen2021incorporating}. 
Ludewig and Jannach~\cite{ludewig2018evaluation} compare the performance of various models for SR and provide an in-depth analysis. 
However, these methods were proposed before 2019, leaving GNNs to be explored.
Wang \textit{et al.}~\cite{wang2021survey, wang2022sequential} clarify the definitions of SR first, then a categorization of existing SR arts is proposed. 
Besides, the challenges and new opportunities in SR are illustrated for future development. 
Although this work provides an overall picture of SR, the detailed technical discussion and comparison between sequential neural networks and GNNs are insufficient. 
Given the impressive pace at which SR research with graph and sequential neural networks is growing, we believe it is necessary and valuable to analyze the characteristics, categorize the existing works in a unified framework, formalize their focused problems, and summarize the general ideas and solutions, the major challenges, and the future directions thoroughly.
\textcolor{black}{Consequently, in this paper, we are devoted to proposing a systematic classification, detailing a comparison of GNNs and sequential neural networks, and providing a comprehensive overview and survey of their application in SR.}
The key contributions of this survey are summarized as follows:    

\begin{itemize}[leftmargin=*]
    \item We standardize the concepts and definitions with regard to SR and introduce various graph structures that are commonly used in SR. Besides, we summarize the features of SR and compare the similarities and distinctions between SR and sequential recommendation. Moreover, we propose unified frameworks to regulate sequential neural networks and GNNs for SR, respectively. 
    \item We propose a systematic category to organize the existing works regarding sequential neural networks and GNNs for SR. Additionally, a comprehensive analysis and comparisons of the properties of these two mainstream methods are presented. We further introduce the overall pipeline and key modules in these methods in detail.  
    \item Finally, we identify open challenges and discuss future directions to inspire more research on this topic.
\end{itemize}

The rest of the survey is organized as follows. Section~\ref{sec:preliminaries} introduces the preliminaries of SR and different graph structures in this paper for easy reading.
Section~\ref{sec:characteristic} introduces the typical features of SR and illustrates the difference between sequential neural networks and GNNs for SR. Moreover, the categorization scheme of relevant works and their characteristics are also introduced. In section~\ref{sec:methodology}, we first formalize unified frameworks of sequential neural networks and GNNs for SR, respectively. Subsequently, a detailed exposition with respect to efficacious modules in those frameworks is also provided. 
\textcolor{black}{In Section~\ref{sec:evaluation_datasets}, we analyze the statistical characteristics of real-world datasets, introduce the commonly used performance evaluation metrics with regard to accuracy and diversity
} Section~\ref{sec:challengeanddirection} outlines this field's challenges and future directions. 
Finally, we conclude the survey in Section~\ref{sec:conclustion}.

\section{Preliminaries}
\label{sec:preliminaries}

In this section, we will first 
clarify some basic definitions in SR, \textit{e.g.}, intra-sessions, inter-sessions, and so forth. Additionally, various typical graph structures and the corresponding concepts are also introduced briefly.  

\eat{
\begin{table}[]
\caption{Key notations in this paper}
\begin{tabular}{ll}
\hline
\multicolumn{1}{l|}{Notations} & Descriptions     \\ \hline
\multicolumn{1}{l|}{$\mathcal{U}$}         & The set of users \\
\multicolumn{1}{l|}{$\mathcal{I}$}         & The set of items                 \\
\multicolumn{1}{l|}{$\mathcal{A}$}          & The set of attributes                 \\
\multicolumn{1}{l|}{$\mathcal{V}$}          & The set of nodes in graph                 \\
\multicolumn{1}{l|}{$\mathcal{E}$}          & The set of edges in graph                 \\
\multicolumn{1}{l|}{$\mathcal{S}$}          & The set of sessions               \\
\multicolumn{1}{l|}{$\mathcal{G}$}          & The graph                 \\
\multicolumn{1}{l|}{$\mathbf{A}$}        & Adjacency matrix of graph \\
\multicolumn{1}{l|}{$\hat{y}$}        &  The probability of predicted items             \\
\multicolumn{1}{l|}{$\mathbf{X}$}        & Embedding            \\
\multicolumn{1}{l|}{$\mathbf{H}$}        & Item representation            \\
\multicolumn{1}{l|}{$\mathbf{S}$}        & Session representation            \\
\multicolumn{1}{l|}{$\mathbf{W}$}        & Learnable parameters            \\
\multicolumn{1}{l|}{$\mathcal{L}$}        & Loss function            \\
\hline
\end{tabular}
\label{tab:notations}
\end{table}
}

\subsection{Session Definitions}
\label{sec:session_definitions}

\begin{figure*}[t]
\centerline{\includegraphics[width=\textwidth]{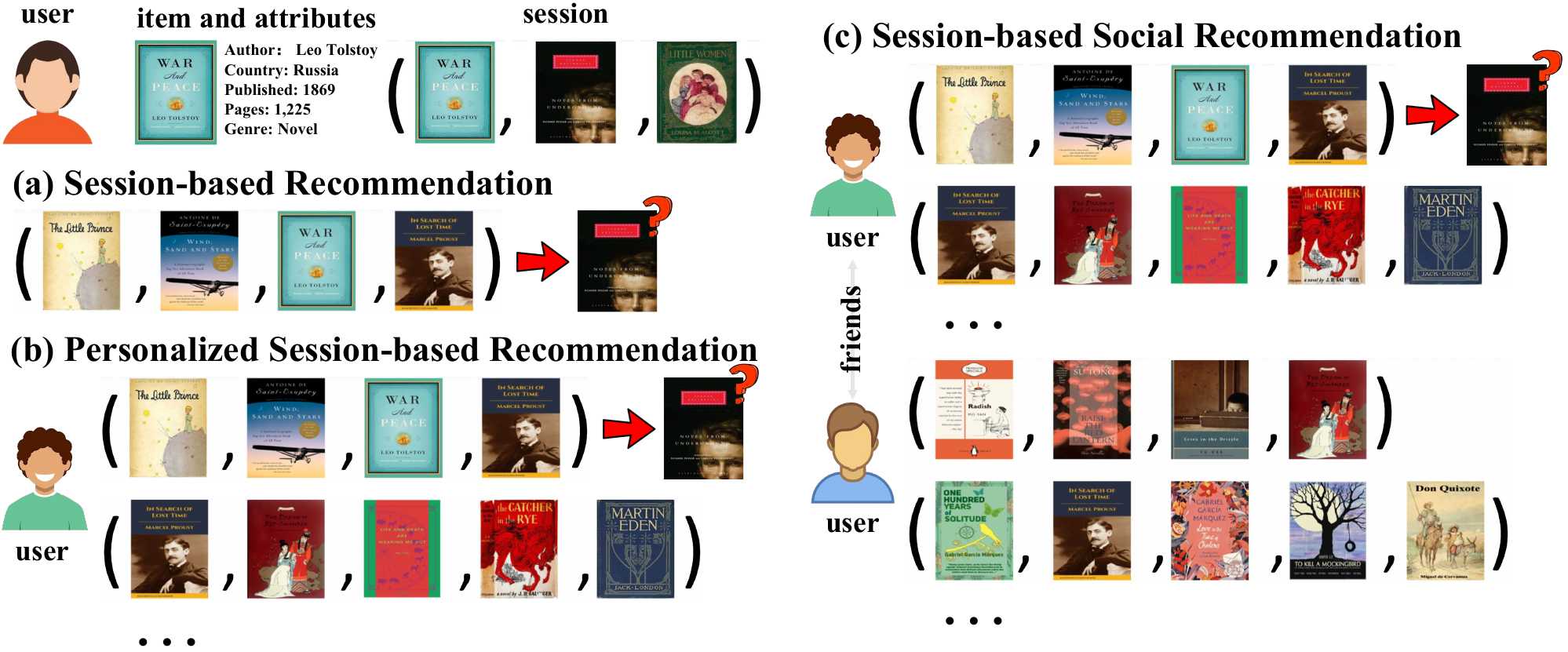}
}
\caption{\textcolor{black}{A toy example of different SR tasks.}
}
\label{fig:SR}
\end{figure*}

According to whether a user's information is anonymous or not, SR can be further divided into personalized session-based recommendation (PSR) and session-based social recommendation (SSR), which are illustrated in Figure~\ref{fig:SR}. In addition, the definitions and the key concepts, \textit{i.e.}, sessions, items, attributes, and social relations are listed below.

\noindent \textbf{Definition 2.1.1. Item} is the object that users interact with. It is usually formalized as an item ID in recommendation system. 

\noindent \textbf{Definition 2.1.2. Attributes} are the associated external information (or side information in~\cite{xie2022decoupled}) of items or interactions. Specifically, the item-oriented attributes contain brand, category, text descriptions, and images. As for interaction-oriented attributes, geographic information, interaction time and order, and behavior types (\textit{e.g.}, search, click, add cart, buy, share, comment, etc) are most commonly used. 
These attributes could serve as auxiliary information for performance improvement.

\noindent \textbf{Definition 2.1.3. Session} refers to a user-interacted list consisting of items or services. The items in sessions are usually organized chronologically\footnote{In most cases this is true, albeit few papers ignore the sequence information in sessions~\cite{hu2017diversifying,wang2017perceiving}}. 

\noindent \textbf{Definition 2.1.4. Session-based Recommendation (SR, depicted in Figure~\ref{fig:SR}, circled by the black dotted box).} Let $\mathcal{I}=\{i_1,i_2,i_3,...i_n\}$ as the set of items, where $n$ is the number of items. Each session $s=[i_1, i_2, i_3,...,i_m]$ consists of a sequence of interactive items $i_k\in I $ from a user. Thus, given a session $s$, the task of SR aims to generate probabilities $\mathbf{\hat{y}}$ for all possible items.
Each element's value in distribution $\mathbf{\hat{y}}$ indicates the recommendation score of the corresponding item. The items with a top-$K$ recommendation score will be recommended.
\textcolor{black}{Although collaborative filtering solutions, including item-based, user-based and content-based methods, aim to predict user's next POI, these methods elaborate to capture the co-occurrence patterns based on the user-item interaction matrix, ignoring the sequential information of each click~\cite{su2009survey}. In contrast, session-based recommendation put efforts on current session modeling, i.e., instead of all the historical records, only the ongoing session and the sequential patterns encapsulated in the sessions required to be considered, thus, it will be more suitable for new users and timely and dynamic recommendation (more detail discussions please ref. Section~\ref{sec:features_sr}).}

\noindent \textbf{Definition 2.1.5. Personalized Session-based Recommendation (PSR, illustrated in Figure~\ref{fig:SR}, circled by the blue dotted box).} Let $\mathcal{U}$ be a set of users, for each user $u\in \mathcal{U}$, denote $\mathcal{S}^u=\{S_i^u\}_{i=1}^{n_u}$ as all the historical sessions of $u$, and $n_u$ stands for the total number of sessions. Let $S^u_i=[i_{i,j}]_{j=1}^{m_i}\in \mathcal{S}^u$ as the $i$-th session of user $u$, and $m_i$ stands for the total number of items in session $S^u_i$. We define $S_c^u$ as the current session of user $u$, the previous sessions in the timeline are historical sessions denoted as $\mathcal{S}_h^u$. Thus, given all the historical sessions $\mathcal{S}_h^u$ of user $u$, the task of PSR is to predict the next interactive item of the current session $S_c^u$. In~\cite{guo2019streaming}, the \textbf{PSR} also named as streaming session-based recommendation.

\noindent \textbf{Definition 2.1.6. Session-based Social Recommendation (SSR, illustrated in Figure~\ref{fig:SR}, circled by the orange dotted box).} Based on PSR, we denote $\{u_k\}_{k=1}^{N(u)}\subseteq \mathcal{U}$ as the neighbors or friends of user $u$
, $N(u)$ is the number of neighbors. Let the sessions of user $u_k$ as $\mathcal{S}^{u_k}$. Thus, given the current session $\mathcal{S}^u$ of user $u$ and all the sessions $\cup_{k=1}^{N(u)} \mathcal{S}^{u_k}$ from user $u$ and his/her neighbors $u_k$, the task of SSR aims to predict the next interactive item of the current session $\mathcal{S}^u$.

\subsection{Graphs Definitions}
\label{sec:graph_define}

Graph-based SR requires us to predefine a graph first, then the GNNs are adopted for information propagation and aggregation. To model more fertile and valuable information from neighbors, researchers endeavor to design dedicated graph structures. We, thereby, introduce five typical graph structures in SR, respectively.

\begin{figure*}[t]
\centerline{\includegraphics[width=0.9\textwidth]
{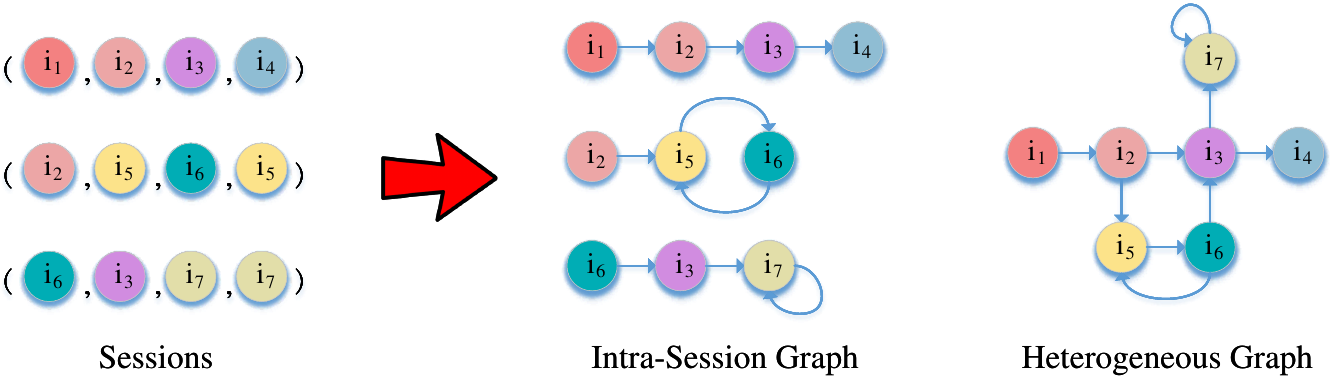}
}
\caption{The diagram of the intra-session graph and the inter-session graph. We represent the items or nodes as solid circles and the edges as arrow lines.}
\label{fig:inter-intra}
\end{figure*}

\noindent \textbf{Definition 2.2.1 Digraph and Undigraph.} Given a session $s$, each item in the session can be presented as a node. Besides, if a user clicks item $i_j$ after item $i_i$ in the session, we add a directed edge $e_{ij}$ from node $i$ point to node $j$. Hence, we could organize those nodes and the directed edges between all adjacent items via a digraph. In contrast, if we add an undirected edge $e_{ij}$ between node $i$ and node $j$, we could construct an undigraph.

\noindent \textbf{Definition 2.2.2 Intra-session graph.} As shown in Figure~\ref{fig:inter-intra} (left), given a series of sessions $\mathcal{S}$, we could construct intra-session graphs $\mathcal{G}=(\mathcal{V},\mathcal{E})$ for each session where the node set are all unique items in $\mathcal{S}$. And $e_{ij}\in \mathcal{E}$ represents an edge where a user clicks item $i_j$ after $i_i$ in sessions. 

\noindent \textbf{Definition 2.2.3 Inter-session graph.}
As shown in Figure~\ref{fig:inter-intra} (right), given a series of sessions $\mathcal{S}$, we could construct a unified inter-session graph $\mathcal{G}=(\mathcal{V, E})$ between all the adjacent items for all the sessions, where $\mathcal{V}$ and $\mathcal{E}$ indicate the set of nodes and edges respectively. For an edge $e_{ij} \in \mathcal{E}$, it 
indicates the edge points from node $v_i$ to node $v_j$. 

\begin{figure}[t]
\centerline{\includegraphics[width=0.9\textwidth]{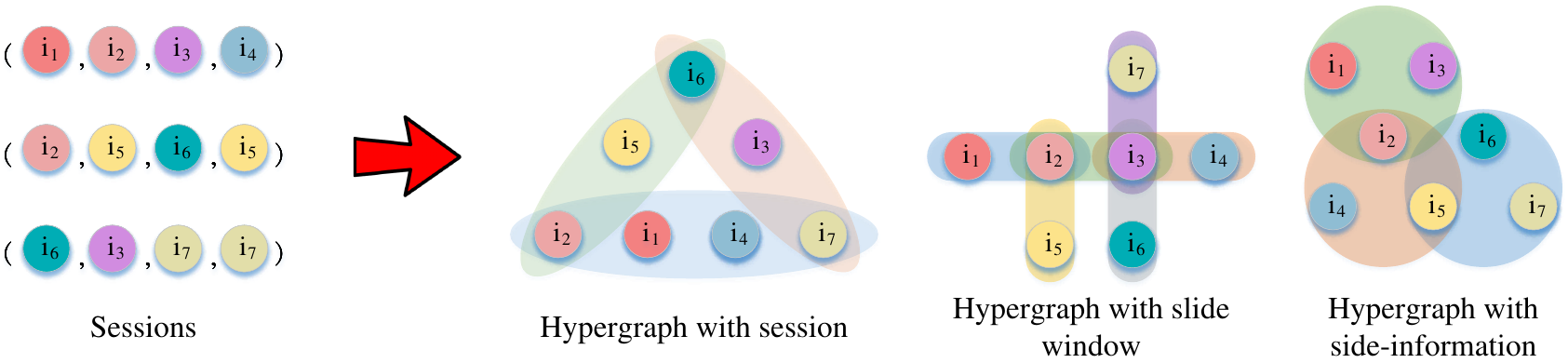}}
\caption{The diagram of hypergraphs. The shades of different colors represent different hyperedges.}
\label{fig:hypergraph}
\end{figure}

\noindent \textbf{Definition 2.2.4 Hypergraph.}Given a series of sessions $\mathcal{S}$, we could construct a hypergraph $\mathcal{G}=(\mathcal{V},\mathcal{E})$, where $\mathcal{V}$ is a node set, which consists of all the unique items $\mathcal{I}$. We further define two adjacent metrics $\mathbf{A}^{n2e}\in \mathbb{R}^{M\times N}$ and $\mathbf{A}^{e2e}\in \mathbb{R}^{M\times M}$, where $M$ and $N$ are the number of hyperedges and items, respectively. Thus, if item $i$ belongs to the hyperedge $j$, we set the element $a_{ij}^{n2e}=1$, otherwise $a_{ij}^{n2e}=0$. In addition, if hyperedge $j$ and hyperedge $k$ share the same items, we set the element $a_{jk}^{e2e}=1$, otherwise $a_{jk}^{e2e}$=0.  Based on the hypergraph, the high-order information in sessions can be captured efficiently. As shown in Figure~\ref{fig:hypergraph}, the hyperedge $e\in \mathcal{E}$ can be defined by: (1) the nodes sharing the same values of attributes (\textit{i.e.}, hypergraph with attributes)~\cite{lai2022attribute}; (2) the items from the same session~\cite{xia2021self} (\textit{i.e.}, hypergraph with session); (3) the item and its incoming items~\cite{li2022enhancing} (\textit{i.e.}, hypergraph with incoming items); (4) belonging to a specific contextual window~\cite{wang2021session} (\textit{i.e.}, hypergraph with slide windows); (5) a specific consecutive intent unit~\cite{guo2022learning}. For instance, in~\cite{zhang2022price} the hyperedge is defined based on the item side information, like item prices. (\textit{i.e.}, hypergraph with side information).

\noindent \textbf{Definition 2.2.5 Heterogeneous graph.} Given a graph $\mathcal{G}=(\mathcal{V, E})$, it consist of a node set $\mathcal{V}$ and an edge set $\mathcal{E}$. We define a node type mapping function: $\phi:\mathcal{V}\rightarrow \mathcal{M}_{node}$ and an edge type mapping function $\psi:\mathcal{E} \rightarrow \mathcal{M}_{edge}$. Denote $\mathcal{M}_{node}$ and $\mathcal{M}_{edge}$ are the sets of node types and edge types. Thus, if $|\mathcal{M}_{node}|+|\mathcal{M}_{edge}|>2$, we define the graph $\mathcal{G}$ as a heterogeneous graph. Attribute to the various types of nodes and edges, the heterogeneous graph is capable of modeling more complicated structures against the homogeneous graph~\cite{wang2019heterogeneous}. In SR, the user-item session graph~\cite{pang2022heterogeneous,chen2021efficient} and the item-attribute knowledge graph~\cite{zhang2021knowledge,meng2020incorporating} are the most commonly heterogeneous graph structures, which can be elucidated below:

\eat{
\begin{figure*}[t]
\centerline{\includegraphics[width=0.9\textwidth]
{images/socialgraph.pdf}
}
\caption{The user-item social session graph.}
\label{fig:user_item}
\end{figure*}
}

\begin{figure*}[t]
\centerline{\includegraphics[width=0.95\textwidth]
{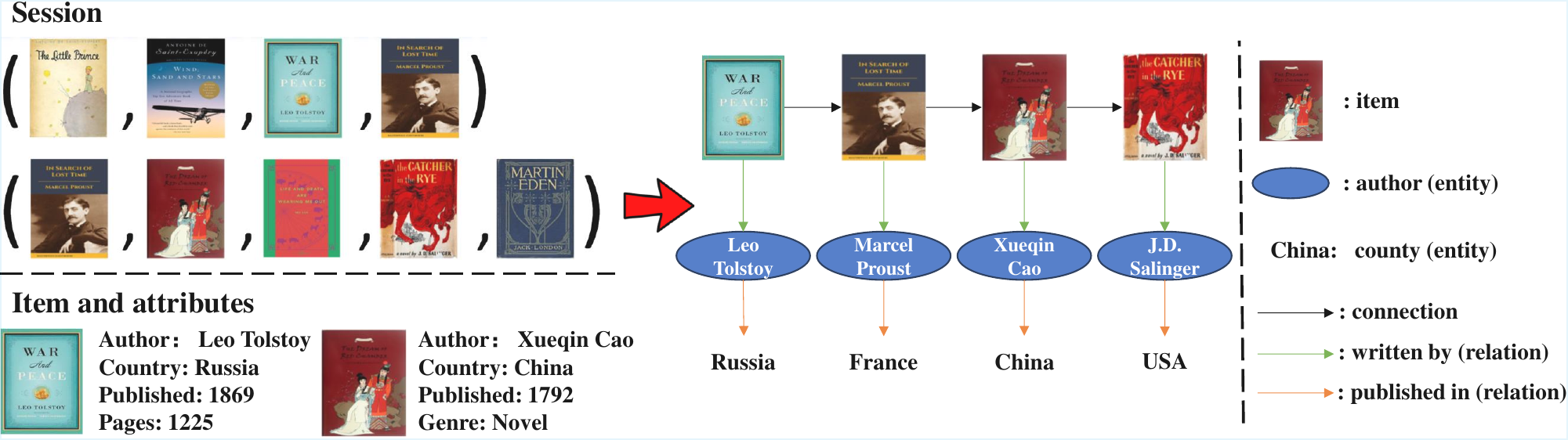}
}
\caption{\textcolor{black}{The item-attributes knowledge graph.}}
\label{fig:KG}
\end{figure*}

\begin{itemize}[leftmargin=*]
    \item \textbf{User-Item Social Graph.} Given a user set $\mathcal{U}$ and the historical interacted sessions of each user, we could construct a user-item session graph $\mathcal{G}=(\mathcal{I},\mathcal{U},\mathcal{E})$ based on user-item interacted records and the user's social network, where $\mathcal{I}$, $\mathcal{U}$ are item nodes and user nodes, respectively. In~\cite{pang2022heterogeneous}, it contains two types of edges, \textit{i.e.}, \textit{item-to-item} and \textit{user-to-item}. Specifically, if item $i_i$ and item $i_j$ are neighbors in sessions, we add an edge between them. Thus, the item transaction relations can be captured via item-to-item edge. Besides, if there are interactive records between user $u$ and item $i$, we add an edge between them. Consequently, the user's historical interests can be modeled via user-item edges. Except for the above two types of edges in a user-item session graph, the social relations are also considered in~\cite{chen2021efficient,wang2022self}. Specifically, if user $u_i$ and $u_j$ are friends or there exists some kind of interactive behavior (\textit{i.e.}, follow, shared, etc) between them, a user-to-user edge can also be added in the graph $\mathcal{G}$. 
    Moreover, if each session is regarded as a node, the user-to-session edge and item-to-session edge can also be created as an extension of the general user-item social graph~\cite{chen2021session},.
    \item \textbf{Item-Attributes Knowledge Graph.}  Given a session set $\mathcal{S}$, an item set $\mathcal{I}$ and its relevant attributes set $\mathcal{A}$, we could construct an item-attributed knowledge graph $\mathcal{G}=(\mathcal{I, A, E})$, which contains two types of nodes (item node and attribute node) and two types of edges (item-to-item and item-to-attribute). 
    If there is an ordered tuple $(i_i, i_j)$, we add an edge from item $i_i$ to $i_j$. Furthermore, if item $i_k$ contains the attribute $a_i$, we add an edge from $i_k$ to $a_i$ with the correspondent relation. As shown in Figure~\ref{fig:KG}, we could construct a knowledge graph for a book recommendation scenario. The entities/nodes contain the book name and the corresponding attributes, \textit{e.g.}, author, country, and genre. Apart from that, we could also define two relations, \textit{i.e.}, \textit{published in} and \textit{written by}, and add edges from items to attributes.
    \item \textbf{User-Behavior Session Graph}. Compared with the inter-session graph or the intra-session graph, the user-behavior session graph further defines the behavior relations, e.g, buy, click, between two adjacent items~\cite{shen2021multi,wang2020beyond}.
    \item \textbf{Spatialtemporal Session Graph}. For some online service platforms, \textit{e.g.}, Meituan and Yelp, both the user’s location as well as time information are significant for the precise recommendation. Consequently, the authors~\cite{li2022spatiotemporal} propose a spatiotemporal graph, which contains item, session, location, and time, four types of nodes, for session-based recommendation.    
\end{itemize}
 
Overall, the connectivity and structure among the above graphs 
can be formalized as an adjacency matrix $\mathbf{A}\in\mathbb{R}^{N\times N}$ with $\mathbf{A}_{ij}\neq 0$ iff $(v_i,v_j)\in \mathcal{E}$ and $\mathbf{A}_{ij}=0$ iff $(v_i,v_j) \notin \mathcal{E}$, where $N$ is the total number of nodes. $\mathcal{N}_i$ indicates the neighbors of node $v_i$ 
. In many works, the adjacency matrix $\mathbf{A}$ will be normalized as the formula: $\mathbf{\Tilde{A}}=\mathbf{A_{i,j}}/ \sum_j \mathbf{A_{i j}}$.

\section{Features and Categorization of SR Approaches}
\label{sec:characteristic}
In this section, we will first summarize the features of SR. Then, a hierarchical categorization will be provided for different research lines organization.

\subsection{The Features of SR}
\label{sec:features_sr}
According to the aforementioned definitions, we summarize the features of SR as below.

\begin{itemize}[leftmargin=*]
    \item \textbf{Session Length.} Compared with the sequential recommendation which organizes the user's whole historical interaction records in chronological order, SR only concentrates on the current ongoing sessions, thus, the length of sessions is quite limited, \textit{i.e.}, the median length of sessions is less than six for most popular public datasets. 
    
    

    \item \textbf{Dynamic and Timely Recommendation.} In general, different from sequential recommendation, SR specializes in the current session. 
    Therefore, it endeavors to model a user's short-term interests instead of the evolutionary process of the user's interests modeling in sequential recommendation. The dynamic and timely recommendation for the ongoing session is the aim of SR.
    \item \textbf{Adjacent Dependency.}
    Although the items in sessions are organized chronologically, there are no obvious order patterns~\cite{hu2017diversifying,wang2018attention, covington2016deep}. Consequently, many works~\cite{xia2021self,guo2022learning,zheng2020dgtn,xu2019graph,huang2021graph} apply GNNs to model the co-occurrence between two items.
    \item \textbf{Beyond Item Information.} Early attempts only adopt the ID as the identification of the item for recommendation. Currently, it becomes ubiquitous to fuse other external information,  \textit{e.g.}, item attributes, and interaction behaviors, to improve the explainability and performance of recommendations~\cite{cui2022intention,meng2020incorporating,liu2020keywords,qiu2020gag,wu2017session} (ref. Section~\ref{sec:embedding} for detail).
    \item \textbf{Anonymous and Non-anonymous.} As we discussed above, SR pays more attention to the current session modeling. Nevertheless, some papers~\cite{chen2021efficient,pang2022heterogeneous,song2019session,qiu2020gag} propose personalized session-based recommendation (PSR) or session-based social recommendation (SSR), which consider the user historical records are non-anonymous. Hence, they utilize a user's whole historical sessions as auxiliary information to improve the performance of current session recommendation.
\end{itemize}

In summary, we could find that the length of sessions is rather limited and there is no obvious dependency between two items in a session, in spite of the items organized in chronological order. Hence, most existing works focus on item correlation modeling with GNNs for a dynamic and timely recommendation. Additionally, although some works apply users' social networks or historical records as the auxiliary information for SR, modeling the current session is still the mainstream of this research venue. Moreover, more and more works emphasize the significance and efficiency of external information for recommendation. Hence, a surge of sophisticated model architectures emerges for side-information fusion. 

\subsection{Classification of SR Methods}
\label{sec:categorizes}

\begin{figure*}[t]
\centerline{\includegraphics[width=\textwidth]{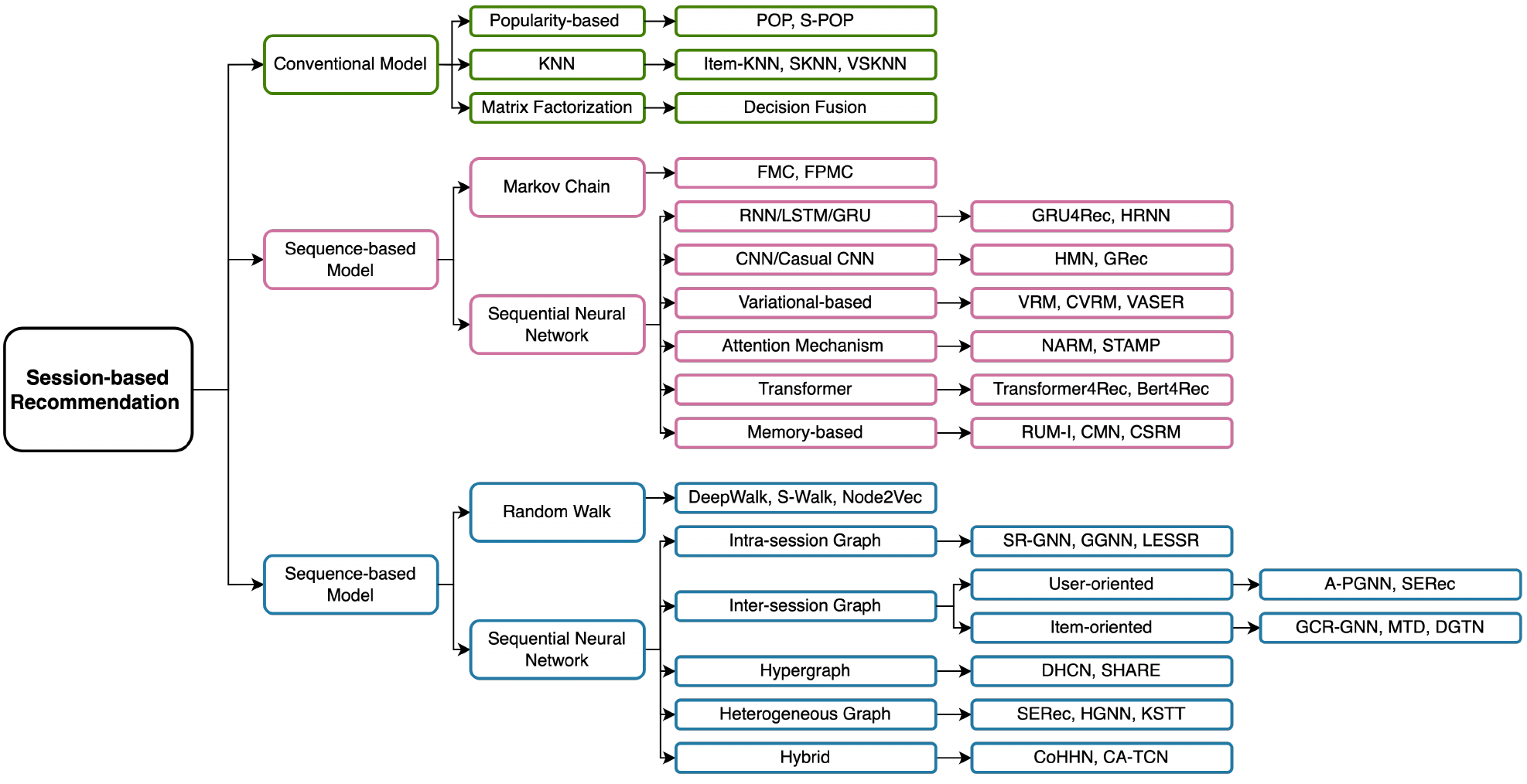}}
\caption{The categorization of SR approaches.  
The gray boxes are representative models for each class.}
\label{fig:category}
\end{figure*}

The taxonomy of existing SR methods is presented in Figure~\ref{fig:category}. In this paper, we summarized the representative solutions as three main categories, \textit{i.e.}, \textit{conventional methods}, \textit{sequence-based methods}, and \textit{graph-based methods}. The conventional methods can be further divided into \textit{popularity-based methods}, KNN, and matrix factorize. To be specific, the idea of \textit{popularity-based methods}, \textit{e.g.}, POP, S-POP, is to recommend the popular items to users while ignoring the cold start items. As the extension of popularity-based methods, frequent pattern or association rule mining approaches, \textit{e.g.}, FP-tree, are also applied, which mine the frequent items and frequent patterns from raw data for recommendation~\cite{forsati2009web,mobasher2001effective,moreno2004using,niranjan2010developing,shao2009music,yap2012effective}. As for KNN-based methods~\cite{linden2003amazon,davidson2010youtube,gharahighehi2020making,ludewig2018evaluation,ludewig2018evaluation}, they rely on similarity calculation for recommendation, which can also be divided into item-oriented KNN and session-oriented KNN. The item-oriented KNN measures the similarity between a target item and candidate items and then recommends the closest (\textit{i.e.}, similar) items to users. In contrast, the session-oriented KNN first selects the most similar sessions via similarity calculation, thus, the candidate items from those sessions are collected for further recommendation. 
Matrix factorization is also a mature method for SR, in which a user's latent feature representation and an item's latent feature representation are learned for the user-item interaction matrix prediction~\cite{rendle2012bpr,lian2013collaborative,liang2016factorization}. 

Apart from that, Markov Chain is also a prominent recipe, which recognizes the next interaction prediction as a Markov Decision Process (MDPs) and learns a state transfer matrix for recommendation~\cite{shani2005mdp,zimdars2001using,chen2012playlist,eirinaki2005web,feng2015personalized,le2016modeling,rendle2010factorizing,wu2013personalized,zhang2007efficient,singha2023scalable,wu2023generic}. For instance, 
FMC~\cite{shani2005mdp} is a pioneering work that extracts sequential patterns to predict the next item based on Markov models. 
FPMC~\cite{rendle2010factorizing} models sequential behavior between every two adjacent items via the personalized probability transition matrix factorization for recommendation.
REKS~\cite{wu2023generic} combines the MDP with a knowledge graph to generate a precise recommendation result and also provide an explanation simultaneously.
Although the Markov chain achieved remarkable success for SR in the early years, the strong assumption that the next clicked item only depends on the previous one confines the development of this method.

Attribute to the powerful feature representation ability of deep learning, sequential neural networks, \textit{e.g.}, RNN, LSTM, and GRU, are carried out consecutively for SR~\cite{hidasi2015session,quadrana2017personalizing,tan2016improved,wang2019collaborative}. For instance, GRU4REC~\cite{hidasi2015session} is the first that applied RNNs to model the session information for the next item recommendation. To be specific, it stacks multiple GRU layers and applies a session-parallel mini-batch training strategy for performance improvements. HRNN~\cite{quadrana2017personalizing} develops hierarchical RNNs with inter-session information to realize personalizing session-based recommendation. As a follow-up study~\cite{tan2016improved}, Tan \textit{et al.} enhance RNNs by leveraging data augmentation. 
Hidasi and Karatzoglou~\cite{hidasi2018recurrent} propose a new class of loss functions combined with modified sampling strategies to improve the performance of SR. Guo \textit{et al.}~\cite{guo2020session} propose Hierarchical Leaping Network (HLN) with Leap Recurrent Unit (LRU) to decide whether the current item should be skipped or not.
Being endowed with the property of global information modeling, the attention mechanism is also applied for SR. NARM~\cite{li2017neural} applies a hybrid encoder with an attention mechanism to model the user's sequential behavior and capture the main intentions in the current session. 
Liu \textit{et al.}~\cite{liu2018stamp} take a similar idea but replace the recurrent neural network with multi-layer perception (MLP) and propose STAMP to enhance the influence of the latest interests in sessions for both long-term and short-term interests capture. 
Furthermore, CNN and Causal CNN are also proposed to capture n-gram multi-scale features for item representation and recommendation~\cite{song2019session_1,yuan2020future}.
Attribute to the remarkable performance achieved by Transformer~\cite{vaswani2017attention} and BERT~\cite{devlin2018bert} in many NLP tasks, some researchers also elaborate on Transformer for SR, \textit{e.g.}, Transformer4Rec~\cite{de2021transformers4rec} and BERT4Rec~\cite{sun2019bert4rec}. Other works, like memory neural networks~\cite{wang2019collaborative,chen2018sequential,ebesu2018collaborative} and Variational Encoder (VAE) are also explored. For example, in~\cite{zhou2019variational,wang2018variational}, the latent variable module is introduced into sequential neural networks to improve the variation of recommendation. To sum up, we categorize the existing sequential neural networks with six aspects, as shown in Table~\ref{tab:Sequential}.

\begin{table}[]
\caption{\textcolor{black}{Comparison of representative sequential neural networks with respect to six aspects including motivation, session, sequential modeling, prediction, loss function, and datasets.
}}
\footnotesize
\begin{tabular}{l|l|l|l|l|l|l}
\hline
\hline
\multicolumn{1}{l|}{\textbf{Motivation}}           & \multicolumn{1}{l|}{\textbf{Session}}  & \multicolumn{1}{l|}{\begin{tabular}[l]{@{}l@{}}\textbf{Sequential}\\ \textbf{Modeling}\end{tabular}} & \multicolumn{1}{l|}{\textbf{Prediction}} & \multicolumn{1}{l|}{\textbf{Loss}} & \multicolumn{1}{l|}{\textbf{Datasets}} & \multicolumn{1}{l}{\textbf{Paper}}                                                                  \\ \hline
Repeat Item                               & Current                                                                          & GRU, Att                                                                           & Inner production+softmax        & CE                        & Dig, Yoo, FM                  & \cite{ren2019repeatnet}                                                   \\ \hline
RNN for SR                                & Current                                                                          & GRU                                                                                & MLP                             & TOP1                      & Yoo, VIDEO                   & \cite{hidasi2015session}                                              \\ \hline
BERT for SR                               & Current                                                                          & Att                                                                                & Inner production+softmax        & CE                        & Mov, Steam, Amz               & \cite{sun2019bert4rec}                                                \\ \hline
\multirow{2}{*}{Vanilla Attention for SR}         & Current                                                                          & Att, GRU                                                                           & Inner production+softmax        & CE                        & Retail, Yoo                   & \cite{li2017neural}                                                 \\
                                          & Current                                                                          & Att                                                                                & Inner production+softmax        & CE                        & Mov, Amz                      & \cite{kang2018self}               \\ \hline
\multirow{2}{*}{Multi Interests}     & Current                                                                          & Variant GRU                                                                        & Inner production+softmax        & CE                        & Yoo, FM                       & \cite{guo2020session}                                               \\
                                          & Current                                                                          & CNN, Att                                                                           & Inner production+softmax        & CE                        & Dig, Yoo                      & \cite{song2019session_1}                                             \\ \hline
\multirow{4}{*}{Long-term and Short-term} & User                                                                             & GRU, Att, Gate                                                                     & Inner production+softmax        & CE                        & Tmall, Oth                    & \cite{chen2019dynamic}                                                 \\
                                          & Current                                                                          & Att                                                                                & Inner production+softmax        & CE                        & Yoo, Oth                      & \cite{de2021transformers4rec}                                           \\
                                          & Current                                                                          & Att                                                                                & Inner production+softmax        & CE                        & Dig, Retail                   & \cite{yuan2021dual}                                                   \\
                                          & Current                                                                          & Att                                                                                & Inner production+softmax        & CE                        & Dig, Yoo                      & \cite{liu2018stamp}                                                     \\ \hline
\multirow{5}{*}{Loss Function}   & Current                                                                          & GRU                                                                                & MLP                             & Ranking-max               & Yoo, VIDEO, Oth              & \cite{hidasi2018recurrent}                                             \\
                                          & Current                                                                          & Att                                                                                & Cosine+softmax                  & RDM                       & Dig, Yoo, Tmall, Now          & \cite{hou2022core}                                            \\
                                          & Current                                                                          & MLP                                                                                & Inner production+softmax        & List-wise Ranking         & Dig                           & \cite{wu2017session}                                                 \\ 
                                          & Current                                                                          & Att                                                                                & Inner production+softmax        & List-wise Ranking         & Dig, Yoo, Oth                           & \cite{wilm2023scaling}                                                 \\ 
                                           & Current                                                                          & Att                                                                                & Inner production+softmax        & Adaptive weight CE         & Tmall, Retail                           & \cite{ouyang2023mining}                                                 \\ \hline
\multirow{9}{*}{Neighbor Sessions}         & User                                                                             & GRU                                                                                & MLP                             & TOP1                      & Xing, VIDEO                   & \cite{quadrana2017personalizing}                                          \\
                                          & User                                                                             & MLP                                                                                & Inner production+softmax        & CE                        & Tmall                         & \cite{hu2017diversifying}                                               \\
                                          & User                                                                             & GRU, Att, MF                                                                       & Inner production+softmax        & CE                        & Gow, FM                       & \cite{guo2019streaming}                                                \\
                                          & Sharing                                                                              & GRU, KNN                                                                           & MLP                             & CE                        & Yoo                          & \cite{jannach2017recurrent}                                              \\
                                          & Time Close                                                                       & GRU, Att, Gate                                                                     & Inner production+softmax        & CE                        & Yoo, FM                       & \cite{wang2019collaborative}                                            \\
                                          & Sim (Jaccard)                                                                    & Att                                                                                & Inner production+softmax        & CE                        & Retail, Yoo                   & \cite{luo2020collaborative}                                             \\
& Sim (Cosine)                                                                     & GRU                                                                                & Inner production+softmax        & CE                        & Dig, Yoo                      & \cite{pan2020intent}                                                   \\

& Sim (Inner Product)                                                                     & RNN, Att                                                                                & Inner production+softmax        & CE+BPR                        & Dig, Yoo                      & \cite{wei2022gsl4rec}\\                                        
                                          \hline
External Info (Position)                  & Current                                                                          & Att                                                                                & Inner production+softmax        & CE                        & Mov, Steam                    & \cite{seol2022exploiting}                                               \\
External Info (Attribute)                 & Current                                                                          & Att                                                                                & Inner production+softmax        & CE                        & Amz, Oth                      & \cite{xie2022decoupled}                                                \\
External Info (Attribute)                 & Current                                                                          & Att                                                                                & Inner production+softmax        & CE                        & Dig, Oth                      & \cite{shalaby2022m2trec}                                                \\
External Info (Attribute)                 & Current                                                                          & Att                                                                                & Inner production+softmax        & CE                        & Amz, Oth                      & \cite{jagatap2023attribert}                                                \\
External Info (Future Interaction)        & Current                                                                          & CausalCNN                                                                          & Inner production+softmax        & CE                        & Mov, Oth                      & \cite{yuan2020future}                                             \\
External Info (Key Words)                 & Current                                                                          & GRU                                                                                & Inner production+softmax        & CE                        & Oth                           & \cite{liu2020keywords}  \\
External Info (Description) & Current & GRU & MLP & CE & Mov & \cite{potter2022gru4recbe} \\
External Info (Social Relations) & Current & Att & Inner production+softmax & CE & Gow, Oth & \cite{ouyang-etal-2022-social}
\\ \hline\hline

\multicolumn{7}{l}{\begin{tabular}[c]{@{}l@{}}
Illustration of Abbreviations:\\
\textbf{(1) Motivation:}\\ 
\textbf{a)} Repeat Item: There is a certain probability that the target item will appear in the current session. Hence, a dedicated module is adopted for repeat item prediction.\\
\textbf{b)} RNN/Attention/BERT for SR: The pioneering works that apply these models for SR.\\ 
\textbf{c)} External Info: External information (\textit{e.g.}, position, attributes of items, interaction information, keywords of items, text description, users' social information)\\ is introduced as auxiliary signals for SR.\\ 
\textbf{d)} Multi Interests: As the user's interests are dynamic and diverse, a single and fixed interest representation vector is insufficient for SR. Therefore, \\multi-interest representations are proposed.\\ 
\textbf{e)} Long-term and Short-term: User's long-term and short-term interests should be captured simultaneously for fine-grained intention recognition.\\ 
\textbf{f)} Loss Function: improving loss functions for SR.\\ 
\textbf{g)} Neighbor Sessions: Similar sessions (\textit{e.g.}, user all historical sessions) are introduced for SR argumentation.\\
\textbf{(2) Neighbor Session:}\\
\textbf{a)} Sharing: the sessions sharing the same items with the current session are considered as neighbor sessions.\\ 
\textbf{b)} User: a user's all historical sessions are selected as neighbor sessions.
\textbf{c)} Current: only consider the current session for SR.\\
\textbf{d)} Sim: calculate the similarity between sessions for neighbor session selection.\\ 
\textbf{e)} Time Close: slice the timeline with the close principle for neighbor session selection.\\
\textbf{(3) Sequential Modeling:}\\ 
\textbf{a)} Att: variant attention mechanisms.
\textbf{b)} Gate: Gate mechanism.
\textbf{c)} MF: Matrix factorization.\\
\textbf{(4)Datasets}:\\ 
\textbf{a)} Dig: Diginetica.
\textbf{b)} Yoo: Yoochoose (Yoochoose 1/4, Yoochoose 1/64 are the most common for SR).
\textbf{c)} Gow: Gowalla. 
\textbf{d)} FM: Last.FM. 
\textbf{e)} Retail: Retailrocket.\\ 
\textbf{f)} Now: Nowplaying.
\textbf{g)} Amz: Amazon.
\textbf{h)} Mov: MovieLens.
\textbf{i)} Oth: some other less common datasets in SR, \textit{e.g.}, Tianchi, G1 news, CLASS, Delicious, OTTO.
\end{tabular}}
\end{tabular}

\label{tab:Sequential}
\end{table}

Although sequential methods achieve satisfactory performance in SR, they endeavor to the session's sequential information learning while ignoring the implicit dependency between items. 
Consequently, graph-based methods, a more flexible solution for transition pattern modeling, are accommodated for SR. 
Owing to the difference in item representation learning, graph-based methods can further be divided into random walk and GNNs. For random walk~\cite{yang2023loam,perozzi2014deepwalk,grover2016node2vec}, in general, given a series of sessions, an item-item adjacent graph is constructed first. Then, different random walk strategies are adopted with unsupervised learning to generate item representations. For instance, DeepWalk~\cite{perozzi2014deepwalk} empirically learns a low-rank transformation of a normalized Laplacian matrix for recommendation. 
Node2vec~\cite{grover2016node2vec} learns node representations based on the word2vec model for recommendation. 
To balance both the accuracy and scalability of SR, S-Walk~\cite{choi2022s} proposes a random walk with a restart strategy to capture inter-session and intra-session relations. 
In contrast, attributed to the process of information propagation and aggregation, GNNs are able to capture multi-hop contextual information between items for representation learning. 
Given that, GNNs demonstrate great superiority against sequential neural networks. As introduced in \textbf{Section}~\ref{sec:graph_define}, based on the variation of graph structures, they can be divided into intra-session graphs, inter-session graphs, hypergraphs, heterogeneous graphs, and hybrids.
A carefully selected representative and state-of-the-art GNN-based method for SR is presented in Tabel~\ref{tab:GNN}.
In the following sections, we will present a comprehensive framework and summarize the technical details of GNN-based methods and sequential neural networks in SR.




\begin{table}[]
\caption{\textcolor{black}{Summarization of representative GNN-based works with motivation, session, graph construction, item representation, session representation, loss function, and datasets.
}}
\footnotesize
\tabcolsep=0.1cm
\begin{tabular}{l|l|l|l|l|l|l|l}
\hline
\hline
\multicolumn{1}{l}{\textbf{Motivation}}                    & \multicolumn{1}{|l}{\textbf{Session}} & \multicolumn{1}{|l}{\textbf{Graph}}                &\multicolumn{1}{|l}{\textbf{Item Rep.}}    & \multicolumn{1}{|l}{\textbf{Session Rep.}}   & \multicolumn{1}{|l}{\textbf{Loss}} & \multicolumn{1}{|l|}{\textbf{Datasets}} & \textbf{Paper}                                                          \\ \hline 
GNN for SR                                        & Sharing                     & Intra                                    & GCN, GRU                                                                              & Att                                                                                  & CE                       & Dig, Yoo                     & ~\cite{wu2019session}                    \\ \hline
\multirow{2}{*}{Historical Info}           & User                        & Intra                                    & GCN,GRU, Att                                                                          & Att                                                                                  & CE                       & Xing, Reddit                 & ~\cite{zhang2020personalized}            \\
                                                  & User                        & Inter, Intra                             & GCN, GGNN                                                                             & Att                                                                                  & CE                       & Dig, Retail, Oth             & ~\cite{zhou2021temporal}                 \\ \hline
\multirow{5}{*}{User and Social Info}      & User                        & Het-Social                               & BiLSTM, GAT                                                                           & Att                                                                                  & CE                       & Dig, Tmall                   & ~\cite{chen2021session}                  \\
                                                  & User                        & Het-Social                               & Att, AvgP, Gate                                                                       & Att                                                                                  & CE                       & Gow, Oth                     & ~\cite{chen2021efficient}                \\
                                                  & User                        & Het-Social                               & GAT                                                                                   & Att, Gate                                                                            & CE                       & FM, Xing, Reddit             & ~\cite{pang2022heterogeneous}            \\
& User\&Friends               & Het-Social                               & LSTM, RNN, Att                                                                        & Att, AvgP, Gate                                                                                 & CE                       & Oth                    & ~\cite{song2019session}                  \\ 
& User\&Friends               & Het-Social                               & GCN, GRU, Att                                                                        & Att,                                                                                  & CE, Oth                       & Gow, Oth                    & ~\cite{wang2022self}                  \\ 
\hline
\multirow{18}{*}{High-order Connection}           & Sim (Duplicate)             & Inter, Intra                             & GCN                                                                                   & Att                                                                                  & CE                       & Dig, Yoo                     & ~\cite{zheng2020dgtn}                    \\
& Sim (Cosine)             & Inter, Hyper                             & GCN, GRU                                                                                   & Att                                                                                  & CE                       & Dig, Yoo, Tmall                     & ~\cite{jia2023smone}                    \\
 & Sim (Last item)             & Inter, Intra                             & GCN, GRU                                                                              & Att                                                                                  & CE                       & Dig, Yoo                     & ~\cite{chen2021incorporating}            \\
& Current                     & Hyper                                    & Att                                                                                   & Att                                                                                  & CE                       & Dig, Yoo, FM                 & ~\cite{wang2021session, guo2022learning} \\
& Current                     & Variant Intra                            & GCN, GRU, Att, Gate                                                                   & Att                                                                                  & CE                       & Dig, Yoo                     & ~\cite{pan2020star}                      \\
& Current                     & Variant Intra                            & GCN, GRU, Att                                                                         & Att                                                                                  & CE                       & Dig, Yoo, FM                 & ~\cite{chen2020handling}                 \\
& Current                     & Variant Intra                                    & GCN                                                                             & Att                                                                                  & CE                       & Dig, Gow, FM                  & ~\cite{yang2023multiple}                      \\
& Current                     & Intra                                    & GCN, GRU                                                                              & Att                                                                                  & CE                       & Dig, Retail                  & ~\cite{xu2019graph}                      \\
    & Current & Hyper & GAT & Att & CE, Oth & Tmall, FM &~\cite{li2022enhancing}    \\
    & Current & Hyper & GAT & Att & CE, Oth & Tmall, FM &~\cite{li2022enhancing}    \\
& Sharing                     & Inter                                    & GCN                                                                                   & AvgP                                                                                 & CE                       & Dig, Yoo, Retail             & ~\cite{huang2021graph}                   \\
& Sharing                     & Inter, Session                           & GCN, CausalCNN                                                                        & Att, GAT                                                                             & CE                       & Dig, Yoo                     & ~\cite{ye2020cross}                      \\
& Sharing                     & Hyper, Session                           & GCN                                                                                   & Att, AvgP                                                                            & CE, InfoNCE              & Dig, Tmall, Now              & ~\cite{xia2021self}                      \\
& Sharing                     & Inter                                    & GCN                                                                                   & Att                                                                                  & CE, InfoNCE              & Tmall, Retail, Dig              & ~\cite{xia2021self_1}                   \\
& Sharing                     & $\epsilon$-Neighbor        & Att                                                                                   & GRU                                                                                  & CE                       & Dig, Yoo, Retail             & ~\cite{zhang2022graph}                   \\
& Sharing                     & Intra, $\epsilon$-Neighbor & Routing, RW                                                                & GRU                                                                                  & CE                       & Dig, Yoo                     & ~\cite{zhang2021graph}                   \\
& Sharing                     & Intra, $\epsilon$-Neighbor & GAT                                                                                   & Att                                                                                  & CE                       & Dig, Tmall, Now              & ~\cite{wang2020global}                   \\ 
& Sharing                     & Intra, Inter & Sparse Att                                                                                   & Sparse Att                                                                                  & CE                      & Dig, Tmall, Now, Yoo              & ~\cite{qiao2023bi}                   \\ 
& Sharing                     & Intra, Inter & Att, GRU                                                                                   & Att                                                                                  & CE, InfoNCE, Oth                       & Dig, Tmall, Now              & ~\cite{su2023enhancing}                   \\ 
\hline
\multirow{5}{*}{External Info (Behavior)} & Current                     & Het-Behavior                            & Att                                                                                   & Att                                                                                  & CE                       & Dig, Gow, FM                 & ~\cite{zhu2022transition}                \\
& Current                     & Het-Behavior                            & AvgP                                                                                  & AvgP                                                                                 & CE                       & Yoo, Oth                     & ~\cite{wang2020beyond}                   \\
& Current                     & Het-Behavior                            & GCN, GRU, Att                                                                                  &  Gate                                                                                & CE                       & Oth                     & ~\cite{yuan2022micro}                   \\
& Current                     & Hyper                                    & GAT, GRU, Att                                                                         & Att                                                                                  & CE                       & Tmall, Yoo, FM               & ~\cite{shen2021multi}                    \\
\multirow{2}{*}{External Info (KG)}                          & Current                      & Het-KG                                   & GAT, TransR                                                                         & Att                                                                                  & CE, BPR              & Dig, Yoo                    & ~\cite{zhang2021knowledge}               \\
& All                      & Het-KG, Intra, Inter                                   & GAT                                                                          & Att                                                                                  & CE               & Amz, Yelp, Oth                     & ~\cite{chen2023knowledge}               \\
External Info (KG, Behavior)              & Current                     & Het-KG                                   & GCN, GRU, TransH                                                                      & Att                                                                                  & CE, BPR                  & Jdata                        & ~\cite{meng2020incorporating}            \\
\multirow{2}{*}{External Info (Att)} & Current & Variant Intra, Hyper & GAT & Att & CE & Dig, Tmall, Oth & ~\cite{lai2022attribute} \\
& Current & Het-Behavior,Het-Attribute & GAT & Att & CE & Amz, Oth & ~\cite{chen2023attribute} \\
External Info (Time, Location) & All & Het-Spatialtemporal & GAT, Att & Att & CE & Oth & ~\cite{li2022spatiotemporal} \\
External Info (Order)                      & Current                     & Intra                  & GAT, GRU                                                                            & Att                                                                            & CE                       & Dig, Yoo                &     ~\cite{qiu2019rethinking}                                                           \\
External Info (Price)                      & Sharing                     & Hyper, Het-Attribute                    & Att, Gate                                                                             & Att, Gate                                                                            & CE                       & Dig, Oth                &     ~\cite{zhang2021knowledge}                                                           \\ 
External Info (Other domain)                                 & Sharing                     & Intra, $\epsilon$-Neighbor & GAT, GRU                                                                              & Att                                                                                  & CE                       & Dig, Yoo, Gow, FM            & ~\cite{chen2021dual}                     \\ \hline
\multirow{3}{*}{Multi Interests}             & Current                     & Intra                                    & GGNN, Att, Gate                                                                       & Att                                                                                  & CE, Oth                  & Retail, Yoo, Jdata           & ~\cite{shen2021temporal}                 \\
                                                  & Current                     & Intra                                    & GCN, GRU                                                                              & Att                                                                                  & CE, Oth                  & Dig, Yoo, Now                & ~\cite{li2022disentangled}               \\
                                                  & Current                     & Intra                                    & GCN, GRU                                                                              & Att                                                                                  & CE                       & Dig, Yoo                     & ~\cite{yu2020tagnn}                      \\ \hline
\multirow{5}{*}{Efficiency}                         & Sharing                     & Inter, Variant Intra                     & RW                                                                          & RW                                                                         & CE, Oth                  & Dig, Retail, Oth             & ~\cite{choi2022s}                        \\
                                                  & Sharing                     & Inter, Intra                             & GCN, GRU, RW                                                                 & Att                                                                                  & CE                       & Dig, Yoo                     & ~\cite{deng2022g}                        \\
                                                         & Sharing                     & Variant Intra                             & GCN                                                               & Att                                                                                  & CE, InfoNCE                       & Tmall, FM, Retail                     & ~\cite{peintner2023spare}                        \\
                                                  & User                        & Intra                                    & GCN                                                                                   & Att                                                                                  & CE                       & FM, Gow                      & ~\cite{qiu2020gag}                       \\ 
                                                   & Current                        & Intra                                    & Att, GRU                                                                                   & Att                                                                                  & CE                       & FM, Gow, Dig                      & ~\cite{zhang2023efficiently}                       \\ 
                                                  \hline\hline
\multicolumn{8}{l}{\begin{tabular}[c]{@{}l@{}}
Illustration of Abbreviations:\\
\textbf{(1) Motivation:}\\ 
\textbf{a)} GNN for SR: A groundbreaking work that applies GNN for SR.\\
\textbf{b)} Historical Info, User and Social Info: Introducing neighbor sessions via different strategies \textit{e.g.}, social network, for SR.\\
\textbf{c)} External Info: Introducing external information or side-information, \textit{e.g.}, interaction behavior types, item's attributes, time and location, knowledge graph,\\ other domain data, and order information for SR.\\
\textbf{d)} Multi Interests: Construct multi-interest representations for each user and SR.\\
\textbf{e)} High-order Connection: One-hop connection is insufficient for SR. Consequently,  sophisticated graph structures (\textit{e.g.}, self-loop, short-cuts) are designed\\ for global information or multi-hops information modeling.\\
\textbf{f)} Efficiency: Focusing on low computational complexity and timely recommendation.\\
\textbf{(2) Session Selection:}\\ 
\textbf{a)} Sharing: the neighbor sessions which contain the same item as the current session. 
\textbf{b)} User: a user's all historical sessions are selected as neighbor sessions.\\ 
\textbf{c)} Current: only consider the current session for SR. 
\textbf{d)} Sim: calculate the similarity between two sessions or observe the last item in sessions\\ for neighbor session selection.
\textbf{e)} All: defines all the training sessions as neighbors.\\
\textbf{(3) Graph Construction:}\\ 
\textbf{a)} Intra: Intra-session graph.
\textbf{b)} Inter: Inter-session graph. 
\textbf{c)} Het-Social: Heterogeneous graph with social relations.\\ 
\textbf{d)} Het-Behavior: Heterogeneous graph with different interaction behavior types.
\textbf{e)} Het-KG: Heterogeneous graph with knowledge graph.\\ 
\textbf{f)} Het-Attribute: Heterogeneous graph with item correspondent attributes.
\textbf{g)} Het-Spatialtemporal: Heterogeneous graph with time and location.\\ 
\textbf{h)} Hyper: Hypergraph.
\textbf{i)} Session: Introduce the session as a node in graph construction.\\ 
\textbf{j)} $\epsilon$-Neighbor: construct a graph based on the $\epsilon$-Neighbor strategy. 
\textbf{k)} Variant Intra: intra-session graph with a virtual star or self-loop edge and short-cut edge.\\
\textbf{(4) Information Propagation and Session Representation:}\\ 
\textbf{a)} Att: various attention mechanisms.
\textbf{b)} AvgP: Average pooling.
\textbf{c)} Gate: Gate mechanism. 
\textbf{d)} RW: Random walk.\\
\textbf{(5) Loss Function:}\\ 
\textbf{a)} CE: Cross-entropy loss.\\ 
\textbf{b)} Oth: auxiliary loss functions for multi-task learning, e.g, link prediction, matrix learning, ELBO loss for item representation distribution reconstruction,\\ loss function with regularization term, etc.\\
\textbf{(6)Datasets}:\\ 
\textbf{a)} Dig: Diginetica. 
\textbf{b)} Yoo: Yoochoose (Yoochoose 1/4, Yoochoose 1/64 are the most common for SR).
\textbf{c)} Gow: Gowalla. 
\textbf{d)} FM: Last.FM. 
\textbf{e)} Retail: Retailrocket.\\ 
\textbf{f)} Now: Nowplaying.
\textbf{j)} Amz: Amazon.
\textbf{h)} Oth: other datasets in SR, \textit{e.g.}, Wechat, Cosmetics, Aotm, 30music, JD, Trivago, Delicious, Foursquare.
\end{tabular}}
\end{tabular}
\label{tab:GNN}
\end{table}

\section{Sequential and Graph Neural Networks for SR}
\label{sec:methodology}

As aforementioned in~\ref{sec:categorizes}, the solutions of SR main include GNN-based methods and sequential neural network-based methods. In this section, we will first give an overview of the pipelines with regard to those two mainstream attempts. Then, the key modules and the technical details are introduced respectively. 
Specifically, subsection~\ref{sec:session_selection}, \ref{sec:embedding} and \ref{sec:loss} are the key modules contained by both two frameworks, the key modules introduced in subsection~\ref{sec:sequence modeling} only belong to the sequential neural networks and the remain subsections are typical modules in GNN.


\eat{
\begin{figure*}[t]
\centerline{\includegraphics[width=\textwidth]{images/section.eps}}
\caption{The organization of subsections in section~\ref{sec:methodology}. The sections circled by blue boxes are shared by both sequential neural networks and GNNs. The sections circled by coral and green boxes will introduce the key modules in sequential neural networks and GNNs, respectively.}
\label{fig:sections}
\end{figure*}
}
\subsection{The Overall Framework of SR with Sequential Neural Networks}
\label{sec:SNN}

Aim to model the sequential information, the sequential neural networks, \textit{e.g.}, RNN, LSTM, GRU, Memory Neural networks, Transformer, and BERT, are proposed successively for SR. 
The completed pipeline of these works can be formalized in Figure~\ref{fig:sequenceNN}, which mainly includes the sequence modeling layer, the session representation layer, and the prediction layer. 
More concretely, given a sequence $s=[i_1,i_2,..,i_m]$ as the input of a sequential neural network, we first generate the embedding $\mathbf{X}$ of each item in $s$ which will undergo a sequential modeling layer $\textrm{SM}(\cdot)$, a session representation layer $\textrm{SR}(\cdot)$ and a prediction layer $\textrm{P}(\cdot)$ in order for the next item prediction. The whole process can be formalized as follows:



\begin{figure*}[t]
\centerline{\includegraphics[width=\textwidth]
{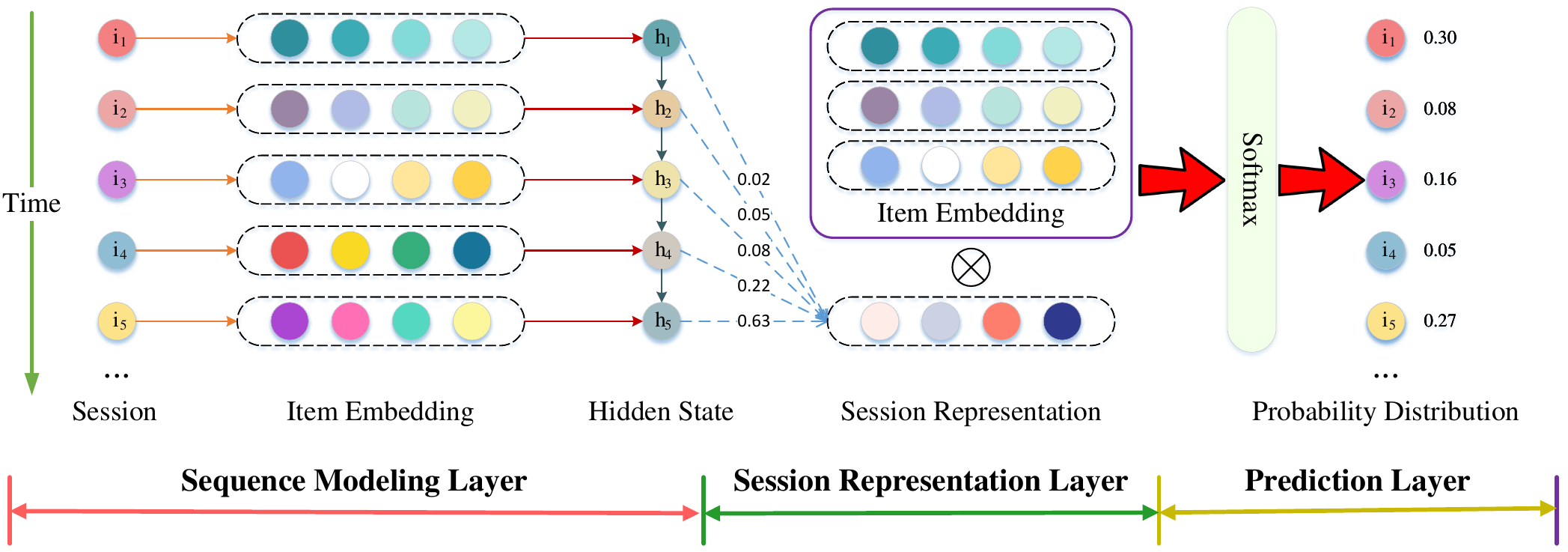}
}
\caption{The framework of sequential neural networks for SR.}
\label{fig:sequenceNN}
\end{figure*}

\begin{equation}
    \begin{split}
        \mathbf{H}&=\textrm{SM}(\mathbf{X})\\
        \mathbf{S}&=\textrm{SR}(\mathbf{H})\\
        \hat{y}&=\textrm{P}(\mathbf{S,X})
    \end{split}
    \label{eq:sequencenn}
\end{equation}
where $\textrm{SM}(\cdot)$ can be RNN, LSTM, GRU~\cite{chen2019dynamic,wang2019collaborative}, attention~\cite{de2021transformers4rec,chen2019dynamic} and memory neural network~\cite{wang2019collaborative} for sequential information modeling. 
$\textrm{SR}(\cdot)$ is the session representation layer, where soft-attention mechanism~\cite{chen2019dynamic,wang2019collaborative,de2021transformers4rec} are commonly used.
$\mathbf{H}$ and $\mathbf{S}$ are the items' representation and session representation respectively.
As for the prediction layer $\textrm{P}(\cdot)$, the inner product between session representation and item embedding is most used for prediction score generation.
After that, the loss function \textit{e.g.}, cross-entropy loss, TOP1-max loss, BPR-max loss, etc, is designed for model training and optimization.

\subsection{The Overall Framework of SR with GNN}
\label{sec:GNN}
\begin{figure*}[t]
\centerline{\includegraphics[width=\textwidth]
{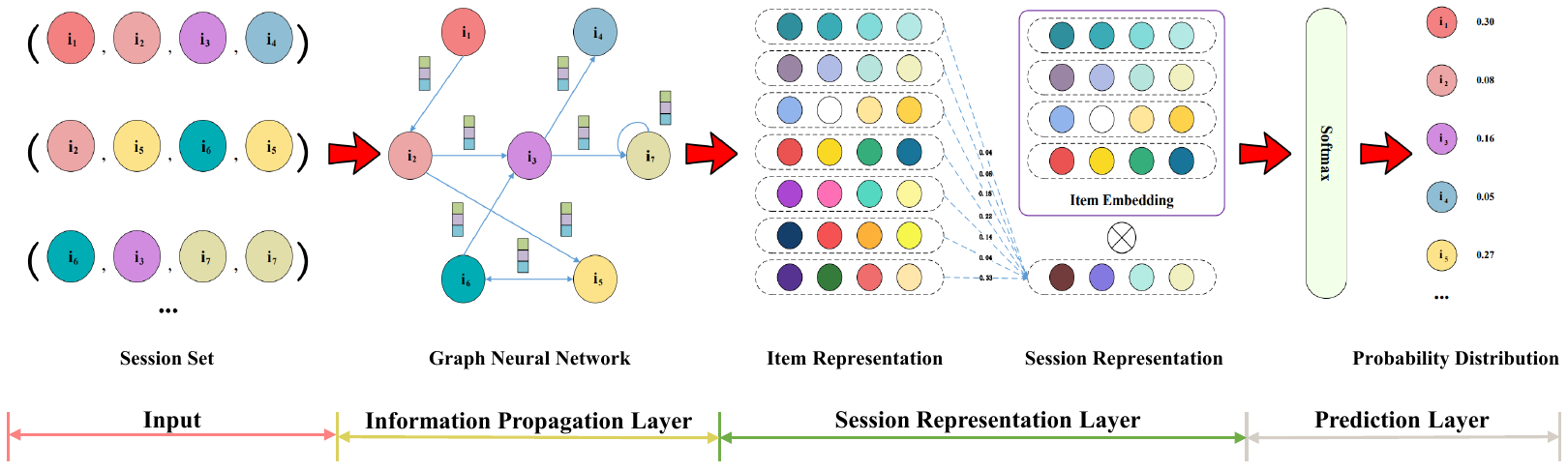}
}
\caption{The framework of GNNs for SR.}
\label{fig:SR-GNN}
\end{figure*}

Encouraged by the powerful expressive ability of complicated graph structure data modeling in many scenarios (\textit{e.g.}, social network, traffic prediction~\cite{jiang2021graph}, and molecular representation for drug discovery~\cite{li2017learning}), 
in 2019, Wu \textit{et al.}~\cite{wu2019session} proposed SR-GNN which is a pioneering work to apply GNN for SR. Since then, a plethora of GNN-based works emerged. In general, the whole process of GNN for SR includes five key modules: session selection, graph construction, information propagation and aggregation, session representation, and target item prediction, which are shown in Figure~\ref{fig:SR-GNN}.

Specifically, given a session set $\mathcal{S}$ or a session $s$, a graph $\mathcal{G}$ should be constructed first. Then, variants GNNs are designed to obtain the item representation via information propagation and aggregation. The item representation $\mathbf{H}$ will be fed into the session representation layer $\textrm{SR}(\cdot)$ for session representation generation. Finally, the predicted item probability $\hat{y}$ can be generated via inner production. This process can be formalized as below.

\begin{equation}
    \begin{split}
        \mathbf{H}&=\textrm{GNN}(\mathbf{X})\\
        \mathbf{S}&=\textrm{SR}(\mathbf{H})\\
        \hat{y}&=\textrm{P}(\mathbf{S,X})
    \end{split}
    \label{eq:GNN}
\end{equation}


\subsection{Neighbor Sessions Selection}
\label{sec:session_selection}

\begin{figure*}[t]
\centerline{\includegraphics[width=\textwidth]
{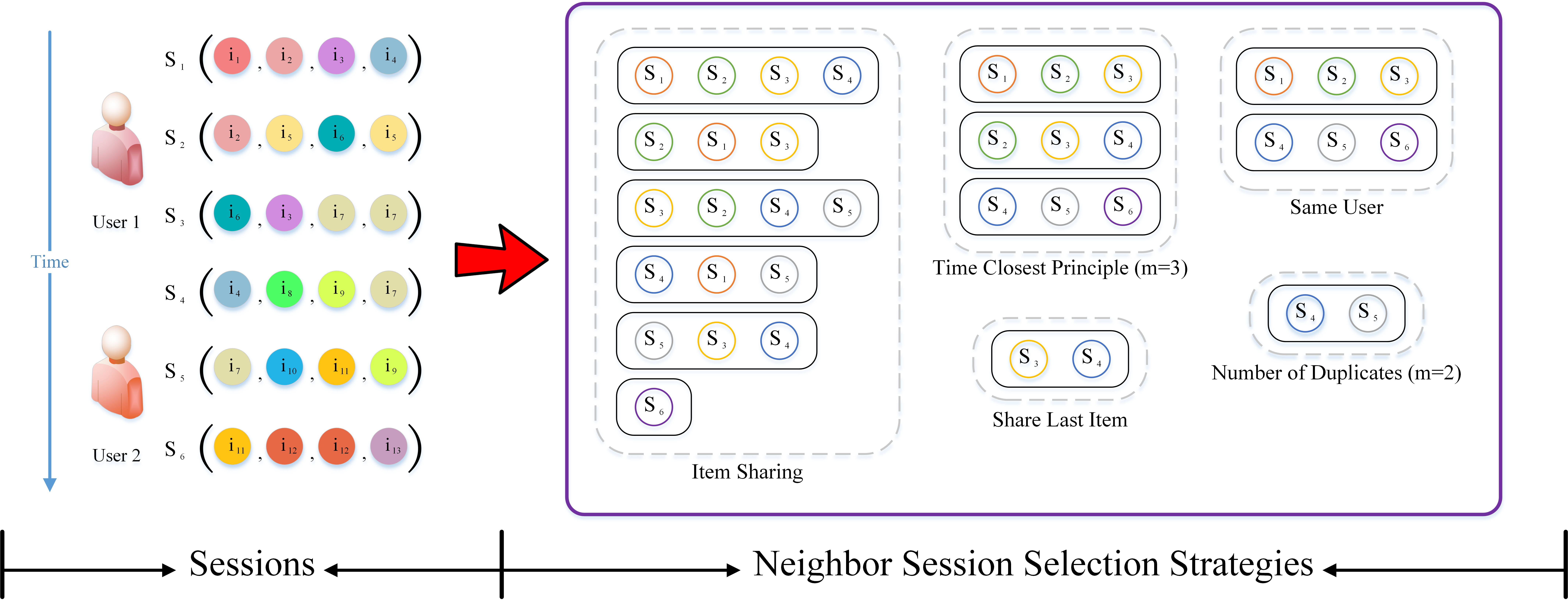}
}
\caption{The categorization of neighbor session selection strategy.}
\label{fig:session_selection_category}
\end{figure*}

Considering the ongoing session for SR is insufficient, as it can only reveal users' current intentions. Therefore, many works tend to collect neighbor sessions as auxiliary information to facilitate the performance of recommendation. Summarizing the relevant works with sequential neural networks and GNNs, neighbor session selection strategies can be divided into four categories (as shown in Figure~\ref{fig:session_selection_category}): adjacent-based, similarity-based, user-based and all together. For adjacent-based methods, they can be further split into space adjacent (\textit{i.e.}, different sessions contain the same items) and time adjacent (\textit{i.e.}, sessions are neighbors on the timeline). In terms of similarity-based methods, similarity calculation based on item ID and session representations are two main strategies. 

\eat{
\begin{figure*}[t]
\centerline{\includegraphics[width=\textwidth]
{images/sessions.pdf}
}
\caption{A toy example of different neighbor session selection strategies.}
\label{fig:session_selection}
\end{figure*}
}

\noindent \textbf{(1) Adjacent-based.} Given a current session $s$, the adjacent-based strategy considers the sessions that contain the same items with $s$ or adjacent to $s$ in the timeline should be neighbors. Specifically, it contains:

\begin{itemize}[leftmargin=*]
    \item \textbf{Item Sharing.} Given a current session $s$, we retrieve the sessions that contain the same item with $s$ as the neighbor sessions~\cite{zhang2022price,xia2021self,xia2021self_1}. 
    \item \textbf{Share Last Item.} In~\cite{chen2021incorporating}, the authors suppose the last clicked item $i_l$ reveals users' short-term or current interests. Thus, the sessions that contain $i_l$ as the current session are selected as neighbor sessions.
    \item \textbf{Time Closest Principle.} Given a current session, \cite{wang2019collaborative} defines the closest $m$ sessions in the timeline before the current session as neighborhood sessions.
\end{itemize}

\noindent \textbf{(2) Similarity-based.} Similarity-based strategies believe the sessions that contain similar patterns as the current session are neighbor sessions. Thus, it is of critical importance to define a proper similarity measurement.

\begin{itemize}[leftmargin=*]
    \item \textbf{Number of Duplicates.} Given the current session $s$, \cite{zheng2020dgtn} counts the number of the same items shared by both the current session and other sessions. Then, top-k sessions with the most same items are collected as the neighbor sessions of $s$.  
    \item \textbf{Binary Cosine Similarity.} Given a current session $s_i$, we denote the interacted item in the current session at time $t$ as $x_t$. First, the sessions that contain the item $x_t$ are searched as candidates $\mathcal{S}_M$. Then, based on the binary cosine similarity, top-$K$ most similar sessions are obtained as the final neighbor sessions~\cite{luo2020collaborative,bonnin2014automated}. The binary cosine similarity is defined as follows.
    \begin{equation}
        \textrm{sim}(s_{i}, s_j)=\frac{|s_{i}\cap s_j|}{\sqrt{|s_{i}|\cdot|s_j|}}\label{eq:jaccard_sim}
    \end{equation}
    where $s_j\in \mathcal{S}_M$. $|s_{i}\cap s_j|$ count the same items between $s_i$ and $s_j$. $|s_i|$ and $|s_j|$ are the total numbers of items in $s_i$ and $s_j$ respectively.
    \item \textbf{Cosine Similarity.} Instead of item ID for similarity calculation as above two strategies, in~\cite{pan2020intent,ye2020cross}, the similarity between sessions is calculated based on the session representations generated by a neural network. Denote $\mathbf{s_i}$ as the current session representation and $\mathbf{s_j}\in \mathcal{S}_M$ as the representation of session $s_j$, where $M$ is a memory block for session reserve. Thus, we could first obtain $K$ sessions in memory $M$ based on the FIFO (first-in-first-out) strategy as the candidate, then calculate their cosine similarities to $s_i$. Finally, top-k most similar sessions are selected as neighbor sessions. The cosine similarity is formalized as below:
    \begin{equation}
        \textrm{sim}(\mathbf{s}_i,\mathbf{s}_j)=\frac{\mathbf{s}_i\cdot\mathbf{s}_j^T}{||\mathbf{s}_i|| ||\mathbf{s}_j||}\label{eq:cosine}
    \end{equation}
    \item \textbf{Inner Production with Projection Layer}. Given any of two sessions $s_i$ and $s_j$, denote $\mathbf{s_i, s_j}$ are the representation of them. In~\cite{wei2022gsl4rec}, the authors calculate the similarity between $\mathbf{s}_i$ and $\mathbf{s}_j$ as follows.
    \begin{equation}
        \textrm{sim}(\mathbf{s}_i,\mathbf{s}_j) = \textrm{ReLU}(\textrm{tanh}(\mathbf{s}_i\cdot\mathbf{s}_j^T)).
    \end{equation}
    \item \textbf{Euclidean Distance with Gaussian Kernel.} Given any of two sessions $s_i$ and $s_j$, denote $\mathbf{s_i, s_j}$ are the representation of them. We calculate the Euclidean distance with the Gaussian kernel to measure the similarity between $s_i$ and $s_j$.
    \begin{equation}
        \textrm{sim}(\textbf{s}_i,\textbf{s}_j)=1-\textrm{exp}\left(\frac{-d_{ij}^2}{2d^{2}_*}\right)
    \end{equation}
    where $d_{ij}$ is the Euclidean distance between $\mathbf{s}_i$ and $\mathbf{s}_j$, \textit{i.e.}, $d_{ij}=||\mathbf{s}_i-\mathbf{s}_i||^2$. $d_*$ is the minimal Euclidean distance between $s_i$ and its neighbors.
\end{itemize}

\noindent\textbf{(3) User-based.} User-based methods can also be classified into two categories:

\begin{itemize}[leftmargin=*]
    \item \textbf{Self-based.} It considers all the historical sessions of this user as neighbor sessions, thus, the user's long-term or global preference~\cite{ranjbar2022fair} can be captured.
    \item \textbf{Self and social connections.} Not only the user's whole historical sessions will be attained, but the sessions from his or her friends are also selected as neighbor sessions~\cite{wang2022self}.
\end{itemize}

\noindent \textbf{(4) All.} It can also be divided as below:

\begin{itemize}[leftmargin=*]
    \item \textbf{All sessions.} In~\cite{li2022spatiotemporal}, all the training sessions are selected as neighbors to construct a global information graph for recommendation.
    \item \textbf{Sessions from the same batch.} Considering leverage all the sessions as neighbors for graph construction will be intractable in the practical scenario, to balance the efficiency and the precision, in~\cite{li2023exploiting}, the authors consider the sessions from the same batch for explicit and implicit graph construction. 
\end{itemize}

\subsection{Item and Side Information Embedding}
\label{sec:embedding}

\begin{figure*}[t]
\centerline{\includegraphics[width=0.9\textwidth]{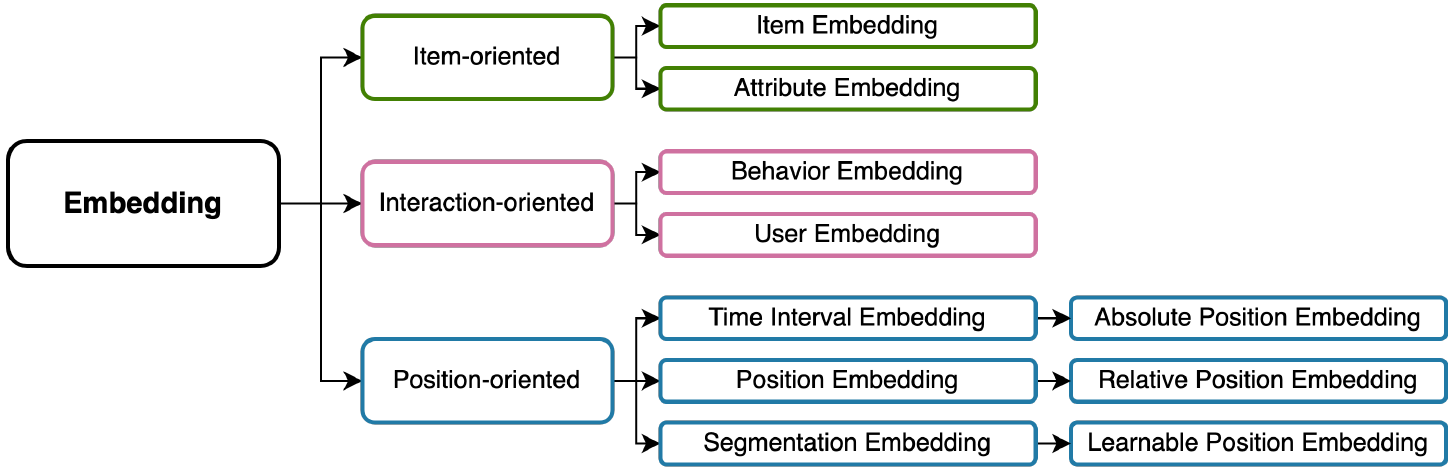}}
\caption{A categorization of various embeddings.}
\label{fig:embedding}
\end{figure*}

Summarizing existing works, existing embeddings can be divided into three categories: item-oriented (\textit{e.g.}, item embedding, attributes embedding), interaction-oriented (\textit{e.g.}, user embedding, interaction behaviors embedding), and position-oriented (\textit{e.g.}, positional embedding, session segment embedding), which are shown in Figure~\ref{fig:embedding}.

\begin{itemize}[leftmargin=*]
    \item \textbf{Item Embedding.} Inspired by word embedding in NLP~\cite{bengio2000neural,collobert2008unified,mikolov2013efficient,jagatap2023attribert}, we could also represent the item ID via an item embedding. Hence, each discrete item ID can be projected into a continuous high-dimensional latent space as the input of neural networks. 
    \item \textbf{Attribute Embedding.} As the side information could also reveal the characteristics of items that will benefit to SR.
    Consequently, some works also consider the items' corresponding attributes, \textit{e.g.}, category, brand, and the title of keywords, for embedding~\cite{cui2022intention,meng2020incorporating,liu2020keywords,shalaby2022m2trec}.
    \item \textbf{Item Description Embedding.} The text information, \textit{e.g.}, item description, title, and user's comments, contains fertile information about items' features and users' preferences. Thus, we could apply language models like BERT to generate an item description embedding~\cite{potter2022gru4recbe} or leverage a pre-trained Word2Vec word embedding for title and meta attributes encoding~\cite{gong2022positive} for a better recommendation. 
    \item \textbf{Interacted Behavior Embedding.} Users' micro-behaviors, \textit{e.g.}, reading comments, adding to a cart, etc, which reveal a fine-grained and deep understanding to the user’s preference. Therefore, a user's interacted behavior embedding can also be introduced for SR~\cite{meng2020incorporating,wu2017session,yuan2022micro}.
    \item \textbf{User Embedding.} To inject the user's identification information into the item embedding, in~\cite{qiu2020gag,wu2017session}, the item embedding concatenates the interacted user embedding for item representation initialization.
    \item \textbf{Time Interval Embedding.} In~\cite{zhang2021knowledge}, the authors believe that the time intervals between two adjacent behaviors could reveal the process of user's interest shift. Thus, the time interval embedding is created. 
    \item \textbf{Positional Embedding.} To supervise the model to capture the order information and the temporal effect of the input, positional embedding is proposed. It mainly includes (1) absolute positional encoding (modeling the item's absolute order in a session) which can be generated via sine and cosine functions~\cite{vaswani2017attention}; (2) relative positional encoding, which encodes the distance information between two items in an order. Different from the absolute encoding, the relative positional encoding can be translation-invariant and free of the length of sessions~\cite{shaw2018self, su2024roformer}; (3) learnable positional encoding, which is the same as item embedding, \textit{i.e.}, based on the position index for embedding generation~\cite{sun2019bert4rec,yuan2021dual}.
    \item \textbf{Session Segment Embedding.} Similar to the segment embedding in NLP to identify two adjacent sentences~\cite {conneau2017supervised}, session segment embedding can be used to indicate the ordinal position of the current session~\cite{seol2022exploiting} to the whole historical sequence.
\end{itemize}

To be specific, given an item $i$ and its correspondent auxiliary information, we could obtain the item initial representation $\mathbf{x}$ via the concatenation, addition operation or gate mechanism~\cite{liu2021non,seol2022exploiting,yuan2021dual,sun2019bert4rec}.

\begin{equation}
\begin{aligned}
    \begin{split}
     \mathbf{x}_i &= \mathbf{Concat}(\mathbf{x}_i^{\textrm{IE}}, \mathbf{x}_i^{\textrm{IOE}},\mathbf{x}_i^\textrm{{NOE}},\mathbf{x}_i^\textrm{{POE}}) \quad&  (\textrm{Concatenation})\\
     \mathbf{x}_i &= \mathbf{x}_i^\textrm{{IE}}+\mathbf{x}_i^\textrm{{IOE}}+\mathbf{x}_i^\textrm{{NOE}}+\mathbf{x}_i^\textrm{{POE}} \quad& (\textrm{Addition})\\
     \mathbf{x}_i &= \sum_{j\in \{\textrm{IE,IOE,NOE,POE}\}}\alpha^{j}\mathbf{x}_i^{j} \quad& (\textrm{Gate}) \\
     \alpha^{j} &= \sigma(\mathbf{x_i}^{j}\mathbf{W})
    \end{split}
    \label{eq:embedding}
\end{aligned}
\end{equation}

where $\mathbf{x}_i^\textrm{{IE}},\mathbf{x}_i^\textrm{{IOE}},\mathbf{x}_i^\textrm{{NOE}},\mathbf{x}_i^\textrm{{POE}}$ are item embedding, item-oriented embedding, interaction-oriented embedding, and position-oriented embedding, respectively. $\sigma$ is an activation function, $\mathbf{W}$ is a learnable parameter matrix.


\subsection{Sequence Modeling Layer on Sequential Neural Networks}
\label{sec:sequence modeling}

Given a sequence $s=[i_1,i_2,..,i_m]$ and the item embedding $\mathbf{X}$, where $\mathbf{X}=[\mathbf{x_1}, \mathbf{x_2},..., \mathbf{x_m}]$. We denote the output of the sequence modeling layer as $\mathbf{H}=[\mathbf{h_1}, \mathbf{h_2},..., \mathbf{h_m}]$, which are the hidden state of LSTM/GRU or the updated item representation generated via Self-attention/MLP or CNN.
Hence, the sequence modeling solution can be summarized below.

\begin{itemize}[leftmargin=*]
    \item \textbf{GRU/LSTM.} Analyzing the existing sequential neural networks in SR, GRU, and LSTM are the most common backbone for sequential information modeling~\cite{chen2019dynamic,wang2019collaborative,li2017neural}. 
    \item \textbf{Self-attention/MLP.} In early works~\cite{chen2019dynamic,hu2017diversifying,wu2017session}, MLP is elaborated for a user's all historical sessions modeling. With the prosperity of Transformer~\cite{vaswani2017attention}, self-attention also proposes to capture the correlation between items~\cite{de2021transformers4rec}.
    \item \textbf{CNN/Causal CNN with Dilated}. In~\cite{song2019session}, a densely connected 1D CNN with multiple convolution blocks is designed to capture n-gram features in the current session. As a variation of 1D CNN, the 1D causal CNN with dilated convolution possesses a wider receptive field, thus, it is applied for long-term dependency modeling in SR~\cite{yuan2020future}.
    \item \textbf{Others.} In~\cite{yuan2021dual, ouyang-etal-2022-social}, the authors argue that the self-attention mechanism will introduce irrelevant information. Thus, a self-attention module with $\alpha$-\textit{entmax} is proposed as an alternative of softmax function for item and session representation. In~\cite{guo2020session}, an extended GRU module with a leap gate, \textit{i.e.}, Leap Recurrent Unit (LPU), is designed to capture users' various preferences in the current session.
    In~\cite{xu2022modeling}, the authors fuse Fast Fourier Transforms (FFT) with Transformer to enhance the session representation for recommendation.
\end{itemize}

\subsection{Graph Construction for GNN}
\label{sec:graph_construction}

\begin{figure*}[t]
\centerline{\includegraphics[width=\textwidth]{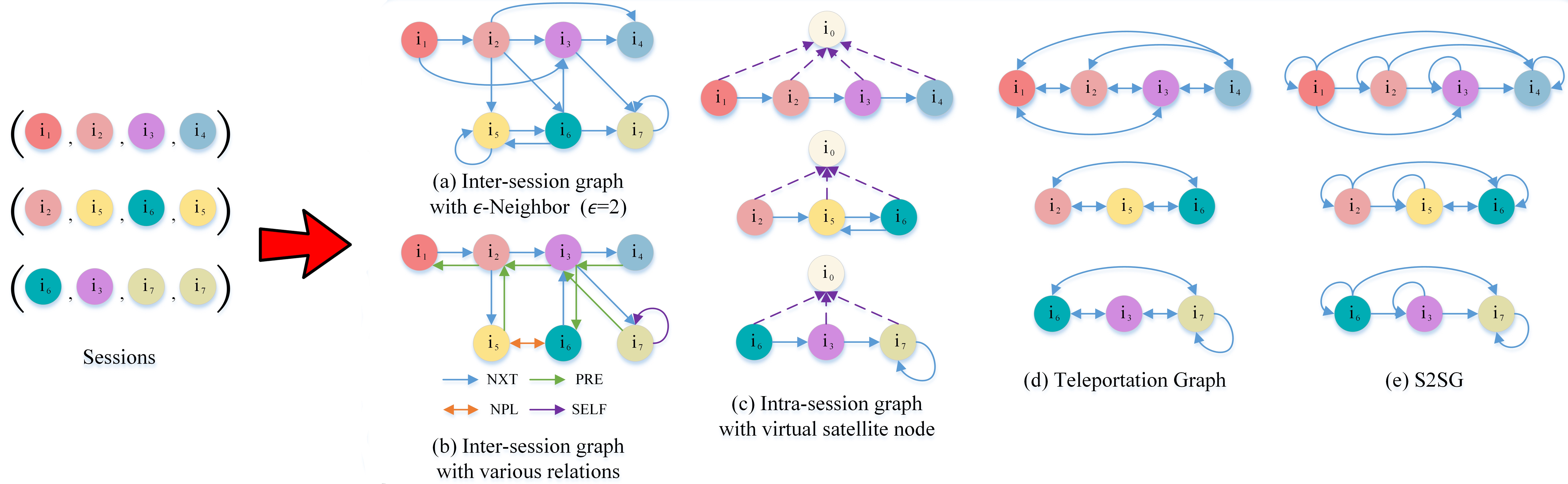}}
\caption{Various graph structures for SR.}
\label{fig:graphs}
\end{figure*}

Given neighbor sessions introduced in Section~\ref{sec:session_selection}, we could construct different graphs for information propagation and aggregation. 
Specifically, apart from the intra-session graph, inter-session graph, hypergraph, and heterogeneous graph, the four standard graph structures introduced in Section~\ref{sec:graph_define}, some other dedicated graphs are proposed to capture more complicated transaction patterns and user preferences for recommendation, as shown in Figure~\ref{fig:graphs}.

\begin{itemize}[leftmargin=*]
    \item \textbf{Inter-session Graph with $\epsilon$-Neighbor.} As shown in Figure~\ref{fig:graphs} (a), aside from the connection between adjacent items, the $\epsilon$-neighbor items are also considered for global graph construction, where $\epsilon$ is the window size~\cite{zhang2021graph,wang2020global}.
    \item \textbf{Intra-session Graph with Various Relations.} In~\cite{zhu2022transition, lai2022attribute}, the authors define four types of relations or edges, \textit{i.e.}, PRE, NET, SELF, NPL (as shown in Figure~\ref{fig:graphs}), for graph construction. Additionally, in~\cite{yuan2022micro, han2022multi}, the authors further consider the user's micro behaviors and item's multi-faceted global relations (\textit{i.e.}, base, sequential, co-occurrence, and incompatible) as external information for graph construction. 
    \item \textbf{Intra-session Graph with Virtual Satellite Node.} In~\cite{pan2020star}, a virtual node is created for each session to aggregate the global information, as shown in Figure~\ref{fig:graphs} (c). Similar to~\cite{pan2020star}, in~\cite{shen2021temporal}, the multi-virtual satellite nodes are created as the user's interests might be various and uncertain.  
    \item \textbf{Teleportation Graph}. As shown in Figure~\ref{fig:graphs} (d), compared with the inter-session graph, the teleportation graph~\cite{choi2022s} adds an edge between any of the two items in each session. Therefore, a fully connected graph is created for global session information extraction. 
    \item \textbf{S2SG}. As shown in Figure~\ref{fig:graphs} (e), based on the intra-session graph, in~\cite{chen2020handling}, the authors add a self-loop and a shortcut from item $a$ to the item $b$, where $a$ is the precursor node of $b$.
\end{itemize}

Overall, the sophisticated graph structures aim to capture global information, fuse more external information, and learn more complicated transaction patterns for SR.

\subsection{Information Propagation and Aggregation Layer for GNN}
\label{sec:node_representation}
Through information propagation and aggregation under different graph structures, GNNs can update the node representation with an iterative approach.
Comparing recent works, we summarize the following typical architectures for information propagation and aggregation. 
 
\begin{itemize}[leftmargin=*]
    \item \textbf{Average Pooling}. Average Pooling is an efficient yet effective method, which is widely used for information propagation and aggregation. Concretely, it can be formulated as:
    \begin{equation}
        \mathbf{h}^{(l+1)}_i=\frac{1}{|\mathcal{N}_i|}\sum_{v_j\in\mathcal{N}_i}\mathbf{W}\mathbf{h}^{(l)}_j\label{eq:pooling}
    \end{equation}
    where $\mathbf{h}_i^{(l+1)}$ are the item $i$ representation derived from $l$ multi-hops aggregation. $\mathcal{N}_i$ is the first-order neighbors of item $i$. $\mathbf{W}$ is the learnable parameters, which can be removed in some works~\cite{chen2021efficient,guo2022learning}.
    \item \textbf{GCN.} Compared with average pooling, GCN~\cite{kipf2016semi} considers the effect of node degree on information propagation and aggregation, which can be formalized below:
    \begin{equation}
    \begin{aligned}
        \mathbf{h}^{(l+1)}_i&=\sigma\left(\frac{1}{d_i+1}\mathbf{W}\mathbf{h}^{(l)}_i+\sum_{\mathbf{h}^{(l)}_j\in\mathcal{N}_i}\frac{1}{\sqrt{(d_i+1)(d_j+1)}}\mathbf{W}\mathbf{h}^{(l)}_j\right)
    \end{aligned}
        \label{eq:GCN}
    \end{equation}
    where $\sigma(\cdot)$ is an activation function, \textit{e.g.}, ReLU or Sigmoid. $d_i$ and $d_j$ are the degrees of node $i$ and node $j$, respectively. In many papers, Equation~(\ref{eq:GCN}) can also be formalized as $\mathbf{H}^{(l+1)}=\sigma(\Tilde{\mathbf{D}}^{-1/2}\Tilde{\mathbf{A}}\Tilde{\mathbf{D}}^{-1/2}\mathbf{H}^{(l)}\mathbf{W}^{(l)})$, where $\Tilde{\mathbf{A}}=\mathbf{A}+\mathbf{I}$, is an adjacency matrix defined by the graph structure. $\Tilde{\mathbf{D}_{ii}}=\sum_{j}\mathbf{\Tilde{A}}_{ij}$. Additionally, the adjacent matrix $\mathbf{A}$ can further be split into $\mathbf{A}_{in}$ and $\mathbf{A}_{out}$ for node representation respectively, where $\mathbf{A}_{in}$ and $\mathbf{A}_{out}$ are the point in and point out, respectively~\cite{qiu2020gag,xu2019graph}. Then, we fuse the representation from these two adjacent matrices via MLP layer or others. Moreover, motivated by the memory decay in reinforcement learning,~\cite{zhou2021temporal} uses an exponential denominator to re-scale the edge weights on the adjacent matrix. Therefore, the historical edge weight impact will become smaller with the increase of time discrepancy between previous interactions and the current interaction. Similar to~\cite{zhou2021temporal}, TMI-GNN~\cite{shen2021temporal} defines an item-level transition interval as the edge weight to model the temporal information in sessions.
    \item \textbf{GAT.} GAT~\cite{velickovic2017graph} proposes an attention mechanism to learn the importance of each neighbor node for information propagation.
    \begin{equation}
        \begin{aligned}     &\mathbf{h}^{(l+1)}_i=\sum_{v_j\in\mathcal{N}_i}\alpha_{ij}\mathbf{W}\mathbf{h}^{(l)}_j\\
            &\alpha_{ij}=\frac{\textrm{exp}(\textrm{LeakyReLU}(\mathbf{a}^T[\mathbf{W}\mathbf{h}_i^{(l)}||\mathbf{W}\mathbf{h}_i^{(T)}]))}{\sum_{v_j\in\mathcal{N}_i}\textrm{exp}(\textrm{LeakReLU}(\mathbf{a}^T[\mathbf{W}\mathbf{h}_i^{(l)}||\mathbf{W}\mathbf{h}_i^{(T)}]))}
        \end{aligned}
        \label{eq:GAT}
    \end{equation}
    where $\mathbf{a}$ and $\mathbf{W}$ are learnable parameter matrices, $\textrm{LeakReLU}(\cdot)$ is a nonlinearity function (with negative input slope $\alpha=0.2$). 
    \item \textbf{Soft-attention}. Different from the average pooling, or GAT, the soft-attention mechanism learns the importance of each first-order neighbor with various functions.
    \begin{equation}
        \begin{aligned}
            &\mathbf{h}^{(l+1)}_i=\sum_{v_j\in\mathcal{N}_i}\alpha_{ij}\mathbf{W}\mathbf{h}^{(l)}_j\\
            &\alpha_{ij}=\frac{\textrm{exp}(\mathbf{W}\mathbf{h_i^{(l)}}\mathbf{W}\mathbf{h_j^{(l)}}^T)}{\sum_{v_j\in\mathcal{N}_i}\textrm{exp}(\mathbf{W}\mathbf{h_i^{(l)}}\mathbf{W}\mathbf{h_j^{(l)}}^T)}\;& \textrm{(inner\;production)}\\
            &\alpha_{ij}=\frac{\textrm{exp}(\sigma(\mathbf{a}^T[\mathbf{W}\mathbf{h}_i^{(l)}||\mathbf{W}\mathbf{h}_i^{(T)}]))}{\sum_{v_j\in\mathcal{N}_i}\textrm{exp}(\sigma(\mathbf{a}^T[\mathbf{W}\mathbf{h}_i^{(l)}||\mathbf{W}\mathbf{h}_i^{(T)}]))}\;& \textrm{(concat)}
        \end{aligned}
        \label{eq:soft_attention_agg}
    \end{equation}
    where $\mathbf{W}$ is a learnable matrix. $\sigma(\cdot)$ is an activation function, \textit{e.g.}, Sigmoid, ReLU. If we implement the LeakyReLU as the activation function of the concat operation, it converts to GAT module. 
    \item \textbf{Routing.} Compared with the attention mechanism, the routing algorithm uses a Squash function for information aggregation without any trainable parameters for a more efficient model training.
    \begin{equation}
        \begin{split}
            &\mathbf{h}_i^{(l+1)} = \textrm{Squash}(\mathbf{h}_i^{(l)}+\sum_{v_j\in\mathcal{N}_i}w_{ij}\mathbf{h}_i^{(l)})\\
            &w_{ij} = \frac{\textrm{exp}(\mathbf{h}_i^{(l)^T}\mathbf{h}_j^{(l)})}{\sum_{h_j\in \mathcal{N}_i}\textrm{exp}(\mathbf{h}_i^{(l)^T}\mathbf{h}_j^{(l)})}
        \end{split}
    \end{equation}
    where \textrm{Squash}($\cdot$), which can be regarded as an activation function, $\textbf{Squash}(\mathbf{x})=\frac{\mathbf{x}}{||\mathbf{x}||}\cdot \frac{||\mathbf{x}||^2}{1+||\mathbf{x}||^2}$. 
    In~\cite{zhang2022graph}, the authors replace the \textrm{Squash} function as a \textit{unit normalization} function \textit{i.e.}, $(\mathbf{h}_i)=\frac{\mathbf{h}_i}{||\mathbf{h}_i||}$ for information propagation and aggregation.
    \item \textbf{Hybrid.} GCN/GAT/Soft-attention with GRU unit is the most common hybrid method for information propagation and aggregation. In~\cite{li2022disentangled,wu2019session,chen2021dual,qiu2019rethinking}, the node representation will first be modeled via a GCN, GAT, or Soft-attention for neighbor information aggregation. Afterwards, they will be fed into a GRU unit and obtain the hidden states as the updated item representation. Formally, the update functions are given as follows:
    \begin{equation}
        \begin{split}
            \mathbf{a}^{(l+1)}_{s,i} &= \textrm{GCN/GAT/Soft-attention}([\mathbf{h}_1^{(l)},...,\mathbf{h}_m^{(l)}])\\
            \mathbf{z}^{(l+1)}_{s,i} &= \sigma(\mathbf{W}_z\mathbf{a}^{(l+1)}_{s,i}+\mathbf{U}_z\mathbf{h}^{(l)}_{i})\\
            \mathbf{r}^{(l+1)}_{s,i} &= \sigma(\mathbf{W}_r\mathbf{a}^{(l+1)}_{s,i}+\mathbf{U}_r\mathbf{h}^{(l)}_{i})\\
            \mathbf{\Tilde{h}}_i^{(l+1)} &= \textrm{tanh}(\mathbf{W}_o\mathbf{a}^{(l+1)}_{s,i}+\mathbf{U}_o(\mathbf{r}^{(l+1)}_{s,i}\odot \mathbf{h}^{(l)}_{i}))\\
            \mathbf{h}^{(l+1)}_{i} &= (1-\mathbf{z}^{(l+1)}_{s,i})\odot \mathbf{h}^{(l)}_{i}+\mathbf{z}^{(l+1)}_{s,i}\odot \mathbf{\Tilde{h}}_i^{(l+1)}
        \end{split}
    \end{equation}
    where $\mathbf{h}_i^{(l)}$ are the representations of items in session $s$ after $l$-th updated. $\mathbf{a}_{s,i}^{(l)}$ is the item $i$ representation after $l$-th graph convolution. $\sigma(\cdot)$ is the Sigomid activation function. $\mathbf{W}_z, \mathbf{U}_z, \mathbf{W}_r, \mathbf{U}_r$, $\mathbf{W}_o, \mathbf{U}_o$ are learnable parameters of GRU module. In~\cite{shen2021multi}, the authors apply GAT with GRU to generate the item representation, then a soft-attention is used for representation fusion.
    \item \textbf{Others.} In~\cite{zhang2021graph,choi2022s}, a random walk with matrix learning is proposed for item representation generation. In~\cite{zhang2022graph}, the authors first propose item entropy to select top-$M$ similar items as neighbors, then use GAT for item representation update. In terms of GAT,~\cite{guo2022learning} applies an element-wise max operation to replace concatenation or mean operations for the multi-heads representation fusion. In~\cite{zhang2020personalized}, the GAT is first applied for item representation in the current session, then the max pooling and self-attention are applied for the user's historical preference modeling. Besides, an addition operation is further utilized to fuse the above two representations. In~\cite{pan2020star}, a virtual node is created on the current intra-session graph to aggregate the information from all the nodes in the session. Thus, based on the gate mechanism, the neighbor nodes and virtual start representations can be fused. In~\cite{guo2022evolutionary}, the authors combine GGNN with an ordinary differential equation (ODE) to capture continuous-time temporal information for session recommendation.
\end{itemize}

\begin{figure}[t]
  \centering
  \includegraphics[width=\linewidth]{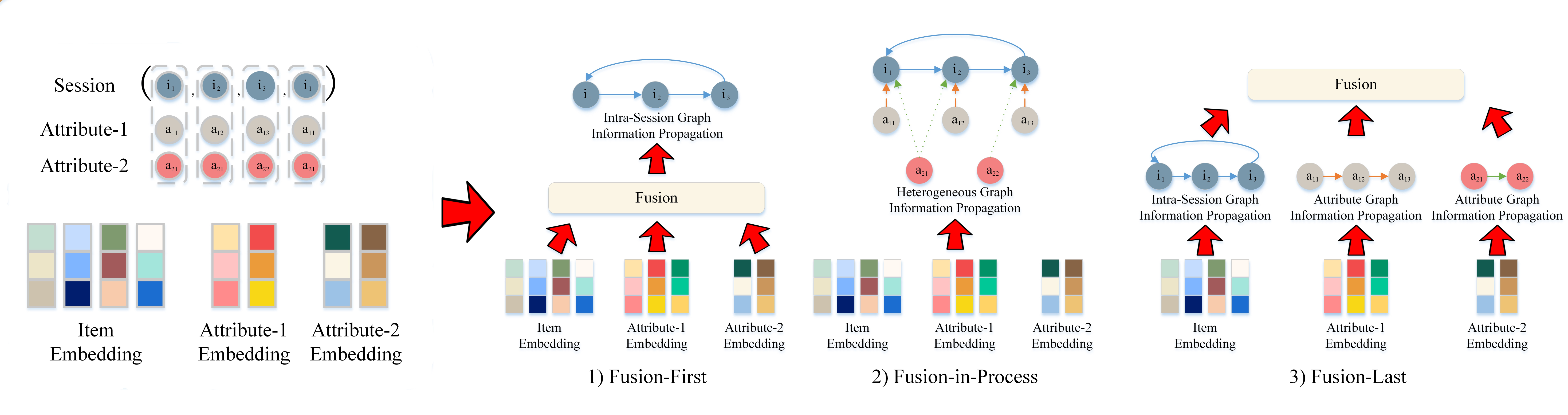}
  \caption{Different stages of external information fusion.}
  \label{fig:fusion}
\end{figure}

It is noteworthy that introducing external information via GNNs is also a desideratum for SR.
Reviewing the existing works, the external information fusion methods mainly include (1) max pooling or average pooling; (2) concatenation operation; (3) gate mechanism; (4) convolution operation. Additionally, as shown in Figure~\ref{fig:fusion}, the fusion stages mainly also include (1) Fusion-First; (2) Fusion-in-Process; and (3) Fusion-Last.  Fusion-First means the external information will be fused at the embedding stage, then update the item embedding via information propagation on a graph. Fusion-in-Process requires us first construct a heterogeneous graph that contains both the item node and the node with regard to external information, then inject the external information into the item representation by the process of information propagation and aggregation~\cite{zhang2022price,zhang2021knowledge,cui2022intention}. 
Fusion-Last requires us to construct various graphs for items and the correspondent external information respectively, then update the representation based on GCN, GRU, or other methods. 
Afterward, the external information and item representation can be fused~\cite{chen2021efficient, zheng2020dgtn,meng2020incorporating, ouyang-etal-2022-social, yuan2022micro}.  

\subsection{Session Representation Layer}

The item representation generated by the sequential neural networks or GNN will be further utilized for session representation. The typical methods include:

\begin{itemize}[leftmargin=*]
    \item \textbf{Last Item Representation.} For a current session $s$, we suppose the last clicked item could reveal the user's current preference. Hence, the last item representations are obtained for session representation generation, \textit{e.g.}, $\mathbf{s}=\mathbf{Wh}_m$, where $\mathbf{W}$ is a linear projection operation, in some works it is omitted~\cite{seol2022exploiting,sun2019bert4rec,xie2022decoupled,kang2018self,yuan2020future,guo2020session}.
    \item \textbf{Mean Pooling/Max Pooling}. Given a session and its item representations, the mean pooling or max pooling can be applied for session representation $\mathbf{s}$~\cite{huang2021graph,chen2021incorporating}. 
    \item \textbf{Concate.} In~\cite{chen2021incorporating}, the item representations after mean-pooling and max-pooling are concate.
    \begin{equation}
        \mathbf{s}=\textrm{Concat}(\frac{1}{m}\sum_{i=1}^m\mathbf{h}_i,\textrm{max}(\mathbf{h}_1,...,\mathbf{h}_m))
    \end{equation}
    \item \textbf{Concatenation with MLP.} To fuse the information of each item, the MLP layer with residual connection is used to generate session representation~\cite{luo2020collaborative}. 
    \begin{equation}
        \begin{split}
            \mathbf{o} &= \textrm{Concat}(\mathbf{h}_1,\mathbf{h}_2,...,\mathbf{h}_m)\\
            \mathbf{s} &= \textrm{ReLU}(\mathbf{oW}_1+\mathbf{b}_1)\mathbf{W}_2+\mathbf{b}_2+\mathbf{o}
        \end{split}
        \label{eq:concat_SR_SQ}
    \end{equation}
    where $\mathbf{W}_1,\mathbf{W}_2,\mathbf{b}_1,\mathbf{b}_2$ are learnable parameters.
     \item \textbf{GRU.} In~\cite{zhang2021graph,zhang2022graph}, the GRU module is applied to learn session representation.
    \begin{equation}
        \mathbf{h}_i' = \textbf{GRU}(\mathbf{h}_i)
    \end{equation}
    where $\mathbf{h}_i$ is item $i$ representation and $\mathbf{h}_i'$ is the correspondent hidden state of GRU. The last hidden state $\mathbf{h}_m$ is attained as the session representation.
    \item \textbf{Soft-attention.} There are numerous methods to generate the session representation via the soft-attention mechanism. For instance, \cite{qiu2020gag} supposes the last item in a session could reveal the user's current or local preference. Hence, we could model the impact of each previous item on the last item via MLP or softmax to capture the user's global preference for session representation generation. 
    \begin{equation}
        \begin{aligned}
            \mathbf{s} &= \sum_{i=1}^{m}\alpha_i \mathbf{h}_i\\
            \alpha_i &= \frac{\textrm{exp}(\mathbf{q}^T\sigma(\mathbf{W}_1\mathbf{h}_i+\mathbf{W}_2\mathbf{h}_l+\mathbf{b}_1))}{\sum_{i=1}^l\textrm{exp}(\mathbf{q}^T\sigma(\mathbf{W}_1\mathbf{h}_i+\mathbf{W}_2\mathbf{h}_l+\mathbf{b}_1))} \quad & \textrm{or}\\
            \alpha_i &= \textrm{MLP}(\mathbf{h}_{l}||\mathbf{h}_i) \quad & \textrm{or}\\
            \alpha_i &= \mathbf{q}^T\sigma(\mathbf{W}_1\mathbf{z}_i+\mathbf{W}_2s'+\mathbf{b}_1)\\ \mathbf{z}_i &= \textrm{tanh}(\mathbf{W}_3\mathbf{h}_i+\mathbf{b}_2),\; \mathbf{s}'=\textrm{Mean Pooling/ATT}(\mathbf{H})\\
        \end{aligned}
        \label{eq:softattention}
    \end{equation}
    where $\mathbf{h}_i\in \mathbf{H}$ and $\mathbf{h}_l$ are the item $i$ representation and last item representation. $\mathbf{q},\mathbf{W}_1,\mathbf{W}_2,\mathbf{W}_3$, $\mathbf{b}_1,\mathbf{b}_2$ are learnable matrices. $\sigma$ is the activation function. 
    \item \textbf{Self-attention with MLP.} 
    In~\cite{xu2019graph,zhang2021knowledge}, the authors use Transformer for session representation.
    \begin{equation}
        \begin{split}
            \mathbf{F}&=\textrm{softmax}(\frac{(\mathbf{HW}^Q)(\mathbf{HW}^K)^T}{\sqrt{d}})(\mathbf{HW}^V)\\
            \mathbf{E}&=\textrm{ReLU}(\mathbf{FW}_1+\mathbf{b}_1)\mathbf{W}_2+\mathbf{b}_2+\mathbf{F}\\
            \mathbf{s}&=\mathbf{E}_{[-1,:]}
        \end{split}
    \end{equation}
    where $\mathbf{W}^Q,\mathbf{W}^K,\mathbf{W}^V,\mathbf{W}_1,\mathbf{W}_2,\mathbf{b}_1,\mathbf{b}_2$ are learnable matrices. The dropout regularization and a residual connection are also applied during training. For a current session $s$, the last dimension of $\mathbf{E}$ is obtained as the global embedding of session $s$.
    \item \textbf{Memory Neural Network (MNN).} In general, a memory neural network is composed of two components: (1) a memory block to store the historical information and (2) a read/write operator to update the information representation. Based on MNN, neighbor sessions (selected by the method introduced in Section~\ref{sec:session_selection}) in memory blocks can be introduced for representation argumentation. 
    Specifically, let $\mathbf{s}^c$ as the current session representation and $\mathbf{s}_i$ as the neighbor session $s_i$'s representation. Therefore, the current session representation $\mathbf{s}^c$ can be updated via soft-attention mechanism~\cite{wang2019collaborative}, which is formalized as below. 
    \begin{equation}
    \begin{aligned}
        &\mathbf{s}^c=\sum_{i=1}^{k}w_i\mathbf{s}_i\\
        &w_i=\frac{\textrm{exp}(\textrm{sim}(\mathbf{s}^c,\mathbf{s}_i))/\tau}{\sum_{j=1}^k \textrm{exp}(\textrm{sim}(\mathbf{s}^c,\mathbf{s}_j)/\tau)}\\
        &\textrm{sim}(\mathbf{s}^c,\mathbf{m}_i)=\frac{\mathbf{s}^c\mathbf{m}_i}{||\mathbf{s}^c||||\mathbf{m}_i||}
    \end{aligned}
    \label{eq:}
    \end{equation}
    where $\mathbf{s}_i$ is the session representation, $s_i \in M$. $\tau$ is a temperature coefficient. 
    \item \textbf{Others.} In some works, the user information representation, position information representation, or the session representation from different modules are also fused via concatenation operation or linear projection for session representation generation~\cite{deng2022g,qiu2020gag,wang2020global,chen2021incorporating}.
\end{itemize}

As the last clicked item could represent the user's current preference or short-term interests, many works will fuse the last item representation $\mathbf{h}_l$ and the global session representation $\mathbf{s}^g$ generated by above methods for final session representation. The fusion methods mainly include concatenation, addition, weight, and so on.

\begin{equation}
    \begin{aligned}
        \mathbf{s}' &= \mathbf{s}^g + \mathbf{s}^l \quad& \textrm{(addition)}\\
        \mathbf{s}' &= w\mathbf{s}^g + (1-w)\mathbf{s}^l\quad& \textrm{(weight)}\\
        \mathbf{s}^l &= \mathbf{h}_l
    \end{aligned}
\end{equation}
where $\mathbf{s}^g$ and $\mathbf{s}^l$ are the global session representation and local session representation respectively. $w$ can be defined as a hyperparameter or learnable parameter, \textit{i.e.}, $\mathbf{w}=\sigma(\mathbf{w}[\mathbf{s}^g||\mathbf{s}^l])$, to balance the information from local and global. 

\subsection{Prediction Layer and Loss Functions}
\label{sec:loss}

In most works, the inner production between item embeddings $\mathbf{x}_i$ and session representation $\mathbf{s}$ is calculated to yield the prediction score, which will be further fed into a softmax function for probability distribution generation. Apart from that, the MLP layer with an activation function (\textit{e.g.}, tanh, sigmoid) is also a popular strategy to generate the scores of predicted items and this method is more efficient against the inner product. 

\begin{equation}
\begin{aligned}
    \begin{split}
        \hat{y}_i &= \textrm{Softmax}(\mathbf{s}^T\mathbf{x}_i)\; & \textrm{(Inner production)}\\
        \hat{y}_i &= \sigma(\textrm{MLP}(\mathbf{s}))\; & \textrm{(MLP)}
    \end{split}
    \label{eq:probability}
\end{aligned}
\end{equation}

As to the loss function, summarizing the existing works it can be divided into three categories: (1) standard loss function and its improvement (\textit{e.g.}, cross-entropy Loss, BPR Loss, BPR-max Loss); (2) auxiliary loss function, which usually integrates a recommendation loss with an auxiliary loss or regularization terms; (3) multi-task loss function (\textit{e.g.}, link prediction and entity representation with knowledge graph). 

\noindent \textbf{(1) Standard Loss Function and Its Improvements.}
\begin{itemize}[leftmargin=*]
    \item \textbf{Cross-entropy Loss.} Cross-entropy loss is the most commonly used loss function for model training in SR. 
    \begin{equation}
        \mathcal{L}_{\textrm{CE}}=-\sum_{i=1}^n\textrm{log}(\hat{y}_i)\label{eq:loss}
    \end{equation}
    where $n$ is the total number of training samples.
    \item \textbf{BPR Loss.} Bayesian Personalized Ranking~\cite{rendle2012bpr} is a standard pairwise ranking loss, which optimizes the difference between the scores of the positive item and sampled negative items.
    \begin{equation}
        \mathcal{L}_{\textrm{bpr}}=-\frac{1}{N_S}\sum_{j=1}^{N_S}\textrm{log}\sigma(\hat{y}_i-\hat{y}_j)\label{eq:bpr}
    \end{equation}
    where $N_S$ is sampled negative items, $\sigma$ is the sigmoid activation function.
    \item \textbf{TOP1 Loss.} Compared with the BPR loss function, TOP1 loss adds a regularization term to force the scores of the negative items to be around zero for a stable training process.
    \begin{equation}
        \mathcal{L}_{\textrm{top1}}=\frac{1}{N_S}\sum_{j=1}^{N_S}\sigma(\hat{y}_j-\hat{y}_i)+\sigma(\hat{y}_j^2)\label{eq:top1}
    \end{equation}
    \item \textbf{BPR-max.} Bayesian Personalized Ranking (BPR) aims to maximize the probability that the target score should be larger than negative sample scores. However, intuitively, if the score of the negative sample is already well below that of the target, there is nothing to learn from that negative sample anymore. Meanwhile, the gradient is always discounted by the total number of negative samples, which will also lead to the vanishing of gradients. To relieve this issue, Hidasi and Karatzoglou~\cite{hidasi2018recurrent} propose a BPR-max loss function, which uses softmax scores on the negative examples. Thus, more relevant negative samples will produce high gradients and obtain more focus on the training process. 
    \begin{equation}
        \mathcal{L}_{\textrm{bpr-max}}=-\textrm{log}\sum_{j\in N_S}s_j\sigma(\hat{y}_i-\hat{y}_j)+\lambda\sum_{j\in N_S}s_j\hat{y}_j^2\label{eq:bpr_max}
    \end{equation}
    where $N_S$ is sampled negative samples. Besides, a softmax weighted $l_2$ regularization over the scores of the negative samples is also added. $\lambda$ is a regularization hyperparameter. 
    \item \textbf{List-wise Ranking Loss.} Compared with cross-entropy loss, list-wise loss only considers the top-k scores instead of all the items' in cross-entropy loss for model training~\cite{wu2017session}. 
    Given a session $s$, denote $\hat{y}_i$ as the target item predicted probability generated by softmax function, and $\hat{y}_j, j\in \mathcal{G}_k$ as the top-k predict scores. The list-wise ranking loss can be formalized as:
    \begin{equation}
        \mathcal{L}_{\textrm{rank}}=-\sum_{j\in \mathcal{G}_k}\hat{y}_i\textrm{log}\hat{y}_j\label{eq:rank_loss}
    \end{equation}
    \item \textbf{Robust Distance Measuring (RDM).} Inspired by contrastive learning, in~\cite{hou2022core}, the authors propose RDM for better alignment and uniformity properties of item embedding.
    \begin{equation}
        \mathcal{L}_{\textrm{RDM}}=-\textrm{log}\frac{\textrm{exp}(\textrm{cos}(\mathbf{s},\mathbf{x^+})/\tau)}{\sum_{i=1}^n\textrm{exp}(\textrm{cos}(\mathbf{s},\mathbf{x}_i)/\tau)}\label{eq:RDM}
    \end{equation}
    where $\mathbf{x^+}$ is the embedding of ground-truth of session $s$. $\mathbf{x_i}\in \mathbf{X}$ is the item $i$ embedding. $\mathbf{s}$ is the session representation. $\tau$ is a hyper-parameter to adjust the distribution yield by the softmax function for different scenarios adaptation. 
    \item \textbf{Adaptive Weight Loss.} Motivated by Focal loss~\cite{lin2017focal}, adaptive weight loss is adopted to arrange different weights for each sample and hinder the gradient of negative samples from dominating the whole training process.
    \begin{equation}
    \begin{aligned}
        \begin{split}
            p_i &=\begin{cases}
            \hat{y}_i & \textrm{if}\; y=1\\
            1-\hat{y}_i & \textrm{otherwise}
            \end{cases}\\
            \mathcal{L}_{\textrm{AW-Loss}} &= -\sum_{i=1}^n(2-2p_i)^\gamma\textrm{log}(p_i)
        \end{split}
        \label{eq:adaptive_loss}
    \end{aligned}
    \end{equation}
    where $\gamma$ is a temperature coefficient to control the effect of modulating factor to the loss, \textit{i.e.}, expand the contribution of hard samples to the loss while diminishing that of simple samples.
    
\end{itemize}

\noindent \textbf{(2) Auxiliary Loss Function.} 
Contrastive Learning with InfoNCE is also a commonly used loss as a regularization term to alleviate the data sparsity issue and improve recommendation performance~\cite{xia2021self_1,xia2021self}.

\begin{equation}
    \mathcal{L}_{InfoNCE}=-\textrm{log}\frac{\sum_{i\in\Theta^+}\textrm{exp}(\phi(\theta,\theta_i)/\tau)}{\sum_{j\in \Theta}\textrm{exp}(\phi(\theta,\theta_j)/\tau)}
\end{equation}
where $\phi(\cdot)$ is the discriminator function that requires two vectors as the input and then scores the agreement between them with inner production or other methods. $(\theta, \theta_i^+)$ is the positive pairs, \textit{e.g.}, the last item representation in the current session and the embedding of top-K highest confidence items predicted by the model, or session representations generated by two GNNs. As a regularization term, contrastive learning tends to optimize uniformity and alignment of representations~\cite{2020arXiv200510242W}. Therefore, the item representations will be uniformly distributed on the unit hypersphere, and the item representations of positive pairs will also close together. \textcolor{black}{Apart from that, other auxiliary loss functions like distillation loss~\cite{yu2023causality} in causal learning SR are also adopted to block the influence of the shortcut path and accentuate the significance of causal dependencies within the session graph. In~\cite{shen2021temporal}, the authors propose TMI-GNN to extract multiple interest representations for users and design an interest-independent loss that hires distance correlation measures as a regularizer to encourage the multi-interests representations to be diverse. }

\noindent \textbf{(3) Multi-task based Loss Functions.} Multi-task learning (MTL) is a satisfied recipe that exploits useful information from other related learning tasks to supervise the model to perform better on the original task~\cite{zhang2018overview,ruder2017overview,shalaby2022m2trec}. There are also some works that apply MTL for SR. 
For instance, 
in~\cite{shen2021multi}, the authors consider multiple interaction types, such as click and purchase, in sessions. Consequently, an interaction session can also be constructed. Afterward, the authors apply two cross-entropy losses for the next behavior type prediction and item recommendation. Furthermore, some works~\cite{meng2020incorporating,zhang2021knowledge} incorporate item-relevant knowledge and construct knowledge graphs for SR. Therefore, the link prediction or entity representation tasks are also introduced for knowledge embedding learning.  

\section{Datasets, Evaluation Metrics and Complexity}
\label{sec:evaluation_datasets}

\textcolor{black}{
In this section, we will first analyze the statistical characteristics of publicly real-world datasets and evaluation metrics s with regard to the accuracy and diversity that are commonly used in SR. 
}

\textcolor{black}{\subsection{Public Datasets}}
\label{sec:datasets}

\textcolor{black}{
According to the existing works in SR, we summarize eight popular public datasets that cover E-commerce, Music and Video, Job Position, and Chick-in scenarios. The statistical characteristics after preprocessing\footnote{Following existing works, we filter out all sessions whose length is $1$ and items appearing less than $5$ times.} are presented in Table~\ref{tab:datasets}. 
}

\textcolor{black}{
\begin{table}[]
\caption{\textcolor{black}{The statistical characteristics of commonly used public datasets after preprocessing for SR. \# means the total numbers, Avg. calculates the mean value.}}
\begin{tabular}{llrrrr}
\hline\hline
\textbf{Domain}                           & \textbf{Dataset}      & \textbf{\# sessions} & \textbf{\# interactions} & \textbf{\# items} & \multicolumn{1}{c}{\begin{tabular}[c]{@{}c@{}}\textbf{Avg. session length}\end{tabular}} \\ \hline 
\multirow{4}{*}{E-commerce}      & Yoochoose\tablefootnote{https://www.kaggle.com/chadgostopp/recsys-challenge-2015
}    & 1,375,128   & 5,426,961       & 28,582   & 3.95                \\
                                 & Tmall\tablefootnote{https://tianchi.aliyun.com/dataset/dataDetail?dataId=42}        & 1,774,729   & 13,418,695      & 425,348  & 7.56                \\
                                 & Diginetica\tablefootnote{https://competitions.codalab.org/competitions/11161
}   & 780,328     & 982,961         & 43,097   & 5.12                \\
                                 & RetailRocket\tablefootnote{https://www.kaggle.com/retailrocket/ecommerce-dataset} & 59,962      & 212,182         & 31,968   & 3.54                \\ \hline
\multirow{2}{*}{Music} & Last.FM\tablefootnote{http://millionsongdataset.com/lastfm/}      & 169,576     & 2,887,349       & 449,037  & 17.03               \\
                                 & NowPlaying\tablefootnote{https://www.kaggle.com/chelseapower/nowplayingrs}   & 27,005      & 271,177         & 75,169   & 10.04         \\ \hline
\multirow{1}{*}{Job Position}                     & Xing\tablefootnote{http://2016.recsyschallenge.com/}         & 91,683      & 546,862         & 59,121   & 5.78                \\ \hline
\multirow{1}{*}{Check-in}                         & Gowalla\tablefootnote{http://snap.stanford.edu/data/loc-gowalla.html}      & 830,893     & 245,157         & 6,871    & 4.32                \\ \hline\hline
\end{tabular}
\label{tab:datasets}
\end{table}
}

\textcolor{black}{
\begin{itemize}[leftmargin=*]
    \item \textbf{Yoochoose} is the dataset of RecSys Challenge 2015, which contains a stream of user clicks and buy events on an online webshop within six months. Since the set of Yoochoose is extremely large, the most recent portions 1/64 and 1/4 subsample of all sessions are usually used as the training set, denoted as "Yoochoose1/64" and "Yoochoose1/4", respectively~\cite{chen2021dual,chen2020handling,liu2018stamp,ren2019repeatnet,wu2019session,wang2020beyond}.
    \item \textbf{Tmall} comes from the IJCAI-15 competition, which contains users’ shopping logs on the Tmall online shopping platform.
    \item \textbf{Digineitca} is a personalized e-commerce research challenge dataset released in CIKM CUP 2016. The dataset contains transition histories, which are suitable for session-based recommendation.  
    \item \textbf{RetailRocket} contains user behavior data and item properties that collected from a real-world e-commerce website.
    \item \textbf{Last.FM} is a music artist recommendation dataset published by Celma \textit{et al}~\cite{celma2009music}.
    \item \textbf{NowPlaying} dataset comes from~\cite{zangerle2014nowplaying} and is created from music-related tweets, which illustrate the music-listening behavior of users.
    \item \textbf{Xing} Recsys Challenge 2016 Dataset~\footnote{http://2016.recsyschallenge.com/} contains user interactions (click, bookmark, reply, and delete) on a job posting platform for 770k users over an 80-day period. 
    \item \textbf{Gowalla} contains the check-in data and social network on a location-based social networking website. It is widely used for point-of-interest recommendation.
\end{itemize}
}
\textcolor{black}{
Observing Table~\ref{tab:datasets}, we could find that the length of sessions in existing public SR datasets is rather limited. As exemplified by the average/median lengths of sessions are $5.12$/$4.0$ and $3.95$/$3.0$ for Diginetica and Yoochoose, the most two popular session-based datasets. 
Therefore, we believe there is no obvious order dependency between any of the two items, indicating that modeling the items' correlation with GNNs rather than sequential information with sequential neural networks is more suitable for SR~\cite{chen2019session}. 
}

\textcolor{black}{\subsection{Evaluation Metrics}}
\label{sec:evaluation}

\noindent \textcolor{black}{
\textbf{Evaluation Metrics for Accuracy.} Accuracy aims to measure the alignment of recommendation results and user interests. In general, HR@K (Hit Rate calculated over top-K items), MRR@N (Mean Reciprocal Rank calculated over top-K items), and NDCG@K (Normalized Discounted Cumulative Gain calculated over top-K items) are widely used evaluation metrics for SR performance comparison, where $K=5, 10, 20$ are the most common settings for evaluation.
}
\textcolor{black}{
\begin{itemize}[leftmargin=*]
    \item \textbf{HR@K}. The HR@K score measures whether the target item is included in the top-K recommendations of the recommended list. 
    \begin{equation}
        \textrm{HR@K}=\frac{n_{hit}}{N}\label{eq:HR}
    \end{equation}
    where $N$ is the number of test sessions in the dataset and $n_{hit}$ counts the number that target items that appear in the top K position of the ranking list. 
    \item\textbf{MRR@K}. The MRR@K is a ranking evaluation matrix. When the target item $\hat{i}$ is not in the top K position, the $Rank(\hat{i})$ is set to 0. It calculates as follows,
    \begin{equation}
        \textrm{MRR@K}=\frac{1}{N}\sum_{\hat{i} \in \mathcal{S}_{test}}\frac{1}{\textrm{Rank}(\hat{i})}\label{eq:MRR}
    \end{equation}
    where $\mathcal{S}_{test}$ are the set of test sessions. $\textrm{Rank}(\hat{i})$ is the position of item $i$ in the recommendation list. The MRR is a normalized ranking of hits. The higher the score, the better the quality of the recommendation, \textit{i.e.}, a higher score indicates a higher ranking position of the target item.
    \item \textbf{NDCG@K}. The NDCG estimates the ranking order of the recommendation list. The same as MRR, if the target item $\hat{i}$ is not in the top K position, the $Rank(\hat{i})$ is set to 0.
    \begin{equation}
        \textrm{NDCG@K}=\frac{1}{N}\sum_{\hat{i} \in \mathcal{S}_{test}}\frac{1}{log_2(\textrm{Rank}(\hat{i})+1)}\label{eq:ndcg}
    \end{equation}
\end{itemize}
}

\noindent\textcolor{black}{
\textbf{Evaluation Metrics for Diversity.} Diversification is first concerned in information retrieval (IR) community where researchers endeavor to disambiguate the input query to cover the user's real intent via diversification optimization~\cite{clarke2008novelty}. In recommendation system, diversity is related to how different the recommended items are with respect to each other~\cite{castells2021novelty}, which aims to alleviate the filter bubble problem. Consequently, the recommendation diversity can be identified as the average pairwise dissimilarity between items in list. In SR, intra-list diversity, coverage, and their variants are the most common metrics for diversity measurement~\cite{yin2023understanding, liu2020long, vargas2011rank}.
}
\textcolor{black}{
\begin{itemize}[leftmargin=*]
    \item \textbf{ILD}. Intra-List Distance (ILD) measures the average distance between every pair of items in recommendation list (RL), which can be formalized as,
    \begin{equation}
        \textrm{ILD}=\frac{\sum_{(i,j)\in RL}d_{ij}}{|RL|\times(|RL|-1)}\label{eq:ILD}
    \end{equation}
    where $d_{ij}$ is the Euclidean distance of category embeddings with respect to item $i$ and item $j$. $|\textrm{RL}|$ is the number of recommendation items in RL.
    \item \textrm{Coverage@K}. Coverage@K measures how many different categories appear in the top-K recommendation items.
    \begin{equation}
        \textrm{Coverage@K}=\frac{\cap_{i\in RL_K} C_i}{K}\label{eq:coverage}
    \end{equation}
    where $C_i$ is the category of item $i$.
\end{itemize}
}

\textcolor{black}{
Apart from ILD and coverage, some other metrics, such as long-tail coverage, which measure the diversity performance in long-tail items, and relevance sensitive expected
intra-list diversity (RR-ILD) which considers the ranks and relevance of top $K$ recommendation simultaneously in ILD calculation. 
}

\section{Challenges and Directions}
\label{sec:challengeanddirection}

GNNs and sequential neural networks have greatly prompted SR research but also are accompanied by some challenging issues. In this section, we outline the following prospective research directions, which we believe are critical to the further development of SR.

\subsection{More External Information}
As we discussed in \textbf{Section}~\ref{sec:embedding}, a plethora of works incline to fuse more external information to explore the user's multi interests or item's complicated transaction patterns for SR. However, comparing the existing efforts, we find that the external information should be selected carefully and not all of them are desiderata for performance improvement. Besides, the fusion method is also significant for SR. 
To sum up, we consider there are two open issues with regard to external information that deserve to be discussed.

\begin{itemize}[leftmargin=*]
    \item \textit{What kind of external information is necessary for SR?} We divide the external information into three categories: item-based, interaction-based, and position-based. Therefore, some of them are costly to model (\textit{e.g.}, item descriptions and user comments), some are not appropriate for SR (\textit{e.g.}, user personal information, since the sessions are expected to be anonymous in SR), some may not be necessary (\textit{e.g.}, position or order information~\cite{seol2022exploiting}, since the length of sessions is limited and there is no obvious order dependency between two items). In addition, for different recommendation scenarios, the required information may also be different. For instance, for news, book, or music recommendation, users will be interested in the categories of the items, while for product recommendations, the item's brand and price probability are more effective for SR. Therefore, what kind of external information should be considered for the practical scenarios is really important.
    \item \textit{How to balance the trade-off between effectiveness and efficiency of fusion methods?} The fusion methods can be various in SR, \textit{e.g.}, embedding addition or concatenation, self-attention, and gate mechanism. Note that as the simplest function for information fusion, the addition operation will not change the model structure or increase numerous external parameters or computations. But at the same time, the external information will not be exploited thoroughly and the representation ability of models will not be enhanced either. 
    Although concatenation could facilitate the representation ability of models, it will also increase the computational complexity and the number of parameters dramatically. In addition, concatenation can not model the implicit correlation and effects between any of the two external information. 
    Recently, self-attention has been proven as an effective method for external information fusion~\cite{xie2022decoupled}, but it will also increase the number of parameters and computation cost. Therefore, more elegant fusion methods are expected to balance the trade-off between the representation ability and the number of parameters for an effective and efficient recommendation.
    \item \textit{Couple v.s Decouple, which is better for SR?}  As shown in Figure~\ref{fig:fusion}, there are three phases for external information fusion. (1) Fusion first. Most of the works consider fusing the external information in the embedding stage. Thus, the external information and item embedding will couple together for SR. (2) Fusion in process. GNN-based methods regard the external information as nodes. Then, a heterogeneous graph is constructed, based on information propagation and aggregation the external information can be fused. (3) Fusion last. The external information and item embedding are decoupled and modeled respectively, then, fuse them in the end. For instance, we could construct two graphs for both item and external information, capitalizing on two GNNs for item and external information representation learning, which will be fused together in the end. In general, by decoupling external information representation with item representation, we could fully exploit the features from different priorities and avoid mixed correlation effects. Conversely, modeling the external information with a unified framework will limit the capacity of models to extract implicit features encapsulated in the heterogeneous side information, while it has a greater advantage in efficiency.
\end{itemize}

\subsection{Session Selection and Graph Construction}
\label{sec:dynamic_session}

As the length of sessions is quite limited, \textit{i.e.}, less than six for most public datasets, many works introduce neighbor sessions for graph construction. Thus, sophisticated transaction patterns can be captured for SR. For neighbor session selection and graph construction, there are also some issues that need to be discussed.

\begin{itemize}[leftmargin=*]
    \item \textit{Scalability Graphs in SR.} Graph structure is pivotal for GNN-based SR. To harness the fertile information from neighbor sessions, large-scale and complicated graphs are constructed, which will include billions of nodes and edges in practical scenarios. Besides, each node contains a variety of features. Hence, it is nontrivial to update and apply such a huge graph for information propagation and prediction straightforwardly. To this end, sampling is a widely adopted solution to reduce the graph size and alleviate the above issue. For instance, we could randomly sample a fixed number of neighbors from original graphs (\textit{e.g.}, Graphsage~\cite{hamilton2017inductive}) or employ the random walk strategy for sampling (\textit{e.g.}, PinSage~\cite{ying2018graph}). However, these methods are accompanied by a high degree of randomness, which may incur an unstable model training. Consequently, how to design a scalable graph structure is vital for SR.
    \item \textit{Dynamic Graphs in SR.} Furthermore, in the real world, the items and their relations are changing over time. To maintain up-to-date recommendation, graph structures should be also adjusted and updated dynamically. Most of the existing studies are based on a static graph structure, few works pay attention to dynamic graphs. Thus, it is a largely under-explored realm and deserves further study.
    \item \textit{Self-learning Graph Structure.} Obtaining a proper graph structure requires considerable effort and this process is also heuristic and problem-specific. Moreover, although recent research~\cite{chen2019dynamic, li2023exploiting} has revealed the necessity of modeling the implicit connections between items for SR, most of the existing graph-based methods can only capture the item relations with a few hops, which can not explore the implication relations between items thoughtfully. 
    We believe the self-learning strategy for graph construction is a reasonable method to alleviate the above problems, and it is quite common in many other tasks~\cite{deng2021graph,wu2020connecting,li2019mad}. Therefore, self-learning graphs can also be look-forwarded for SR.
\end{itemize}

\subsection{Diverse and Uncertain Representation of User Interests}
Despite the superior performance of existing methods for SR, most of the works concentrate on a single and fixed user interest representation for recommendation and fail to disentangle multiple interests of users. Considering the diversity of user interests, a fixed representation is insufficient and leads to sub-optimal results. In addition, users' sequential behaviors are uncertain. 
Therefore, compared with one single fixed representation, it is worthwhile to capture interest diversity, inject uncertainties, and provide more flexibility for SR.

\begin{itemize}[leftmargin=*]
    \item \textit{Diverse Representation of User Interests.} Rather than learning a single and fixed user interests representation, some works endeavor to extend such one-fold vector to multiple vectors with capsule networks or attention mechanism~\cite{sabour2017dynamic} for multi-interest, short-term and long-term interest, and interest diversity representation. For instance, Tian \textit{et al.}~\cite{tian2022multi} apply GCN and capsule network to capture user's multi-level and multi-interests representation for recommendation. Guo \textit{et al.}~\cite{guo2020session} propose a Hierarchical Leaping Network (HLN), which extracts various subsequences from the current session. Thus, by learning the representation of each subsequence, the user's multi-interest representation can be captured. Li \textit{et al.}~\cite{li2022disentangled} split the item embedding into several chunks and apply GGNN for each chunk to learn user interests representation with multiple factors for SR. \cite{turgut2023prod2vec} uses clustering technology for similar products to improve the diversity of recommendations.
    Aside from multi-interest representation, Li \textit{et al.}~\cite{li2024disentangle} believe the user interest can be split into interest trend and interest diversity. The former one is determined by her education level, income and occupation, while the latter is easier impacted on the advertisement or marketing. Therefore, the authors tailored two modules for user interest trend and diversity modeling.
    Although there are some works that focus on multiple interests representation, it is still in a preliminary stage and many issues (\textit{e.g.}, how to determine the number of different interests for each user in an adaptive fashion; how to fuse those interests representation for SR) need to be further explored.
    \item \textit{Uncertain Representation of User Interests.} Compared with vector representation, distribution representation could inject uncertainties and provide more flexibility, it has been attracting interest from the research community~\cite{bojchevski2017deep,he2015learning,sun2018gaussian,vilnis2014word}. For the recommendation system, the item embedding can also be initialized as a Multi Gaussian Distribution governed by a mean vector and a covariance vector. The mean vector could reveal the user's various basic preferences, while the covariance vector injects a potential uncertainty. Fan \textit{et al.} elaborate to apply distribution representation and \textit{Wasserstein Distance} for recommendation ~\cite{fan2022sequential,fan2021modeling}. However, for SR, there are very few works focused on this issue. It will be interesting and valuable to explore uncertain representation in SR.
\end{itemize} 

\subsection{\textcolor{black}{Explainability and Privacy Production for SR}}

\textcolor{black}{Apart from the accuracy, the explainability, security, and privacy production of outputs are also significant and expected for a good recommendation system. In general, the explanation of recommendation aims to answer “why”, that is, why the items are recommended, an explainable recommendation can improve the transparency and persuasiveness of systems, boosting the satisfaction and stickiness of users~\cite{zhang2020explainable}. With the prosperity of deep learning, sequential and graph-based neural networks behave as black boxes, leaving this research area more challenging. Investigating the recent efforts, these methods can be divided into two categories: (1) the explanation generation based on language models~\cite{li2021personalized, chen2021generate, yang2021explanation, geng2022recommendation, li2023personalized, liu2023chatgpt} knowledge graphs~\cite{fu2020fairness, xian2019reinforcement} or image visualizations~\cite{chen2019personalized}, and (2) the explainability of deep learning model structures~\cite{zhang2014explicit, chen2019personalized}. However, very few works~\cite{wu2023causality, narwariya2023x4sr} focus on the explainability in session-based recommendation. As for privacy production in recommendation, the user's historical interaction records encompass privacy information such as gender, age, and even political orientation, which can be inferred by the recommender system~\cite{zhu2023membership}. Thereby, it is inappropriate to request such private specific historical interactions for recommendation. Recently attempts resort to unlearning strategies that eradicate the sensitive data in model training to tackle this problem~\cite{xin2024effectiveness, bourtoule2021machine}. Overall, we believe it is an interesting and promising topic for future research in session-based recommendation.}

\subsection{Streaming or Online SR}
As discussed in~\ref{sec:dynamic_session}, in a real-life situation, sessions are dynamically produced as a stream, while most of the relevant work elaborates on training the recommendation system with the historical sessions which preserve the users' long-term static interests. Hence, it might be inappropriate to apply a static model for new coming sessions as users' preferences are changing over time. To alleviate this issue, the model should be online updated with the latest sessions. 
Therefore, it is a challenge to effectively learn users’ dynamic and real-time preferences for better SR. Existing research~\cite{qiu2020gag,guo2019streaming} maintains a reservoir to update the model for an online recommendation. To be specific, the reservoir can be the incoming sessions that contain new items or new users. Then, for each session, a sample probability is generated via Wasserstein distance (also known as Earth Mover's Distance~\cite{rubner2000earth}) or ranking-based distance~\cite{wang2018streaming}. After that, the sampling distribution is applied to update the reservoir. Thus, the model can be updated timely based on the updated reservoir. However, the above methods heavily rely on the sampling strategy and distance computation, which is also a cost-expensive process. Rendering it difficult to balance the trade-off between effectiveness and efficiency. \textcolor{black}{To overcome this issue, some works~\cite{xia2023towards, li2021lightweight, xia2023efficient} propose lightweight model architectures based on model compression techniques, such as low-rank decomposition, hash coding, and quantization. However, these solutions require a pre-defined fixed compressed ratio to retrain, leading to sub-optimal results when the ratio is inappropriate.} 
How to effectively capture users' dynamic preference change and recommendation in real-time is a promising direction. 
   
\subsection{Causal Debias and Denoise in SR}
Due to the exposure mechanism, popularity effects, and the feedback loop in recommendation system, bias and noise problems become serious, and heavily deteriorate the recommendation effectiveness~\cite{chen2020bias}. To explain the relations between a cause and effect, causality-based methods, \textit{e.g.}, causal-inference, and causal graph, are a major solution to debias and provide explanations of the recommendation system~\cite{wang2021deconfounded,liu2021mitigating,wei2021model,xu2021deconfounded,yang2021top,zhang2021causal,zheng2021disentangling, yu2023causality}. However, there are very few works specializing in this issue on SR. We believe the causal model will bring SR research into a new frontier. 

\subsection{Reinforcement Learning for SR} 
Different from supervised and unsupervised learning, reinforcement learning~\cite{kiumarsi2017optimal} focuses on goal-directed learning that maximizes the total reward achieved by an agent when interacting with its environment. Hence, it is a potential and prominent solution to model interactions between the user and agent, capture rapid changes in users' preferences, and realize dynamic recommendation.  
Recent years have witnessed significant progress of reinforcement learning in recommendation systems~\cite {chen2021survey,lin2021survey}. Specifically, modeling the user's interaction behaviors as a decision-making process, we could maintain and update a corresponding recommendation policy in real time. For instance, in~\cite{shih2018automatic}, the authors treat music playlist generation as a language modeling process, Thus, an attention-based language model with the policy gradient is applied for recommendation. 
Wang \textit{et al.}~\cite{wang2020kerl} propose a method named knowledge-guided reinforcement learning (KERL), which integrates knowledge graphs into reinforcement learning. 
To be specific, KERL adopts TransE~\cite{bordes2013translating} with the MLP layer to predict future knowledge of user preferences and recommendation. 
Overall, the above works with reinforcement learning could be summarized as (1) simulating users' interaction in SR for dynamic and timely recommendation; (2) capturing the interest shift of users; (3) filtering the noise in sessions; and (4) selecting the valuable items for SR. 
However, model-free deep reinforcement learning requires a significant number of samples as there is no guarantee that the received state is useful. In SR, the length of sessions is very short, accompanied by an extremely large action space (\textit{i.e.}, number of items), thus it will require more high-quality samples to cover the exploration space, which hinders the further development of reinforcement learning in SR.

\subsection{SR with Language Model and Diffusion Model}
Encouraged by the remarkable success of large language models (LLM) in NLP, utilizing language models for recommendation has become a cutting-edge very recently. 
Reviewing existing research, prompt or in-context learning, and parameter-efficient fine-tuning (PEFT) are prominent solutions in this venue.
For instance, Wang and Lim~\cite{wang2023zero} design different prompting strategies to investigate the performance of GPT-3~\cite{kojima2022large} for next-item prediction. 
Along this line of research, Hou \textit{et al.}~\cite{hou2023large} devise various prompting templates and formalize the sequential recommendation as a conditional ranking task. Analyzing the zero-shot learning capability of GPT-3.5. 
Apart from that, TALLRec~\cite{bao2023tallrec}, applies LoRA~\cite{hu2021lora} to effectively fine-tune LLaMA~\cite{touvron2023llama} on recommendation datasets. 
M6-Rec~\cite{cui2022m6} obtains M6~\cite{lin2021m6}, a visual-linguistic pre-trained model, as the backbone and proposes an improved prompt tuning, named option tuning, for task-specific parameter fine-tuning.
\textcolor{black}{To alleviate the hallucination and improve the quality of outputs, retrieval-augmented generation (RAG) technology has become ubiquitous in LLM~\cite{gao2023retrieval}, where the external knowledge can be retrieved as auxiliary information to guide the LLM for generation. In recommendation with LLM, based on the user's historical interacted items, RAG is employed as a retriever for candidate selection or reranking~\cite{deldjoo2024review, zeng2024federated}. \cite{contal2024ragsys,carraro2024enhancing} show that RAG is facilitated to address the cold-start problem and enhance the diversity of recommendation.}
However, there are few efforts that explore the application of LLM and RAG in SR, leaving this research direction to be cultivated.
Unlike the auto-regressive generation methods used in LLMs, diffusion models have established a new paradigm for generative tasks and have achieved remarkable success across a broad spectrum of applications~\cite{croitoru2023diffusion, yang2023diffusion, li2023diffurec, chen2024temporal}. Overall, diffusion model can be split into two stages, the diffusion stage aiming to corrupt the original input as a Gaussian distribution, and the reverse stage where to recover the data from a Gaussian noise iteratively conditioned on the input. Diffurec~\cite{li2023diffurec} is the first work that applies diffusion for sequential recommendation.
Besides, some works explore the diffusion model in CTR prediction~\cite{wang2023diffusion}, multi-scenario recommendation~\cite{wang2024diff}. However, no works investigate the performance of diffusion model in SR.
The discrete embedding space and the time cost in the reverse stage also impede the widespread application in on-line recommendation. The diffusion model is an uphill research area and I believe it has a promising prospect in SR.

\section{Conclusion}
\label{sec:conclustion}

This paper systematically investigated session-based recommendation with GNNs and sequential neural networks. Specifically, we first clarify the corresponding definitions, and concepts, analyzing the features of SR. Then, we review more than 150 papers and provide a systematic taxonomy to organize the existing works with regard to their key motivations and typical models. Besides, we conduct a unified framework for SR with GNNs and sequential neural networks respectively. 
After that, a detailed introduction to the key modules including neighbor session selection, graph construction, embedding, information propagation and aggregation, sequence modeling, session representation, prediction, and loss function, are also provided. 
Finally, we introduce the challenges and point out new potential directions for the research on SR. 
It is our hope that this survey can provide readers with a comprehensive understanding of the key aspects, main challenges, and notable progress in this area, and shed some light on future research.

\bibliographystyle{ACM-Reference-Format}
\bibliography{csur_sim}


\begin{thebibliography}{244}


\ifx \showCODEN    \undefined \def \showCODEN     #1{\unskip}     \fi
\ifx \showDOI      \undefined \def \showDOI       #1{#1}\fi
\ifx \showISBNx    \undefined \def \showISBNx     #1{\unskip}     \fi
\ifx \showISBNxiii \undefined \def \showISBNxiii  #1{\unskip}     \fi
\ifx \showISSN     \undefined \def \showISSN      #1{\unskip}     \fi
\ifx \showLCCN     \undefined \def \showLCCN      #1{\unskip}     \fi
\ifx \shownote     \undefined \def \shownote      #1{#1}          \fi
\ifx \showarticletitle \undefined \def \showarticletitle #1{#1}   \fi
\ifx \showURL      \undefined \def \showURL       {\relax}        \fi
\providecommand\bibfield[2]{#2}
\providecommand\bibinfo[2]{#2}
\providecommand\natexlab[1]{#1}
\providecommand\showeprint[2][]{arXiv:#2}

\bibitem[Adomavicius and Tuzhilin(2005)]%
        {adomavicius2005toward}
\bibfield{author}{\bibinfo{person}{Gediminas Adomavicius} {and} \bibinfo{person}{Alexander Tuzhilin}.} \bibinfo{year}{2005}\natexlab{}.
\newblock \showarticletitle{Toward the next generation of recommender systems: A survey of the state-of-the-art and possible extensions}.
\newblock \bibinfo{journal}{\emph{IEEE transactions on knowledge and data engineering}} (\bibinfo{year}{2005}).
\newblock


\bibitem[Balabanovi{\'c} and Shoham(1997)]%
        {balabanovic1997fab}
\bibfield{author}{\bibinfo{person}{Marko Balabanovi{\'c}} {and} \bibinfo{person}{Yoav Shoham}.} \bibinfo{year}{1997}\natexlab{}.
\newblock \showarticletitle{Fab: content-based, collaborative recommendation}.
\newblock \bibinfo{journal}{\emph{Commun. ACM}} (\bibinfo{year}{1997}).
\newblock


\bibitem[Bao et~al\mbox{.}(2023)]%
        {bao2023tallrec}
\bibfield{author}{\bibinfo{person}{Keqin Bao}, \bibinfo{person}{Jizhi Zhang}, \bibinfo{person}{Yang Zhang}, \bibinfo{person}{Wenjie Wang}, \bibinfo{person}{Fuli Feng}, {and} \bibinfo{person}{Xiangnan He}.} \bibinfo{year}{2023}\natexlab{}.
\newblock \showarticletitle{Tallrec: An effective and efficient tuning framework to align large language model with recommendation}. In \bibinfo{booktitle}{\emph{Proceedings of the 17th ACM Conference on Recommender Systems}}. \bibinfo{pages}{1007--1014}.
\newblock


\bibitem[Bengio et~al\mbox{.}(2000)]%
        {bengio2000neural}
\bibfield{author}{\bibinfo{person}{Yoshua Bengio}, \bibinfo{person}{R{\'e}jean Ducharme}, {and} \bibinfo{person}{Pascal Vincent}.} \bibinfo{year}{2000}\natexlab{}.
\newblock \showarticletitle{A neural probabilistic language model}.
\newblock \bibinfo{journal}{\emph{Proc. of NeurIPS}} (\bibinfo{year}{2000}).
\newblock


\bibitem[Bojchevski and G{\"u}nnemann(2018)]%
        {bojchevski2017deep}
\bibfield{author}{\bibinfo{person}{Aleksandar Bojchevski} {and} \bibinfo{person}{Stephan G{\"u}nnemann}.} \bibinfo{year}{2018}\natexlab{}.
\newblock \showarticletitle{Deep Gaussian Embedding of Graphs: Unsupervised Inductive Learning via Ranking}. In \bibinfo{booktitle}{\emph{Proc. of ICLR}}.
\newblock


\bibitem[Bonnin and Jannach(2014)]%
        {bonnin2014automated}
\bibfield{author}{\bibinfo{person}{Geoffray Bonnin} {and} \bibinfo{person}{Dietmar Jannach}.} \bibinfo{year}{2014}\natexlab{}.
\newblock \showarticletitle{Automated generation of music playlists: Survey and experiments}.
\newblock \bibinfo{journal}{\emph{ACM Computing Surveys (CSUR)}} (\bibinfo{year}{2014}).
\newblock


\bibitem[Bordes et~al\mbox{.}(2013)]%
        {bordes2013translating}
\bibfield{author}{\bibinfo{person}{Antoine Bordes}, \bibinfo{person}{Nicolas Usunier}, \bibinfo{person}{Alberto Garcia-Duran}, \bibinfo{person}{Jason Weston}, {and} \bibinfo{person}{Oksana Yakhnenko}.} \bibinfo{year}{2013}\natexlab{}.
\newblock \showarticletitle{Translating embeddings for modeling multi-relational data}.
\newblock \bibinfo{journal}{\emph{Proc. of NeurIPS}} (\bibinfo{year}{2013}).
\newblock


\bibitem[Bourtoule et~al\mbox{.}(2021)]%
        {bourtoule2021machine}
\bibfield{author}{\bibinfo{person}{Lucas Bourtoule}, \bibinfo{person}{Varun Chandrasekaran}, \bibinfo{person}{Christopher~A Choquette-Choo}, \bibinfo{person}{Hengrui Jia}, \bibinfo{person}{Adelin Travers}, \bibinfo{person}{Baiwu Zhang}, \bibinfo{person}{David Lie}, {and} \bibinfo{person}{Nicolas Papernot}.} \bibinfo{year}{2021}\natexlab{}.
\newblock \showarticletitle{Machine unlearning}. In \bibinfo{booktitle}{\emph{Proc. of SP}}. \bibinfo{pages}{141--159}.
\newblock


\bibitem[Carraro and Bridge(2024)]%
        {carraro2024enhancing}
\bibfield{author}{\bibinfo{person}{Diego Carraro} {and} \bibinfo{person}{Derek Bridge}.} \bibinfo{year}{2024}\natexlab{}.
\newblock \showarticletitle{Enhancing Recommendation Diversity by Re-ranking with Large Language Models}.
\newblock \bibinfo{journal}{\emph{arXiv preprint arXiv:2401.11506}} (\bibinfo{year}{2024}).
\newblock


\bibitem[Castells et~al\mbox{.}(2021)]%
        {castells2021novelty}
\bibfield{author}{\bibinfo{person}{Pablo Castells}, \bibinfo{person}{Neil Hurley}, {and} \bibinfo{person}{Saul Vargas}.} \bibinfo{year}{2021}\natexlab{}.
\newblock \showarticletitle{Novelty and diversity in recommender systems}.
\newblock In \bibinfo{booktitle}{\emph{Recommender systems handbook}}. \bibinfo{pages}{603--646}.
\newblock


\bibitem[Celma~Herrada et~al\mbox{.}(2009)]%
        {celma2009music}
\bibfield{author}{\bibinfo{person}{{\`O}scar Celma~Herrada} {et~al\mbox{.}}} \bibinfo{year}{2009}\natexlab{}.
\newblock \bibinfo{booktitle}{\emph{Music recommendation and discovery in the long tail}}.
\newblock \bibinfo{publisher}{Universitat Pompeu Fabra}.
\newblock


\bibitem[Chen et~al\mbox{.}(2021b)]%
        {chen2021dual}
\bibfield{author}{\bibinfo{person}{Chen Chen}, \bibinfo{person}{Jie Guo}, {and} \bibinfo{person}{Bin Song}.} \bibinfo{year}{2021}\natexlab{b}.
\newblock \showarticletitle{Dual attention transfer in session-based recommendation with multi-dimensional integration}. In \bibinfo{booktitle}{\emph{Proc. of SIGIR}}.
\newblock


\bibitem[Chen et~al\mbox{.}(2021a)]%
        {chen2021generate}
\bibfield{author}{\bibinfo{person}{Hanxiong Chen}, \bibinfo{person}{Xu Chen}, \bibinfo{person}{Shaoyun Shi}, {and} \bibinfo{person}{Yongfeng Zhang}.} \bibinfo{year}{2021}\natexlab{a}.
\newblock \showarticletitle{Generate natural language explanations for recommendation}.
\newblock \bibinfo{journal}{\emph{ArXiv preprint}} (\bibinfo{year}{2021}).
\newblock


\bibitem[Chen et~al\mbox{.}(2023a)]%
        {chen2020bias}
\bibfield{author}{\bibinfo{person}{Jiawei Chen}, \bibinfo{person}{Hande Dong}, \bibinfo{person}{Xiang Wang}, \bibinfo{person}{Fuli Feng}, \bibinfo{person}{Meng Wang}, {and} \bibinfo{person}{Xiangnan He}.} \bibinfo{year}{2023}\natexlab{a}.
\newblock \showarticletitle{Bias and debias in recommender system: A survey and future directions}.
\newblock \bibinfo{journal}{\emph{ACM Transactions on Information Systems}} (\bibinfo{year}{2023}), \bibinfo{pages}{1--39}.
\newblock


\bibitem[Chen et~al\mbox{.}(2021c)]%
        {chen2021session}
\bibfield{author}{\bibinfo{person}{Jinpeng Chen}, \bibinfo{person}{Haiyang Li}, \bibinfo{person}{Fan Zhang}, \bibinfo{person}{Senzhang Wang}, {and} \bibinfo{person}{Kaimin Wei}.} \bibinfo{year}{2021}\natexlab{c}.
\newblock \showarticletitle{Session-based Recommendation with Heterogeneous Graph Neural Network}.
\newblock \bibinfo{journal}{\emph{arXiv preprint arXiv:2108.05641}} (\bibinfo{year}{2021}).
\newblock


\bibitem[Chen and Zheng(2021)]%
        {chen2021incorporating}
\bibfield{author}{\bibinfo{person}{Minghao Chen} {and} \bibinfo{person}{Jiale Zheng}.} \bibinfo{year}{2021}\natexlab{}.
\newblock \showarticletitle{Incorporating Adjacent User Modeling into Session-based Recommendation with Graph Neural Networks}. In \bibinfo{booktitle}{\emph{Proc. of ICDM}}.
\newblock


\bibitem[Chen et~al\mbox{.}(2023b)]%
        {chen2023knowledge}
\bibfield{author}{\bibinfo{person}{Qian Chen}, \bibinfo{person}{Zhiqiang Guo}, \bibinfo{person}{Jianjun Li}, {and} \bibinfo{person}{Guohui Li}.} \bibinfo{year}{2023}\natexlab{b}.
\newblock \showarticletitle{Knowledge-enhanced Multi-View Graph Neural Networks for Session-based Recommendation}. In \bibinfo{booktitle}{\emph{Proc. of SIGIR}}.
\newblock


\bibitem[Chen et~al\mbox{.}(2023c)]%
        {chen2023attribute}
\bibfield{author}{\bibinfo{person}{Qian Chen}, \bibinfo{person}{Jianjun Li}, \bibinfo{person}{Zhiqiang Guo}, \bibinfo{person}{Guohui Li}, {and} \bibinfo{person}{Zhiying Deng}.} \bibinfo{year}{2023}\natexlab{c}.
\newblock \showarticletitle{Attribute-enhanced dual channel representation learning for session-based recommendation}. In \bibinfo{booktitle}{\emph{Proc. of CIKM}}. \bibinfo{pages}{3793--3797}.
\newblock


\bibitem[Chen et~al\mbox{.}(2012)]%
        {chen2012playlist}
\bibfield{author}{\bibinfo{person}{Shuo Chen}, \bibinfo{person}{Josh~L Moore}, \bibinfo{person}{Douglas Turnbull}, {and} \bibinfo{person}{Thorsten Joachims}.} \bibinfo{year}{2012}\natexlab{}.
\newblock \showarticletitle{Playlist prediction via metric embedding}. In \bibinfo{booktitle}{\emph{Proc. of KDD}}.
\newblock


\bibitem[Chen and Wong(2019)]%
        {chen2019session}
\bibfield{author}{\bibinfo{person}{Tianwen Chen} {and} \bibinfo{person}{Raymond Chi-Wing Wong}.} \bibinfo{year}{2019}\natexlab{}.
\newblock \showarticletitle{Session-based recommendation with local invariance}. In \bibinfo{booktitle}{\emph{Proc. of ICDM}}.
\newblock


\bibitem[Chen and Wong(2020)]%
        {chen2020handling}
\bibfield{author}{\bibinfo{person}{Tianwen Chen} {and} \bibinfo{person}{Raymond Chi-Wing Wong}.} \bibinfo{year}{2020}\natexlab{}.
\newblock \showarticletitle{Handling information loss of graph neural networks for session-based recommendation}. In \bibinfo{booktitle}{\emph{Proc. of KDD}}.
\newblock


\bibitem[Chen and Wong(2021)]%
        {chen2021efficient}
\bibfield{author}{\bibinfo{person}{Tianwen Chen} {and} \bibinfo{person}{Raymond Chi-Wing Wong}.} \bibinfo{year}{2021}\natexlab{}.
\newblock \showarticletitle{An efficient and effective framework for session-based social recommendation}. In \bibinfo{booktitle}{\emph{Proc. of WSDM}}.
\newblock


\bibitem[Chen et~al\mbox{.}(2019a)]%
        {chen2019dynamic}
\bibfield{author}{\bibinfo{person}{Wanyu Chen}, \bibinfo{person}{Fei Cai}, \bibinfo{person}{Honghui Chen}, {and} \bibinfo{person}{Maarten de Rijke}.} \bibinfo{year}{2019}\natexlab{a}.
\newblock \showarticletitle{A dynamic co-attention network for session-based recommendation}. In \bibinfo{booktitle}{\emph{Proc. of CIKM}}.
\newblock


\bibitem[Chen et~al\mbox{.}(2019b)]%
        {chen2019personalized}
\bibfield{author}{\bibinfo{person}{Xu Chen}, \bibinfo{person}{Hanxiong Chen}, \bibinfo{person}{Hongteng Xu}, \bibinfo{person}{Yongfeng Zhang}, \bibinfo{person}{Yixin Cao}, \bibinfo{person}{Zheng Qin}, {and} \bibinfo{person}{Hongyuan Zha}.} \bibinfo{year}{2019}\natexlab{b}.
\newblock \showarticletitle{Personalized Fashion Recommendation with Visual Explanations based on Multimodal Attention Network: Towards Visually Explainable Recommendation}. In \bibinfo{booktitle}{\emph{Proc. of SIGIR}}. \bibinfo{pages}{765--774}.
\newblock


\bibitem[Chen et~al\mbox{.}(2018)]%
        {chen2018sequential}
\bibfield{author}{\bibinfo{person}{Xu Chen}, \bibinfo{person}{Hongteng Xu}, \bibinfo{person}{Yongfeng Zhang}, \bibinfo{person}{Jiaxi Tang}, \bibinfo{person}{Yixin Cao}, \bibinfo{person}{Zheng Qin}, {and} \bibinfo{person}{Hongyuan Zha}.} \bibinfo{year}{2018}\natexlab{}.
\newblock \showarticletitle{Sequential recommendation with user memory networks}. In \bibinfo{booktitle}{\emph{Proc. of WSDM}}.
\newblock


\bibitem[Chen et~al\mbox{.}(2021d)]%
        {chen2021survey}
\bibfield{author}{\bibinfo{person}{Xiaocong Chen}, \bibinfo{person}{Lina Yao}, \bibinfo{person}{Julian McAuley}, \bibinfo{person}{Guanglin Zhou}, {and} \bibinfo{person}{Xianzhi Wang}.} \bibinfo{year}{2021}\natexlab{d}.
\newblock \showarticletitle{A survey of deep reinforcement learning in recommender systems: A systematic review and future directions}.
\newblock \bibinfo{journal}{\emph{arXiv preprint arXiv:2109.03540}} (\bibinfo{year}{2021}).
\newblock


\bibitem[Chen et~al\mbox{.}(2024)]%
        {chen2024temporal}
\bibfield{author}{\bibinfo{person}{Yakun Chen}, \bibinfo{person}{Kaize Shi}, \bibinfo{person}{Zhangkai Wu}, \bibinfo{person}{Juan Chen}, \bibinfo{person}{Xianzhi Wang}, \bibinfo{person}{Julian McAuley}, \bibinfo{person}{Guandong Xu}, {and} \bibinfo{person}{Shui Yu}.} \bibinfo{year}{2024}\natexlab{}.
\newblock \showarticletitle{Temporal Disentangled Contrastive Diffusion Model for Spatiotemporal Imputation}.
\newblock \bibinfo{journal}{\emph{arXiv preprint arXiv:2402.11558}} (\bibinfo{year}{2024}).
\newblock


\bibitem[Cheng et~al\mbox{.}(2013)]%
        {cheng2013you}
\bibfield{author}{\bibinfo{person}{Chen Cheng}, \bibinfo{person}{Haiqin Yang}, \bibinfo{person}{Michael~R Lyu}, {and} \bibinfo{person}{Irwin King}.} \bibinfo{year}{2013}\natexlab{}.
\newblock \showarticletitle{Where you like to go next: Successive point-of-interest recommendation}. In \bibinfo{booktitle}{\emph{Proc. of IJCAI}}.
\newblock


\bibitem[Choi et~al\mbox{.}(2022)]%
        {choi2022s}
\bibfield{author}{\bibinfo{person}{Minjin Choi}, \bibinfo{person}{Jinhong Kim}, \bibinfo{person}{Joonseok Lee}, \bibinfo{person}{Hyunjung Shim}, {and} \bibinfo{person}{Jongwuk Lee}.} \bibinfo{year}{2022}\natexlab{}.
\newblock \showarticletitle{S-Walk: Accurate and scalable session-based recommendation with random walks}. In \bibinfo{booktitle}{\emph{Proc. of WSDM}}. \bibinfo{pages}{150--160}.
\newblock


\bibitem[Clarke et~al\mbox{.}(2008)]%
        {clarke2008novelty}
\bibfield{author}{\bibinfo{person}{Charles~LA Clarke}, \bibinfo{person}{Maheedhar Kolla}, \bibinfo{person}{Gordon~V Cormack}, \bibinfo{person}{Olga Vechtomova}, \bibinfo{person}{Azin Ashkan}, \bibinfo{person}{Stefan B{\"u}ttcher}, {and} \bibinfo{person}{Ian MacKinnon}.} \bibinfo{year}{2008}\natexlab{}.
\newblock \showarticletitle{Novelty and diversity in information retrieval evaluation}. In \bibinfo{booktitle}{\emph{Proc. of SIGIR}}. \bibinfo{pages}{659--666}.
\newblock


\bibitem[Collobert and Weston(2008)]%
        {collobert2008unified}
\bibfield{author}{\bibinfo{person}{Ronan Collobert} {and} \bibinfo{person}{Jason Weston}.} \bibinfo{year}{2008}\natexlab{}.
\newblock \showarticletitle{A unified architecture for natural language processing: Deep neural networks with multitask learning}. In \bibinfo{booktitle}{\emph{Proc. of ICML}}.
\newblock


\bibitem[Conneau et~al\mbox{.}(2017)]%
        {conneau2017supervised}
\bibfield{author}{\bibinfo{person}{Alexis Conneau}, \bibinfo{person}{Douwe Kiela}, \bibinfo{person}{Holger Schwenk}, \bibinfo{person}{Loic Barrault}, {and} \bibinfo{person}{Antoine Bordes}.} \bibinfo{year}{2017}\natexlab{}.
\newblock \showarticletitle{Supervised learning of universal sentence representations from natural language inference data}.
\newblock \bibinfo{journal}{\emph{arXiv preprint arXiv:1705.02364}} (\bibinfo{year}{2017}).
\newblock


\bibitem[Contal and McGoldrick(2024)]%
        {contal2024ragsys}
\bibfield{author}{\bibinfo{person}{Emile Contal} {and} \bibinfo{person}{Garrin McGoldrick}.} \bibinfo{year}{2024}\natexlab{}.
\newblock \showarticletitle{RAGSys: Item-Cold-Start Recommender as RAG System}.
\newblock \bibinfo{journal}{\emph{arXiv preprint arXiv:2405.17587}} (\bibinfo{year}{2024}).
\newblock


\bibitem[Covington et~al\mbox{.}(2016)]%
        {covington2016deep}
\bibfield{author}{\bibinfo{person}{Paul Covington}, \bibinfo{person}{Jay Adams}, {and} \bibinfo{person}{Emre Sargin}.} \bibinfo{year}{2016}\natexlab{}.
\newblock \showarticletitle{Deep neural networks for youtube recommendations}. In \bibinfo{booktitle}{\emph{Proc. of RecSys}}.
\newblock


\bibitem[Croitoru et~al\mbox{.}(2023)]%
        {croitoru2023diffusion}
\bibfield{author}{\bibinfo{person}{Florinel-Alin Croitoru}, \bibinfo{person}{Vlad Hondru}, \bibinfo{person}{Radu~Tudor Ionescu}, {and} \bibinfo{person}{Mubarak Shah}.} \bibinfo{year}{2023}\natexlab{}.
\newblock \showarticletitle{Diffusion models in vision: A survey}.
\newblock \bibinfo{journal}{\emph{IEEE Transactions on Pattern Analysis and Machine Intelligence}} (\bibinfo{year}{2023}), \bibinfo{pages}{10850--10869}.
\newblock


\bibitem[Cui et~al\mbox{.}(2022b)]%
        {cui2022intention}
\bibfield{author}{\bibinfo{person}{Chuan Cui}, \bibinfo{person}{Qi Shen}, \bibinfo{person}{Shixuan Zhu}, \bibinfo{person}{Yitong Pang}, \bibinfo{person}{Yiming Zhang}, \bibinfo{person}{Hanning Gao}, {and} \bibinfo{person}{Zhihua Wei}.} \bibinfo{year}{2022}\natexlab{b}.
\newblock \showarticletitle{Intention Adaptive Graph Neural Network for Category-Aware Session-Based Recommendation}. In \bibinfo{booktitle}{\emph{Proc. of DASFAA}}.
\newblock


\bibitem[Cui et~al\mbox{.}(2022a)]%
        {cui2022m6}
\bibfield{author}{\bibinfo{person}{Zeyu Cui}, \bibinfo{person}{Jianxin Ma}, \bibinfo{person}{Chang Zhou}, \bibinfo{person}{Jingren Zhou}, {and} \bibinfo{person}{Hongxia Yang}.} \bibinfo{year}{2022}\natexlab{a}.
\newblock \showarticletitle{M6-rec: Generative pretrained language models are open-ended recommender systems}.
\newblock \bibinfo{journal}{\emph{arXiv preprint arXiv:2205.08084}} (\bibinfo{year}{2022}).
\newblock


\bibitem[Davidson et~al\mbox{.}(2010)]%
        {davidson2010youtube}
\bibfield{author}{\bibinfo{person}{James Davidson}, \bibinfo{person}{Benjamin Liebald}, \bibinfo{person}{Junning Liu}, \bibinfo{person}{Palash Nandy}, \bibinfo{person}{Taylor Van~Vleet}, \bibinfo{person}{Ullas Gargi}, \bibinfo{person}{Sujoy Gupta}, \bibinfo{person}{Yu He}, \bibinfo{person}{Mike Lambert}, \bibinfo{person}{Blake Livingston}, {et~al\mbox{.}}} \bibinfo{year}{2010}\natexlab{}.
\newblock \showarticletitle{The YouTube video recommendation system}. In \bibinfo{booktitle}{\emph{Proc. of RecSys}}.
\newblock


\bibitem[de~Souza Pereira~Moreira et~al\mbox{.}(2021)]%
        {de2021transformers4rec}
\bibfield{author}{\bibinfo{person}{Gabriel de Souza Pereira~Moreira}, \bibinfo{person}{Sara Rabhi}, \bibinfo{person}{Jeong~Min Lee}, \bibinfo{person}{Ronay Ak}, {and} \bibinfo{person}{Even Oldridge}.} \bibinfo{year}{2021}\natexlab{}.
\newblock \showarticletitle{Transformers4Rec: Bridging the Gap between NLP and Sequential/Session-Based Recommendation}. In \bibinfo{booktitle}{\emph{Proc. of RecSys}}.
\newblock


\bibitem[Deldjoo et~al\mbox{.}(2024)]%
        {deldjoo2024review}
\bibfield{author}{\bibinfo{person}{Yashar Deldjoo}, \bibinfo{person}{Zhankui He}, \bibinfo{person}{Julian McAuley}, \bibinfo{person}{Anton Korikov}, \bibinfo{person}{Scott Sanner}, \bibinfo{person}{Arnau Ramisa}, \bibinfo{person}{Ren{\'e} Vidal}, \bibinfo{person}{Maheswaran Sathiamoorthy}, \bibinfo{person}{Atoosa Kasirzadeh}, {and} \bibinfo{person}{Silvia Milano}.} \bibinfo{year}{2024}\natexlab{}.
\newblock \showarticletitle{A Review of Modern Recommender Systems Using Generative Models (Gen-RecSys)}.
\newblock \bibinfo{journal}{\emph{arXiv preprint arXiv:2404.00579}} (\bibinfo{year}{2024}).
\newblock


\bibitem[Deng and Hooi(2021)]%
        {deng2021graph}
\bibfield{author}{\bibinfo{person}{Ailin Deng} {and} \bibinfo{person}{Bryan Hooi}.} \bibinfo{year}{2021}\natexlab{}.
\newblock \showarticletitle{Graph neural network-based anomaly detection in multivariate time series}. In \bibinfo{booktitle}{\emph{Proc. of AAAI}}.
\newblock


\bibitem[Deng et~al\mbox{.}(2022)]%
        {deng2022g}
\bibfield{author}{\bibinfo{person}{Zhi-Hong Deng}, \bibinfo{person}{Chang-Dong Wang}, \bibinfo{person}{Ling Huang}, \bibinfo{person}{Jian-Huang Lai}, {and} \bibinfo{person}{S~Yu Philip}.} \bibinfo{year}{2022}\natexlab{}.
\newblock \showarticletitle{G\^{} 3SR: Global Graph Guided Session-Based Recommendation}.
\newblock \bibinfo{journal}{\emph{IEEE Transactions on Neural Networks and Learning Systems}} (\bibinfo{year}{2022}).
\newblock


\bibitem[Ebesu et~al\mbox{.}(2018)]%
        {ebesu2018collaborative}
\bibfield{author}{\bibinfo{person}{Travis Ebesu}, \bibinfo{person}{Bin Shen}, {and} \bibinfo{person}{Yi Fang}.} \bibinfo{year}{2018}\natexlab{}.
\newblock \showarticletitle{Collaborative memory network for recommendation systems}. In \bibinfo{booktitle}{\emph{Proc. of SIGIR}}.
\newblock


\bibitem[Eirinaki et~al\mbox{.}(2005)]%
        {eirinaki2005web}
\bibfield{author}{\bibinfo{person}{Magdalini Eirinaki}, \bibinfo{person}{Michalis Vazirgiannis}, {and} \bibinfo{person}{Dimitris Kapogiannis}.} \bibinfo{year}{2005}\natexlab{}.
\newblock \showarticletitle{Web path recommendations based on page ranking and markov models}. In \bibinfo{booktitle}{\emph{Proceedings of the 7th annual ACM international workshop on Web information and data management}}.
\newblock


\bibitem[Elahi et~al\mbox{.}(2024)]%
        {elahi2024knowledge}
\bibfield{author}{\bibinfo{person}{Ehsan Elahi}, \bibinfo{person}{Sajid Anwar}, \bibinfo{person}{Babar Shah}, \bibinfo{person}{Zahid Halim}, \bibinfo{person}{Abrar Ullah}, \bibinfo{person}{Imad Rida}, {and} \bibinfo{person}{Muhammad Waqas}.} \bibinfo{year}{2024}\natexlab{}.
\newblock \showarticletitle{Knowledge Graph Enhanced Contextualized Attention-Based Network for Responsible User-Specific Recommendation}.
\newblock \bibinfo{journal}{\emph{ACM Transactions on Intelligent Systems and Technology}} (\bibinfo{year}{2024}).
\newblock


\bibitem[Fan et~al\mbox{.}(2021)]%
        {fan2021modeling}
\bibfield{author}{\bibinfo{person}{Ziwei Fan}, \bibinfo{person}{Zhiwei Liu}, \bibinfo{person}{Shen Wang}, \bibinfo{person}{Lei Zheng}, {and} \bibinfo{person}{Philip~S Yu}.} \bibinfo{year}{2021}\natexlab{}.
\newblock \showarticletitle{Modeling Sequences as Distributions with Uncertainty for Sequential Recommendation}. In \bibinfo{booktitle}{\emph{Proc. of CIKM}}.
\newblock


\bibitem[Fan et~al\mbox{.}(2022)]%
        {fan2022sequential}
\bibfield{author}{\bibinfo{person}{Ziwei Fan}, \bibinfo{person}{Zhiwei Liu}, \bibinfo{person}{Yu Wang}, \bibinfo{person}{Alice Wang}, \bibinfo{person}{Zahra Nazari}, \bibinfo{person}{Lei Zheng}, \bibinfo{person}{Hao Peng}, {and} \bibinfo{person}{Philip~S Yu}.} \bibinfo{year}{2022}\natexlab{}.
\newblock \showarticletitle{Sequential recommendation via stochastic self-attention}. In \bibinfo{booktitle}{\emph{Proc. of WWW}}. \bibinfo{pages}{2036--2047}.
\newblock


\bibitem[Fang et~al\mbox{.}(2020)]%
        {fang2020deep}
\bibfield{author}{\bibinfo{person}{Hui Fang}, \bibinfo{person}{Danning Zhang}, \bibinfo{person}{Yiheng Shu}, {and} \bibinfo{person}{Guibing Guo}.} \bibinfo{year}{2020}\natexlab{}.
\newblock \showarticletitle{Deep learning for sequential recommendation: Algorithms, influential factors, and evaluations}.
\newblock \bibinfo{journal}{\emph{ACM Transactions on Information Systems}} (\bibinfo{year}{2020}).
\newblock


\bibitem[Feng et~al\mbox{.}(2015)]%
        {feng2015personalized}
\bibfield{author}{\bibinfo{person}{Shanshan Feng}, \bibinfo{person}{Xutao Li}, \bibinfo{person}{Yifeng Zeng}, \bibinfo{person}{Gao Cong}, \bibinfo{person}{Yeow~Meng Chee}, {and} \bibinfo{person}{Quan Yuan}.} \bibinfo{year}{2015}\natexlab{}.
\newblock \showarticletitle{Personalized ranking metric embedding for next new poi recommendation}. In \bibinfo{booktitle}{\emph{Proc. of IJCAI}}.
\newblock


\bibitem[Forsati et~al\mbox{.}(2009)]%
        {forsati2009web}
\bibfield{author}{\bibinfo{person}{Rana Forsati}, \bibinfo{person}{Mohammad~Reza Meybodi}, {and} \bibinfo{person}{A~Ghari Neiat}.} \bibinfo{year}{2009}\natexlab{}.
\newblock \showarticletitle{Web page personalization based on weighted association rules}. In \bibinfo{booktitle}{\emph{2009 International Conference on Electronic Computer Technology}}.
\newblock


\bibitem[Fu et~al\mbox{.}(2020)]%
        {fu2020fairness}
\bibfield{author}{\bibinfo{person}{Zuohui Fu}, \bibinfo{person}{Yikun Xian}, \bibinfo{person}{Ruoyuan Gao}, \bibinfo{person}{Jieyu Zhao}, \bibinfo{person}{Qiaoying Huang}, \bibinfo{person}{Yingqiang Ge}, \bibinfo{person}{Shuyuan Xu}, \bibinfo{person}{Shijie Geng}, \bibinfo{person}{Chirag Shah}, \bibinfo{person}{Yongfeng Zhang}, {and} \bibinfo{person}{Gerard de Melo}.} \bibinfo{year}{2020}\natexlab{}.
\newblock \showarticletitle{Fairness-Aware Explainable Recommendation over Knowledge Graphs}. In \bibinfo{booktitle}{\emph{Proc. of SIGIR}}. \bibinfo{pages}{69--78}.
\newblock


\bibitem[Gao et~al\mbox{.}(2023)]%
        {gao2023retrieval}
\bibfield{author}{\bibinfo{person}{Yunfan Gao}, \bibinfo{person}{Yun Xiong}, \bibinfo{person}{Xinyu Gao}, \bibinfo{person}{Kangxiang Jia}, \bibinfo{person}{Jinliu Pan}, \bibinfo{person}{Yuxi Bi}, \bibinfo{person}{Yi Dai}, \bibinfo{person}{Jiawei Sun}, {and} \bibinfo{person}{Haofen Wang}.} \bibinfo{year}{2023}\natexlab{}.
\newblock \showarticletitle{Retrieval-augmented generation for large language models: A survey}.
\newblock \bibinfo{journal}{\emph{arXiv preprint arXiv:2312.10997}} (\bibinfo{year}{2023}).
\newblock


\bibitem[Geng et~al\mbox{.}(2022)]%
        {geng2022recommendation}
\bibfield{author}{\bibinfo{person}{Shijie Geng}, \bibinfo{person}{Shuchang Liu}, \bibinfo{person}{Zuohui Fu}, \bibinfo{person}{Yingqiang Ge}, {and} \bibinfo{person}{Yongfeng Zhang}.} \bibinfo{year}{2022}\natexlab{}.
\newblock \showarticletitle{Recommendation as language processing (rlp): A unified pretrain, personalized prompt \& predict paradigm (p5)}. In \bibinfo{booktitle}{\emph{Proceedings of the 16th ACM Conference on Recommender Systems}}. \bibinfo{pages}{299--315}.
\newblock


\bibitem[Gharahighehi and Vens(2020)]%
        {gharahighehi2020making}
\bibfield{author}{\bibinfo{person}{Alireza Gharahighehi} {and} \bibinfo{person}{Celine Vens}.} \bibinfo{year}{2020}\natexlab{}.
\newblock \showarticletitle{Making session-based news recommenders diversity-aware}. In \bibinfo{booktitle}{\emph{Proceedings of the Workshop on Online Misinformation-and Harm-Aware Recommender Systems}}.
\newblock


\bibitem[Gong and Zhu(2022)]%
        {gong2022positive}
\bibfield{author}{\bibinfo{person}{Shansan Gong} {and} \bibinfo{person}{Kenny~Q Zhu}.} \bibinfo{year}{2022}\natexlab{}.
\newblock \showarticletitle{Positive, Negative and Neutral: Modeling Implicit Feedback in Session-based News Recommendation}. In \bibinfo{booktitle}{\emph{Proc. of SIGIR}}.
\newblock


\bibitem[Grover and Leskovec(2016)]%
        {grover2016node2vec}
\bibfield{author}{\bibinfo{person}{Aditya Grover} {and} \bibinfo{person}{Jure Leskovec}.} \bibinfo{year}{2016}\natexlab{}.
\newblock \showarticletitle{node2vec: Scalable feature learning for networks}. In \bibinfo{booktitle}{\emph{Proc. of KDD}}.
\newblock


\bibitem[Guo et~al\mbox{.}(2020)]%
        {guo2020session}
\bibfield{author}{\bibinfo{person}{Cheng Guo}, \bibinfo{person}{Mengfei Zhang}, \bibinfo{person}{Jinyun Fang}, \bibinfo{person}{Jiaqi Jin}, {and} \bibinfo{person}{Mao Pan}.} \bibinfo{year}{2020}\natexlab{}.
\newblock \showarticletitle{Session-based recommendation with hierarchical leaping networks}. In \bibinfo{booktitle}{\emph{Proc. of SIGIR}}.
\newblock


\bibitem[Guo et~al\mbox{.}(2022a)]%
        {guo2022learning}
\bibfield{author}{\bibinfo{person}{Jiayan Guo}, \bibinfo{person}{Yaming Yang}, \bibinfo{person}{Xiangchen Song}, \bibinfo{person}{Yuan Zhang}, \bibinfo{person}{Yujing Wang}, \bibinfo{person}{Jing Bai}, {and} \bibinfo{person}{Yan Zhang}.} \bibinfo{year}{2022}\natexlab{a}.
\newblock \showarticletitle{Learning Multi-granularity Consecutive User Intent Unit for Session-based Recommendation}. In \bibinfo{booktitle}{\emph{Proc. of WSDM}}.
\newblock


\bibitem[Guo et~al\mbox{.}(2022b)]%
        {guo2022evolutionary}
\bibfield{author}{\bibinfo{person}{Jiayan Guo}, \bibinfo{person}{Peiyan Zhang}, \bibinfo{person}{Chaozhuo Li}, \bibinfo{person}{Xing Xie}, \bibinfo{person}{Yan Zhang}, {and} \bibinfo{person}{Sunghun Kim}.} \bibinfo{year}{2022}\natexlab{b}.
\newblock \showarticletitle{Evolutionary Preference Learning via Graph Nested GRU ODE for Session-based Recommendation}. In \bibinfo{booktitle}{\emph{Proc. of CIKM}}.
\newblock


\bibitem[Guo et~al\mbox{.}(2019)]%
        {guo2019streaming}
\bibfield{author}{\bibinfo{person}{Lei Guo}, \bibinfo{person}{Hongzhi Yin}, \bibinfo{person}{Qinyong Wang}, \bibinfo{person}{Tong Chen}, \bibinfo{person}{Alexander Zhou}, {and} \bibinfo{person}{Nguyen Quoc Viet~Hung}.} \bibinfo{year}{2019}\natexlab{}.
\newblock \showarticletitle{Streaming session-based recommendation}. In \bibinfo{booktitle}{\emph{Proc. of KDD}}.
\newblock


\bibitem[Hamilton et~al\mbox{.}(2017)]%
        {hamilton2017inductive}
\bibfield{author}{\bibinfo{person}{Will Hamilton}, \bibinfo{person}{Zhitao Ying}, {and} \bibinfo{person}{Jure Leskovec}.} \bibinfo{year}{2017}\natexlab{}.
\newblock \showarticletitle{Inductive representation learning on large graphs}.
\newblock \bibinfo{journal}{\emph{Proc. of NeurIPS}} (\bibinfo{year}{2017}).
\newblock


\bibitem[Han et~al\mbox{.}(2022)]%
        {han2022multi}
\bibfield{author}{\bibinfo{person}{Qilong Han}, \bibinfo{person}{Chi Zhang}, \bibinfo{person}{Rui Chen}, \bibinfo{person}{Riwei Lai}, \bibinfo{person}{Hongtao Song}, {and} \bibinfo{person}{Li Li}.} \bibinfo{year}{2022}\natexlab{}.
\newblock \showarticletitle{Multi-Faceted Global Item Relation Learning for Session-Based Recommendation}. In \bibinfo{booktitle}{\emph{Proc. of SIGIR}}.
\newblock


\bibitem[He et~al\mbox{.}(2015)]%
        {he2015learning}
\bibfield{author}{\bibinfo{person}{Shizhu He}, \bibinfo{person}{Kang Liu}, \bibinfo{person}{Guoliang Ji}, {and} \bibinfo{person}{Jun Zhao}.} \bibinfo{year}{2015}\natexlab{}.
\newblock \showarticletitle{Learning to represent knowledge graphs with gaussian embedding}. In \bibinfo{booktitle}{\emph{Proceedings of the 24th ACM international on conference on information and knowledge management}}.
\newblock


\bibitem[Hidasi and Karatzoglou(2018)]%
        {hidasi2018recurrent}
\bibfield{author}{\bibinfo{person}{Bal{\'a}zs Hidasi} {and} \bibinfo{person}{Alexandros Karatzoglou}.} \bibinfo{year}{2018}\natexlab{}.
\newblock \showarticletitle{Recurrent neural networks with top-k gains for session-based recommendations}. In \bibinfo{booktitle}{\emph{Proc. of CIKM}}.
\newblock


\bibitem[Hidasi et~al\mbox{.}(2015)]%
        {hidasi2015session}
\bibfield{author}{\bibinfo{person}{Bal{\'a}zs Hidasi}, \bibinfo{person}{Alexandros Karatzoglou}, \bibinfo{person}{Linas Baltrunas}, {and} \bibinfo{person}{Domonkos Tikk}.} \bibinfo{year}{2015}\natexlab{}.
\newblock \showarticletitle{Session-based recommendations with recurrent neural networks}.
\newblock \bibinfo{journal}{\emph{arXiv preprint arXiv:1511.06939}} (\bibinfo{year}{2015}).
\newblock


\bibitem[Hou et~al\mbox{.}(2022)]%
        {hou2022core}
\bibfield{author}{\bibinfo{person}{Yupeng Hou}, \bibinfo{person}{Binbin Hu}, \bibinfo{person}{Zhiqiang Zhang}, {and} \bibinfo{person}{Wayne~Xin Zhao}.} \bibinfo{year}{2022}\natexlab{}.
\newblock \showarticletitle{Core: simple and effective session-based recommendation within consistent representation space}. In \bibinfo{booktitle}{\emph{Proc. of SIGIR}}. \bibinfo{pages}{1796--1801}.
\newblock


\bibitem[Hou et~al\mbox{.}(2024)]%
        {hou2023large}
\bibfield{author}{\bibinfo{person}{Yupeng Hou}, \bibinfo{person}{Junjie Zhang}, \bibinfo{person}{Zihan Lin}, \bibinfo{person}{Hongyu Lu}, \bibinfo{person}{Ruobing Xie}, \bibinfo{person}{Julian McAuley}, {and} \bibinfo{person}{Wayne~Xin Zhao}.} \bibinfo{year}{2024}\natexlab{}.
\newblock \showarticletitle{Large language models are zero-shot rankers for recommender systems}. In \bibinfo{booktitle}{\emph{European Conference on Information Retrieval}}. \bibinfo{pages}{364--381}.
\newblock


\bibitem[Hu et~al\mbox{.}(2021)]%
        {hu2021lora}
\bibfield{author}{\bibinfo{person}{Edward~J Hu}, \bibinfo{person}{Yelong Shen}, \bibinfo{person}{Phillip Wallis}, \bibinfo{person}{Zeyuan Allen-Zhu}, \bibinfo{person}{Yuanzhi Li}, \bibinfo{person}{Shean Wang}, \bibinfo{person}{Lu Wang}, {and} \bibinfo{person}{Weizhu Chen}.} \bibinfo{year}{2021}\natexlab{}.
\newblock \showarticletitle{Lora: Low-rank adaptation of large language models}.
\newblock \bibinfo{journal}{\emph{arXiv preprint arXiv:2106.09685}} (\bibinfo{year}{2021}).
\newblock


\bibitem[Hu et~al\mbox{.}(2017)]%
        {hu2017diversifying}
\bibfield{author}{\bibinfo{person}{Liang Hu}, \bibinfo{person}{Longbing Cao}, \bibinfo{person}{Shoujin Wang}, \bibinfo{person}{Guandong Xu}, \bibinfo{person}{Jian Cao}, {and} \bibinfo{person}{Zhiping Gu}.} \bibinfo{year}{2017}\natexlab{}.
\newblock \showarticletitle{Diversifying Personalized Recommendation with User-session Context.}. In \bibinfo{booktitle}{\emph{Proc. of IJCAI}}.
\newblock


\bibitem[Huang et~al\mbox{.}(2021)]%
        {huang2021graph}
\bibfield{author}{\bibinfo{person}{Chao Huang}, \bibinfo{person}{Jiahui Chen}, \bibinfo{person}{Lianghao Xia}, \bibinfo{person}{Yong Xu}, \bibinfo{person}{Peng Dai}, \bibinfo{person}{Yanqing Chen}, \bibinfo{person}{Liefeng Bo}, \bibinfo{person}{Jiashu Zhao}, {and} \bibinfo{person}{Jimmy~Xiangji Huang}.} \bibinfo{year}{2021}\natexlab{}.
\newblock \showarticletitle{Graph-enhanced multi-task learning of multi-level transition dynamics for session-based recommendation}. In \bibinfo{booktitle}{\emph{Proc. of AAAI}}.
\newblock


\bibitem[Jagatap et~al\mbox{.}(2023)]%
        {jagatap2023attribert}
\bibfield{author}{\bibinfo{person}{Akshay Jagatap}, \bibinfo{person}{Nikki Gupta}, \bibinfo{person}{Sachin Farfade}, {and} \bibinfo{person}{Prakash~Mandayam Comar}.} \bibinfo{year}{2023}\natexlab{}.
\newblock \showarticletitle{AttriBERT: Session-based product attribute recommendation with BERT}.
\newblock  (\bibinfo{year}{2023}).
\newblock


\bibitem[Jannach and Ludewig(2017)]%
        {jannach2017recurrent}
\bibfield{author}{\bibinfo{person}{Dietmar Jannach} {and} \bibinfo{person}{Malte Ludewig}.} \bibinfo{year}{2017}\natexlab{}.
\newblock \showarticletitle{When recurrent neural networks meet the neighborhood for session-based recommendation}. In \bibinfo{booktitle}{\emph{Proc. of RecSys}}.
\newblock


\bibitem[Jia et~al\mbox{.}(2023)]%
        {jia2023smone}
\bibfield{author}{\bibinfo{person}{Bohan Jia}, \bibinfo{person}{Jian Cao}, \bibinfo{person}{Shiyou Qian}, \bibinfo{person}{Nengjun Zhu}, \bibinfo{person}{Xin Dong}, \bibinfo{person}{Liang Zhang}, \bibinfo{person}{Lei Cheng}, {and} \bibinfo{person}{Linjian Mo}.} \bibinfo{year}{2023}\natexlab{}.
\newblock \showarticletitle{SMONE: A Session-based Recommendation Model Based on Neighbor Sessions with Similar Probabilistic Intentions}.
\newblock \bibinfo{journal}{\emph{ACM Transactions on Knowledge Discovery from Data}} (\bibinfo{year}{2023}).
\newblock


\bibitem[Jiang and Luo(2022)]%
        {jiang2021graph}
\bibfield{author}{\bibinfo{person}{Weiwei Jiang} {and} \bibinfo{person}{Jiayun Luo}.} \bibinfo{year}{2022}\natexlab{}.
\newblock \showarticletitle{Graph neural network for traffic forecasting: A survey}.
\newblock \bibinfo{journal}{\emph{Expert systems with applications}} (\bibinfo{year}{2022}), \bibinfo{pages}{117921}.
\newblock


\bibitem[Kang and McAuley(2018)]%
        {kang2018self}
\bibfield{author}{\bibinfo{person}{Wang-Cheng Kang} {and} \bibinfo{person}{Julian McAuley}.} \bibinfo{year}{2018}\natexlab{}.
\newblock \showarticletitle{Self-attentive sequential recommendation}. In \bibinfo{booktitle}{\emph{Proc. of ICDM}}.
\newblock


\bibitem[Kenton and Toutanova(2019)]%
        {devlin2018bert}
\bibfield{author}{\bibinfo{person}{Jacob Devlin Ming-Wei~Chang Kenton} {and} \bibinfo{person}{Lee~Kristina Toutanova}.} \bibinfo{year}{2019}\natexlab{}.
\newblock \showarticletitle{BERT: Pre-training of Deep Bidirectional Transformers for Language Understanding}. In \bibinfo{booktitle}{\emph{Proc. of AACL}}. \bibinfo{pages}{4171--4186}.
\newblock


\bibitem[Kipf and Welling(2022)]%
        {kipf2016semi}
\bibfield{author}{\bibinfo{person}{Thomas~N Kipf} {and} \bibinfo{person}{Max Welling}.} \bibinfo{year}{2022}\natexlab{}.
\newblock \showarticletitle{Semi-Supervised Classification with Graph Convolutional Networks}. In \bibinfo{booktitle}{\emph{Proc. of ICLR}}.
\newblock


\bibitem[Kiumarsi et~al\mbox{.}(2017)]%
        {kiumarsi2017optimal}
\bibfield{author}{\bibinfo{person}{Bahare Kiumarsi}, \bibinfo{person}{Kyriakos~G Vamvoudakis}, \bibinfo{person}{Hamidreza Modares}, {and} \bibinfo{person}{Frank~L Lewis}.} \bibinfo{year}{2017}\natexlab{}.
\newblock \showarticletitle{Optimal and autonomous control using reinforcement learning: A survey}.
\newblock \bibinfo{journal}{\emph{IEEE transactions on neural networks and learning systems}} (\bibinfo{year}{2017}).
\newblock


\bibitem[Kojima et~al\mbox{.}(2022)]%
        {kojima2022large}
\bibfield{author}{\bibinfo{person}{Takeshi Kojima}, \bibinfo{person}{Shixiang~Shane Gu}, \bibinfo{person}{Machel Reid}, \bibinfo{person}{Yutaka Matsuo}, {and} \bibinfo{person}{Yusuke Iwasawa}.} \bibinfo{year}{2022}\natexlab{}.
\newblock \showarticletitle{Large language models are zero-shot reasoners}.
\newblock \bibinfo{journal}{\emph{Proc. of NeurIPS}} (\bibinfo{year}{2022}).
\newblock


\bibitem[Koren(2008)]%
        {koren2008factorization}
\bibfield{author}{\bibinfo{person}{Yehuda Koren}.} \bibinfo{year}{2008}\natexlab{}.
\newblock \showarticletitle{Factorization meets the neighborhood: a multifaceted collaborative filtering model}. In \bibinfo{booktitle}{\emph{Proc. of KDD}}.
\newblock


\bibitem[Koren et~al\mbox{.}(2009)]%
        {koren2009matrix}
\bibfield{author}{\bibinfo{person}{Yehuda Koren}, \bibinfo{person}{Robert Bell}, {and} \bibinfo{person}{Chris Volinsky}.} \bibinfo{year}{2009}\natexlab{}.
\newblock \showarticletitle{Matrix factorization techniques for recommender systems}.
\newblock \bibinfo{journal}{\emph{Computer}} (\bibinfo{year}{2009}).
\newblock


\bibitem[Lai et~al\mbox{.}(2022)]%
        {lai2022attribute}
\bibfield{author}{\bibinfo{person}{Siqi Lai}, \bibinfo{person}{Erli Meng}, \bibinfo{person}{Fan Zhang}, \bibinfo{person}{Chenliang Li}, \bibinfo{person}{Bin Wang}, {and} \bibinfo{person}{Aixin Sun}.} \bibinfo{year}{2022}\natexlab{}.
\newblock \showarticletitle{An Attribute-Driven Mirror Graph Network for Session-based Recommendation}. In \bibinfo{booktitle}{\emph{Proc. of SIGIR}}.
\newblock


\bibitem[Le et~al\mbox{.}(2016)]%
        {le2016modeling}
\bibfield{author}{\bibinfo{person}{Duc-Trong Le}, \bibinfo{person}{Yuan Fang}, {and} \bibinfo{person}{Hady~W Lauw}.} \bibinfo{year}{2016}\natexlab{}.
\newblock \showarticletitle{Modeling sequential preferences with dynamic user and context factors}. In \bibinfo{booktitle}{\emph{Proc. of ECML}}.
\newblock


\bibitem[Li et~al\mbox{.}(2022a)]%
        {li2022disentangled}
\bibfield{author}{\bibinfo{person}{Ansong Li}, \bibinfo{person}{Zhiyong Cheng}, \bibinfo{person}{Fan Liu}, \bibinfo{person}{Zan Gao}, \bibinfo{person}{Weili Guan}, {and} \bibinfo{person}{Yuxin Peng}.} \bibinfo{year}{2022}\natexlab{a}.
\newblock \showarticletitle{Disentangled graph neural networks for session-based recommendation}.
\newblock \bibinfo{journal}{\emph{IEEE Transactions on Knowledge and Data Engineering}} (\bibinfo{year}{2022}), \bibinfo{pages}{7870--7882}.
\newblock


\bibitem[Li et~al\mbox{.}(2019)]%
        {li2019mad}
\bibfield{author}{\bibinfo{person}{Dan Li}, \bibinfo{person}{Dacheng Chen}, \bibinfo{person}{Baihong Jin}, \bibinfo{person}{Lei Shi}, \bibinfo{person}{Jonathan Goh}, {and} \bibinfo{person}{See-Kiong Ng}.} \bibinfo{year}{2019}\natexlab{}.
\newblock \showarticletitle{MAD-GAN: Multivariate anomaly detection for time series data with generative adversarial networks}. In \bibinfo{booktitle}{\emph{Proc. of ICANN}}.
\newblock


\bibitem[Li et~al\mbox{.}(2017a)]%
        {li2017learning}
\bibfield{author}{\bibinfo{person}{Junying Li}, \bibinfo{person}{Deng Cai}, {and} \bibinfo{person}{Xiaofei He}.} \bibinfo{year}{2017}\natexlab{a}.
\newblock \showarticletitle{Learning graph-level representation for drug discovery}.
\newblock \bibinfo{journal}{\emph{arXiv preprint arXiv:1709.03741}} (\bibinfo{year}{2017}).
\newblock


\bibitem[Li et~al\mbox{.}(2017b)]%
        {li2017neural}
\bibfield{author}{\bibinfo{person}{Jing Li}, \bibinfo{person}{Pengjie Ren}, \bibinfo{person}{Zhumin Chen}, \bibinfo{person}{Zhaochun Ren}, \bibinfo{person}{Tao Lian}, {and} \bibinfo{person}{Jun Ma}.} \bibinfo{year}{2017}\natexlab{b}.
\newblock \showarticletitle{Neural attentive session-based recommendation}. In \bibinfo{booktitle}{\emph{Proceedings of the 2017 ACM on Conference on Information and Knowledge Management}}.
\newblock


\bibitem[Li et~al\mbox{.}(2021b)]%
        {li2021personalized}
\bibfield{author}{\bibinfo{person}{Lei Li}, \bibinfo{person}{Yongfeng Zhang}, {and} \bibinfo{person}{Li Chen}.} \bibinfo{year}{2021}\natexlab{b}.
\newblock \showarticletitle{Personalized Transformer for Explainable Recommendation}. In \bibinfo{booktitle}{\emph{Proc. of ACL}}. \bibinfo{pages}{4947--4957}.
\newblock


\bibitem[Li et~al\mbox{.}(2023c)]%
        {li2023personalized}
\bibfield{author}{\bibinfo{person}{Lei Li}, \bibinfo{person}{Yongfeng Zhang}, {and} \bibinfo{person}{Li Chen}.} \bibinfo{year}{2023}\natexlab{c}.
\newblock \showarticletitle{Personalized prompt learning for explainable recommendation}.
\newblock \bibinfo{journal}{\emph{ACM Transactions on Information Systems}} (\bibinfo{year}{2023}), \bibinfo{pages}{1--26}.
\newblock


\bibitem[Li et~al\mbox{.}(2021a)]%
        {li2021lightweight}
\bibfield{author}{\bibinfo{person}{Yang Li}, \bibinfo{person}{Tong Chen}, \bibinfo{person}{Peng-Fei Zhang}, {and} \bibinfo{person}{Hongzhi Yin}.} \bibinfo{year}{2021}\natexlab{a}.
\newblock \showarticletitle{Lightweight self-attentive sequential recommendation}. In \bibinfo{booktitle}{\emph{Proc. of CIKM}}. \bibinfo{pages}{967--977}.
\newblock


\bibitem[Li et~al\mbox{.}(2022b)]%
        {li2022spatiotemporal}
\bibfield{author}{\bibinfo{person}{Yinfeng Li}, \bibinfo{person}{Chen Gao}, \bibinfo{person}{Xiaoyi Du}, \bibinfo{person}{Huazhou Wei}, \bibinfo{person}{Hengliang Luo}, \bibinfo{person}{Depeng Jin}, {and} \bibinfo{person}{Yong Li}.} \bibinfo{year}{2022}\natexlab{b}.
\newblock \showarticletitle{Spatiotemporal-aware Session-based Recommendation with Graph Neural Networks}. In \bibinfo{booktitle}{\emph{Proc. of CIKM}}.
\newblock


\bibitem[Li et~al\mbox{.}(2022c)]%
        {li2022enhancing}
\bibfield{author}{\bibinfo{person}{Yinfeng Li}, \bibinfo{person}{Chen Gao}, \bibinfo{person}{Hengliang Luo}, \bibinfo{person}{Depeng Jin}, {and} \bibinfo{person}{Yong Li}.} \bibinfo{year}{2022}\natexlab{c}.
\newblock \showarticletitle{Enhancing Hypergraph Neural Networks with Intent Disentanglement for Session-based Recommendation}. In \bibinfo{booktitle}{\emph{Proc. of SIGIR}}.
\newblock


\bibitem[Li et~al\mbox{.}(2023a)]%
        {li2023diffurec}
\bibfield{author}{\bibinfo{person}{Zihao Li}, \bibinfo{person}{Aixin Sun}, {and} \bibinfo{person}{Chenliang Li}.} \bibinfo{year}{2023}\natexlab{a}.
\newblock \showarticletitle{Diffurec: A diffusion model for sequential recommendation}.
\newblock \bibinfo{journal}{\emph{ACM Transactions on Information Systems}} (\bibinfo{year}{2023}), \bibinfo{pages}{1--28}.
\newblock


\bibitem[Li et~al\mbox{.}(2023b)]%
        {li2023exploiting}
\bibfield{author}{\bibinfo{person}{Zihao Li}, \bibinfo{person}{Xianzhi Wang}, \bibinfo{person}{Chao Yang}, \bibinfo{person}{Lina Yao}, \bibinfo{person}{Julian McAuley}, {and} \bibinfo{person}{Guandong Xu}.} \bibinfo{year}{2023}\natexlab{b}.
\newblock \showarticletitle{Exploiting Explicit and Implicit Item relationships for Session-based Recommendation}. In \bibinfo{booktitle}{\emph{Proc. of WSDM}}.
\newblock


\bibitem[Li et~al\mbox{.}(2024)]%
        {li2024disentangle}
\bibfield{author}{\bibinfo{person}{Zihao Li}, \bibinfo{person}{Yunfan Xie}, \bibinfo{person}{Wei~Emma Zhang}, \bibinfo{person}{Pengfei Wang}, \bibinfo{person}{Lixin Zou}, \bibinfo{person}{Fei Li}, \bibinfo{person}{Xiangyang Luo}, {and} \bibinfo{person}{Chenliang Li}.} \bibinfo{year}{2024}\natexlab{}.
\newblock \showarticletitle{Disentangle interest trend and diversity for sequential recommendation}.
\newblock \bibinfo{journal}{\emph{Information Processing \& Management}} \bibinfo{volume}{61}, \bibinfo{number}{3} (\bibinfo{year}{2024}), \bibinfo{pages}{103619}.
\newblock


\bibitem[Lian et~al\mbox{.}(2013)]%
        {lian2013collaborative}
\bibfield{author}{\bibinfo{person}{Defu Lian}, \bibinfo{person}{Vincent~W Zheng}, {and} \bibinfo{person}{Xing Xie}.} \bibinfo{year}{2013}\natexlab{}.
\newblock \showarticletitle{Collaborative filtering meets next check-in location prediction}. In \bibinfo{booktitle}{\emph{Proc. of WWW}}.
\newblock


\bibitem[Liang et~al\mbox{.}(2016)]%
        {liang2016factorization}
\bibfield{author}{\bibinfo{person}{Dawen Liang}, \bibinfo{person}{Jaan Altosaar}, \bibinfo{person}{Laurent Charlin}, {and} \bibinfo{person}{David~M Blei}.} \bibinfo{year}{2016}\natexlab{}.
\newblock \showarticletitle{Factorization meets the item embedding: Regularizing matrix factorization with item co-occurrence}. In \bibinfo{booktitle}{\emph{Proc. of RecSys}}.
\newblock


\bibitem[Lin et~al\mbox{.}(2021)]%
        {lin2021m6}
\bibfield{author}{\bibinfo{person}{Junyang Lin}, \bibinfo{person}{Rui Men}, \bibinfo{person}{An Yang}, \bibinfo{person}{Chang Zhou}, \bibinfo{person}{Yichang Zhang}, \bibinfo{person}{Peng Wang}, \bibinfo{person}{Jingren Zhou}, \bibinfo{person}{Jie Tang}, {and} \bibinfo{person}{Hongxia Yang}.} \bibinfo{year}{2021}\natexlab{}.
\newblock \showarticletitle{M6: Multi-modality-to-multi-modality multitask mega-transformer for unified pretraining}. In \bibinfo{booktitle}{\emph{Proc. of KDD}}.
\newblock


\bibitem[Lin et~al\mbox{.}(2017)]%
        {lin2017focal}
\bibfield{author}{\bibinfo{person}{Tsung-Yi Lin}, \bibinfo{person}{Priya Goyal}, \bibinfo{person}{Ross Girshick}, \bibinfo{person}{Kaiming He}, {and} \bibinfo{person}{Piotr Doll{\'a}r}.} \bibinfo{year}{2017}\natexlab{}.
\newblock \showarticletitle{Focal loss for dense object detection}. In \bibinfo{booktitle}{\emph{Proc. of ICCV}}.
\newblock


\bibitem[Lin et~al\mbox{.}(2023)]%
        {lin2021survey}
\bibfield{author}{\bibinfo{person}{Yuanguo Lin}, \bibinfo{person}{Yong Liu}, \bibinfo{person}{Fan Lin}, \bibinfo{person}{Lixin Zou}, \bibinfo{person}{Pengcheng Wu}, \bibinfo{person}{Wenhua Zeng}, \bibinfo{person}{Huanhuan Chen}, {and} \bibinfo{person}{Chunyan Miao}.} \bibinfo{year}{2023}\natexlab{}.
\newblock \showarticletitle{A survey on reinforcement learning for recommender systems}.
\newblock \bibinfo{journal}{\emph{IEEE Transactions on Neural Networks and Learning Systems}} (\bibinfo{year}{2023}).
\newblock


\bibitem[Linden et~al\mbox{.}(2003)]%
        {linden2003amazon}
\bibfield{author}{\bibinfo{person}{Greg Linden}, \bibinfo{person}{Brent Smith}, {and} \bibinfo{person}{Jeremy York}.} \bibinfo{year}{2003}\natexlab{}.
\newblock \showarticletitle{Amazon. com recommendations: Item-to-item collaborative filtering}.
\newblock \bibinfo{journal}{\emph{IEEE Internet computing}} (\bibinfo{year}{2003}).
\newblock


\bibitem[Liu et~al\mbox{.}(2021b)]%
        {liu2021non}
\bibfield{author}{\bibinfo{person}{Chang Liu}, \bibinfo{person}{Xiaoguang Li}, \bibinfo{person}{Guohao Cai}, \bibinfo{person}{Zhenhua Dong}, \bibinfo{person}{Hong Zhu}, {and} \bibinfo{person}{Lifeng Shang}.} \bibinfo{year}{2021}\natexlab{b}.
\newblock \showarticletitle{Noninvasive self-attention for side information fusion in sequential recommendation}. In \bibinfo{booktitle}{\emph{Proc. of AAAI}}. \bibinfo{pages}{4249--4256}.
\newblock


\bibitem[Liu et~al\mbox{.}(2021a)]%
        {liu2021mitigating}
\bibfield{author}{\bibinfo{person}{Dugang Liu}, \bibinfo{person}{Pengxiang Cheng}, \bibinfo{person}{Hong Zhu}, \bibinfo{person}{Zhenhua Dong}, \bibinfo{person}{Xiuqiang He}, \bibinfo{person}{Weike Pan}, {and} \bibinfo{person}{Zhong Ming}.} \bibinfo{year}{2021}\natexlab{a}.
\newblock \showarticletitle{Mitigating Confounding Bias in Recommendation via Information Bottleneck}. In \bibinfo{booktitle}{\emph{Proc. of RecSys}}.
\newblock


\bibitem[Liu et~al\mbox{.}(2023)]%
        {liu2023chatgpt}
\bibfield{author}{\bibinfo{person}{Junling Liu}, \bibinfo{person}{Chao Liu}, \bibinfo{person}{Renjie Lv}, \bibinfo{person}{Kang Zhou}, {and} \bibinfo{person}{Yan Zhang}.} \bibinfo{year}{2023}\natexlab{}.
\newblock \showarticletitle{Is chatgpt a good recommender? a preliminary study}.
\newblock \bibinfo{journal}{\emph{ArXiv preprint}} (\bibinfo{year}{2023}).
\newblock


\bibitem[Liu et~al\mbox{.}(2018)]%
        {liu2018stamp}
\bibfield{author}{\bibinfo{person}{Qiao Liu}, \bibinfo{person}{Yifu Zeng}, \bibinfo{person}{Refuoe Mokhosi}, {and} \bibinfo{person}{Haibin Zhang}.} \bibinfo{year}{2018}\natexlab{}.
\newblock \showarticletitle{STAMP: short-term attention/memory priority model for session-based recommendation}. In \bibinfo{booktitle}{\emph{Proc. of KDD}}.
\newblock


\bibitem[Liu and Zheng(2020)]%
        {liu2020long}
\bibfield{author}{\bibinfo{person}{Siyi Liu} {and} \bibinfo{person}{Yujia Zheng}.} \bibinfo{year}{2020}\natexlab{}.
\newblock \showarticletitle{Long-tail session-based recommendation}. In \bibinfo{booktitle}{\emph{Proceedings of the 14th ACM conference on recommender systems}}. \bibinfo{pages}{509--514}.
\newblock


\bibitem[Liu et~al\mbox{.}(2020)]%
        {liu2020keywords}
\bibfield{author}{\bibinfo{person}{Yuanxing Liu}, \bibinfo{person}{Zhaochun Ren}, \bibinfo{person}{Wei-Nan Zhang}, \bibinfo{person}{Wanxiang Che}, \bibinfo{person}{Ting Liu}, {and} \bibinfo{person}{Dawei Yin}.} \bibinfo{year}{2020}\natexlab{}.
\newblock \showarticletitle{Keywords generation improves e-commerce session-based recommendation}. In \bibinfo{booktitle}{\emph{Proc. of WWW}}.
\newblock


\bibitem[Lu et~al\mbox{.}(2015)]%
        {lu2015content}
\bibfield{author}{\bibinfo{person}{Zhongqi Lu}, \bibinfo{person}{Zhicheng Dou}, \bibinfo{person}{Jianxun Lian}, \bibinfo{person}{Xing Xie}, {and} \bibinfo{person}{Qiang Yang}.} \bibinfo{year}{2015}\natexlab{}.
\newblock \showarticletitle{Content-based collaborative filtering for news topic recommendation}. In \bibinfo{booktitle}{\emph{Proc. of AAAI}}.
\newblock


\bibitem[Ludewig and Jannach(2018)]%
        {ludewig2018evaluation}
\bibfield{author}{\bibinfo{person}{Malte Ludewig} {and} \bibinfo{person}{Dietmar Jannach}.} \bibinfo{year}{2018}\natexlab{}.
\newblock \showarticletitle{Evaluation of session-based recommendation algorithms}.
\newblock \bibinfo{journal}{\emph{User Modeling and User-Adapted Interaction}} (\bibinfo{year}{2018}).
\newblock


\bibitem[Luo et~al\mbox{.}(2020)]%
        {luo2020collaborative}
\bibfield{author}{\bibinfo{person}{Anjing Luo}, \bibinfo{person}{Pengpeng Zhao}, \bibinfo{person}{Yanchi Liu}, \bibinfo{person}{Fuzhen Zhuang}, \bibinfo{person}{Deqing Wang}, \bibinfo{person}{Jiajie Xu}, \bibinfo{person}{Junhua Fang}, {and} \bibinfo{person}{Victor~S Sheng}.} \bibinfo{year}{2020}\natexlab{}.
\newblock \showarticletitle{Collaborative Self-Attention Network for Session-based Recommendation.}. In \bibinfo{booktitle}{\emph{Proc. of IJCAI}}.
\newblock


\bibitem[Ma et~al\mbox{.}(2008)]%
        {ma2008sorec}
\bibfield{author}{\bibinfo{person}{Hao Ma}, \bibinfo{person}{Haixuan Yang}, \bibinfo{person}{Michael~R Lyu}, {and} \bibinfo{person}{Irwin King}.} \bibinfo{year}{2008}\natexlab{}.
\newblock \showarticletitle{Sorec: social recommendation using probabilistic matrix factorization}. In \bibinfo{booktitle}{\emph{Proceedings of the 17th ACM conference on Information and knowledge management}}.
\newblock


\bibitem[Meng et~al\mbox{.}(2020)]%
        {meng2020incorporating}
\bibfield{author}{\bibinfo{person}{Wenjing Meng}, \bibinfo{person}{Deqing Yang}, {and} \bibinfo{person}{Yanghua Xiao}.} \bibinfo{year}{2020}\natexlab{}.
\newblock \showarticletitle{Incorporating user micro-behaviors and item knowledge into multi-task learning for session-based recommendation}. In \bibinfo{booktitle}{\emph{Proc. of SIGIR}}.
\newblock


\bibitem[Mikolov et~al\mbox{.}(2013)]%
        {mikolov2013efficient}
\bibfield{author}{\bibinfo{person}{Tomas Mikolov}, \bibinfo{person}{Kai Chen}, \bibinfo{person}{Greg Corrado}, {and} \bibinfo{person}{Jeffrey Dean}.} \bibinfo{year}{2013}\natexlab{}.
\newblock \showarticletitle{Efficient estimation of word representations in vector space}.
\newblock \bibinfo{journal}{\emph{arXiv preprint arXiv:1301.3781}} (\bibinfo{year}{2013}).
\newblock


\bibitem[Miyahara and Pazzani(2000)]%
        {miyahara2000collaborative}
\bibfield{author}{\bibinfo{person}{Koji Miyahara} {and} \bibinfo{person}{Michael~J Pazzani}.} \bibinfo{year}{2000}\natexlab{}.
\newblock \showarticletitle{Collaborative filtering with the simple bayesian classifier}. In \bibinfo{booktitle}{\emph{Proc. of PRICAI}}.
\newblock


\bibitem[Mobasher et~al\mbox{.}(2001)]%
        {mobasher2001effective}
\bibfield{author}{\bibinfo{person}{Bamshad Mobasher}, \bibinfo{person}{Honghua Dai}, \bibinfo{person}{Tao Luo}, {and} \bibinfo{person}{Miki Nakagawa}.} \bibinfo{year}{2001}\natexlab{}.
\newblock \showarticletitle{Effective personalization based on association rule discovery from web usage data}. In \bibinfo{booktitle}{\emph{Proceedings of the 3rd international workshop on Web information and data management}}.
\newblock


\bibitem[Modani et~al\mbox{.}(2002)]%
        {modani2002series}
\bibfield{author}{\bibinfo{person}{Natwar Modani}, \bibinfo{person}{Parul~A Mittal}, \bibinfo{person}{Amit~A Nanavati}, {and} \bibinfo{person}{Biplav Srivastava}.} \bibinfo{year}{2002}\natexlab{}.
\newblock \showarticletitle{Series of dynamic targeted recommendations}. In \bibinfo{booktitle}{\emph{International Conference on Electronic Commerce and Web Technologies}}.
\newblock


\bibitem[Modani et~al\mbox{.}(2005)]%
        {modani2005framework}
\bibfield{author}{\bibinfo{person}{Natwar Modani}, \bibinfo{person}{Yogish Sabharwal}, {and} \bibinfo{person}{S Karthik}.} \bibinfo{year}{2005}\natexlab{}.
\newblock \showarticletitle{A framework for session based recommendations}. In \bibinfo{booktitle}{\emph{International Conference on Electronic Commerce and Web Technologies}}.
\newblock


\bibitem[Moreno et~al\mbox{.}(2004)]%
        {moreno2004using}
\bibfield{author}{\bibinfo{person}{Mar{\'\i}a~N Moreno}, \bibinfo{person}{Francisco~J Garc{\'\i}a}, \bibinfo{person}{M~Jos{\'e} Polo}, {and} \bibinfo{person}{Vivian~F L{\'o}pez}.} \bibinfo{year}{2004}\natexlab{}.
\newblock \showarticletitle{Using association analysis of web data in recommender systems}. In \bibinfo{booktitle}{\emph{International Conference on Electronic Commerce and Web Technologies}}.
\newblock


\bibitem[Narwariya et~al\mbox{.}(2023)]%
        {narwariya2023x4sr}
\bibfield{author}{\bibinfo{person}{Jyoti Narwariya}, \bibinfo{person}{Priyanka Gupta}, \bibinfo{person}{Garima Gupta}, \bibinfo{person}{Lovekesh Vig}, {and} \bibinfo{person}{Gautam Shroff}.} \bibinfo{year}{2023}\natexlab{}.
\newblock \showarticletitle{X4SR: Post-Hoc Explanations for Session-based Recommendations.}. In \bibinfo{booktitle}{\emph{Proc. of SIGIR}}.
\newblock


\bibitem[Niranjan et~al\mbox{.}(2010)]%
        {niranjan2010developing}
\bibfield{author}{\bibinfo{person}{Utpala Niranjan}, \bibinfo{person}{RBV Subramanyam}, {and} \bibinfo{person}{V Khanaa}.} \bibinfo{year}{2010}\natexlab{}.
\newblock \showarticletitle{Developing a web recommendation system based on closed sequential patterns}. In \bibinfo{booktitle}{\emph{International Conference on Advances in Information and Communication Technologies}}.
\newblock


\bibitem[Ouyang et~al\mbox{.}(2023)]%
        {ouyang2023mining}
\bibfield{author}{\bibinfo{person}{Kai Ouyang}, \bibinfo{person}{Xianghong Xu}, \bibinfo{person}{Miaoxin Chen}, \bibinfo{person}{Zuotong Xie}, \bibinfo{person}{Hai-Tao Zheng}, \bibinfo{person}{Shuangyong Song}, {and} \bibinfo{person}{Yu Zhao}.} \bibinfo{year}{2023}\natexlab{}.
\newblock \showarticletitle{Mining interest trends and adaptively assigning sample weight for session-based recommendation}. In \bibinfo{booktitle}{\emph{Proc. of SIGIR}}. \bibinfo{pages}{2174--2178}.
\newblock


\bibitem[Ouyang et~al\mbox{.}(2022)]%
        {ouyang-etal-2022-social}
\bibfield{author}{\bibinfo{person}{Kai Ouyang}, \bibinfo{person}{Xianghong Xu}, \bibinfo{person}{Chen Tang}, \bibinfo{person}{Wang Chen}, {and} \bibinfo{person}{Haitao Zheng}.} \bibinfo{year}{2022}\natexlab{}.
\newblock \showarticletitle{Social-aware Sparse Attention Network for Session-based Social Recommendation}. In \bibinfo{booktitle}{\emph{Proc. of EMNLP Findings}}.
\newblock


\bibitem[Pan et~al\mbox{.}(2020a)]%
        {pan2020star}
\bibfield{author}{\bibinfo{person}{Zhiqiang Pan}, \bibinfo{person}{Fei Cai}, \bibinfo{person}{Wanyu Chen}, \bibinfo{person}{Honghui Chen}, {and} \bibinfo{person}{Maarten de Rijke}.} \bibinfo{year}{2020}\natexlab{a}.
\newblock \showarticletitle{Star graph neural networks for session-based recommendation}. In \bibinfo{booktitle}{\emph{Proc. of CIKM}}.
\newblock


\bibitem[Pan et~al\mbox{.}(2020b)]%
        {pan2020intent}
\bibfield{author}{\bibinfo{person}{Zhiqiang Pan}, \bibinfo{person}{Fei Cai}, \bibinfo{person}{Yanxiang Ling}, {and} \bibinfo{person}{Maarten de Rijke}.} \bibinfo{year}{2020}\natexlab{b}.
\newblock \showarticletitle{An intent-guided collaborative machine for session-based recommendation}. In \bibinfo{booktitle}{\emph{Proc. of SIGIR}}.
\newblock


\bibitem[Pang et~al\mbox{.}(2022)]%
        {pang2022heterogeneous}
\bibfield{author}{\bibinfo{person}{Yitong Pang}, \bibinfo{person}{Lingfei Wu}, \bibinfo{person}{Qi Shen}, \bibinfo{person}{Yiming Zhang}, \bibinfo{person}{Zhihua Wei}, \bibinfo{person}{Fangli Xu}, \bibinfo{person}{Ethan Chang}, \bibinfo{person}{Bo Long}, {and} \bibinfo{person}{Jian Pei}.} \bibinfo{year}{2022}\natexlab{}.
\newblock \showarticletitle{Heterogeneous global graph neural networks for personalized session-based recommendation}. In \bibinfo{booktitle}{\emph{Proc. of WSDM}}.
\newblock


\bibitem[Peintner et~al\mbox{.}(2023)]%
        {peintner2023spare}
\bibfield{author}{\bibinfo{person}{Andreas Peintner}, \bibinfo{person}{Amir~Reza Mohammadi}, {and} \bibinfo{person}{Eva Zangerle}.} \bibinfo{year}{2023}\natexlab{}.
\newblock \showarticletitle{SPARE: Shortest Path Global Item Relations for Efficient Session-based Recommendation}. In \bibinfo{booktitle}{\emph{Proc. of RecSys}}.
\newblock


\bibitem[Perozzi et~al\mbox{.}(2014)]%
        {perozzi2014deepwalk}
\bibfield{author}{\bibinfo{person}{Bryan Perozzi}, \bibinfo{person}{Rami Al-Rfou}, {and} \bibinfo{person}{Steven Skiena}.} \bibinfo{year}{2014}\natexlab{}.
\newblock \showarticletitle{Deepwalk: Online learning of social representations}. In \bibinfo{booktitle}{\emph{Proc. of KDD}}.
\newblock


\bibitem[Potter et~al\mbox{.}(2022)]%
        {potter2022gru4recbe}
\bibfield{author}{\bibinfo{person}{Michael Potter}, \bibinfo{person}{Hamlin Liu}, \bibinfo{person}{Yash Lala}, \bibinfo{person}{Christian Loanzon}, {and} \bibinfo{person}{Yizhou Sun}.} \bibinfo{year}{2022}\natexlab{}.
\newblock \showarticletitle{GRU4RecBE: A Hybrid Session-Based Movie Recommendation System (Student Abstract)}. In \bibinfo{booktitle}{\emph{Proc. of AAAI}}.
\newblock


\bibitem[Qiao et~al\mbox{.}(2023)]%
        {qiao2023bi}
\bibfield{author}{\bibinfo{person}{Shutong Qiao}, \bibinfo{person}{Wei Zhou}, \bibinfo{person}{Junhao Wen}, \bibinfo{person}{Hongyu Zhang}, {and} \bibinfo{person}{Min Gao}.} \bibinfo{year}{2023}\natexlab{}.
\newblock \showarticletitle{Bi-channel Multiple Sparse Graph Attention Networks for Session-based Recommendation}. In \bibinfo{booktitle}{\emph{Proc. of CIKM}}. \bibinfo{pages}{2075--2084}.
\newblock


\bibitem[Qiu et~al\mbox{.}(2019)]%
        {qiu2019rethinking}
\bibfield{author}{\bibinfo{person}{Ruihong Qiu}, \bibinfo{person}{Jingjing Li}, \bibinfo{person}{Zi Huang}, {and} \bibinfo{person}{Hongzhi Yin}.} \bibinfo{year}{2019}\natexlab{}.
\newblock \showarticletitle{Rethinking the item order in session-based recommendation with graph neural networks}. In \bibinfo{booktitle}{\emph{Proc. of CIKM}}.
\newblock


\bibitem[Qiu et~al\mbox{.}(2020)]%
        {qiu2020gag}
\bibfield{author}{\bibinfo{person}{Ruihong Qiu}, \bibinfo{person}{Hongzhi Yin}, \bibinfo{person}{Zi Huang}, {and} \bibinfo{person}{Tong Chen}.} \bibinfo{year}{2020}\natexlab{}.
\newblock \showarticletitle{Gag: Global attributed graph neural network for streaming session-based recommendation}. In \bibinfo{booktitle}{\emph{Proc. of SIGIR}}.
\newblock


\bibitem[Quadrana et~al\mbox{.}(2018)]%
        {quadrana2018sequence}
\bibfield{author}{\bibinfo{person}{Massimo Quadrana}, \bibinfo{person}{Paolo Cremonesi}, {and} \bibinfo{person}{Dietmar Jannach}.} \bibinfo{year}{2018}\natexlab{}.
\newblock \showarticletitle{Sequence-aware recommender systems}.
\newblock \bibinfo{journal}{\emph{ACM Computing Surveys (CSUR)}} (\bibinfo{year}{2018}).
\newblock


\bibitem[Quadrana et~al\mbox{.}(2017)]%
        {quadrana2017personalizing}
\bibfield{author}{\bibinfo{person}{Massimo Quadrana}, \bibinfo{person}{Alexandros Karatzoglou}, \bibinfo{person}{Bal{\'a}zs Hidasi}, {and} \bibinfo{person}{Paolo Cremonesi}.} \bibinfo{year}{2017}\natexlab{}.
\newblock \showarticletitle{Personalizing session-based recommendations with hierarchical recurrent neural networks}. In \bibinfo{booktitle}{\emph{Proc. of RecSys}}.
\newblock


\bibitem[Ranjbar~Kermany et~al\mbox{.}(2022)]%
        {ranjbar2022fair}
\bibfield{author}{\bibinfo{person}{Naime Ranjbar~Kermany}, \bibinfo{person}{Jian Yang}, \bibinfo{person}{Jia Wu}, {and} \bibinfo{person}{Luiz Pizzato}.} \bibinfo{year}{2022}\natexlab{}.
\newblock \showarticletitle{Fair-srs: a fair session-based recommendation system}. In \bibinfo{booktitle}{\emph{Proc. of WSDM}}.
\newblock


\bibitem[Ren et~al\mbox{.}(2019)]%
        {ren2019repeatnet}
\bibfield{author}{\bibinfo{person}{Pengjie Ren}, \bibinfo{person}{Zhumin Chen}, \bibinfo{person}{Jing Li}, \bibinfo{person}{Zhaochun Ren}, \bibinfo{person}{Jun Ma}, {and} \bibinfo{person}{Maarten De~Rijke}.} \bibinfo{year}{2019}\natexlab{}.
\newblock \showarticletitle{Repeatnet: A repeat aware neural recommendation machine for session-based recommendation}. In \bibinfo{booktitle}{\emph{Proc. of AAAI}}.
\newblock


\bibitem[Rendle et~al\mbox{.}(2009)]%
        {rendle2012bpr}
\bibfield{author}{\bibinfo{person}{Steffen Rendle}, \bibinfo{person}{Christoph Freudenthaler}, \bibinfo{person}{Zeno Gantner}, {and} \bibinfo{person}{Lars Schmidt-Thieme}.} \bibinfo{year}{2009}\natexlab{}.
\newblock \showarticletitle{BPR: Bayesian personalized ranking from implicit feedback}. In \bibinfo{booktitle}{\emph{Proc. of UAI}}. \bibinfo{pages}{452--461}.
\newblock


\bibitem[Rendle et~al\mbox{.}(2010)]%
        {rendle2010factorizing}
\bibfield{author}{\bibinfo{person}{Steffen Rendle}, \bibinfo{person}{Christoph Freudenthaler}, {and} \bibinfo{person}{Lars Schmidt-Thieme}.} \bibinfo{year}{2010}\natexlab{}.
\newblock \showarticletitle{Factorizing personalized markov chains for next-basket recommendation}. In \bibinfo{booktitle}{\emph{Proc. of WWW}}.
\newblock


\bibitem[Rubner et~al\mbox{.}(2000)]%
        {rubner2000earth}
\bibfield{author}{\bibinfo{person}{Yossi Rubner}, \bibinfo{person}{Carlo Tomasi}, {and} \bibinfo{person}{Leonidas~J Guibas}.} \bibinfo{year}{2000}\natexlab{}.
\newblock \showarticletitle{The earth mover's distance as a metric for image retrieval}.
\newblock \bibinfo{journal}{\emph{International journal of computer vision}} (\bibinfo{year}{2000}).
\newblock


\bibitem[Ruder(2017)]%
        {ruder2017overview}
\bibfield{author}{\bibinfo{person}{Sebastian Ruder}.} \bibinfo{year}{2017}\natexlab{}.
\newblock \showarticletitle{An overview of multi-task learning in deep neural networks}.
\newblock \bibinfo{journal}{\emph{arXiv preprint arXiv:1706.05098}} (\bibinfo{year}{2017}).
\newblock


\bibitem[Saadat et~al\mbox{.}(2022)]%
        {saadat2022knowledge}
\bibfield{author}{\bibinfo{person}{Hajira Saadat}, \bibinfo{person}{Babar Shah}, \bibinfo{person}{Zahid Halim}, {and} \bibinfo{person}{Sajid Anwar}.} \bibinfo{year}{2022}\natexlab{}.
\newblock \showarticletitle{Knowledge graph-based convolutional network coupled with sentiment analysis towards enhanced drug recommendation}.
\newblock \bibinfo{journal}{\emph{IEEE/ACM Transactions on Computational Biology and Bioinformatics}} (\bibinfo{year}{2022}).
\newblock


\bibitem[Sabour et~al\mbox{.}(2017)]%
        {sabour2017dynamic}
\bibfield{author}{\bibinfo{person}{Sara Sabour}, \bibinfo{person}{Nicholas Frosst}, {and} \bibinfo{person}{Geoffrey~E Hinton}.} \bibinfo{year}{2017}\natexlab{}.
\newblock \showarticletitle{Dynamic routing between capsules}.
\newblock \bibinfo{journal}{\emph{Proc. of NeurIPS}} (\bibinfo{year}{2017}).
\newblock


\bibitem[Salakhutdinov et~al\mbox{.}(2007)]%
        {salakhutdinov2007restricted}
\bibfield{author}{\bibinfo{person}{Ruslan Salakhutdinov}, \bibinfo{person}{Andriy Mnih}, {and} \bibinfo{person}{Geoffrey Hinton}.} \bibinfo{year}{2007}\natexlab{}.
\newblock \showarticletitle{Restricted Boltzmann machines for collaborative filtering}. In \bibinfo{booktitle}{\emph{Proc. of ICML}}.
\newblock


\bibitem[Sarwar et~al\mbox{.}(2001)]%
        {sarwar2001item}
\bibfield{author}{\bibinfo{person}{Badrul Sarwar}, \bibinfo{person}{George Karypis}, \bibinfo{person}{Joseph Konstan}, {and} \bibinfo{person}{John Riedl}.} \bibinfo{year}{2001}\natexlab{}.
\newblock \showarticletitle{Item-based collaborative filtering recommendation algorithms}. In \bibinfo{booktitle}{\emph{Proc. of WWW}}.
\newblock


\bibitem[Seol et~al\mbox{.}(2022)]%
        {seol2022exploiting}
\bibfield{author}{\bibinfo{person}{Jinseok~Jamie Seol}, \bibinfo{person}{Youngrok Ko}, {and} \bibinfo{person}{Sang-goo Lee}.} \bibinfo{year}{2022}\natexlab{}.
\newblock \showarticletitle{Exploiting session information in BERT-based session-aware sequential recommendation}. In \bibinfo{booktitle}{\emph{Proc. of SIGIR}}. \bibinfo{pages}{2639--2644}.
\newblock


\bibitem[Shalaby et~al\mbox{.}(2022)]%
        {shalaby2022m2trec}
\bibfield{author}{\bibinfo{person}{Walid Shalaby}, \bibinfo{person}{Sejoon Oh}, \bibinfo{person}{Amir Afsharinejad}, \bibinfo{person}{Srijan Kumar}, {and} \bibinfo{person}{Xiquan Cui}.} \bibinfo{year}{2022}\natexlab{}.
\newblock \showarticletitle{M2TRec: Metadata-aware Multi-task Transformer for Large-scale and Cold-start free Session-based Recommendations}. In \bibinfo{booktitle}{\emph{Proc. of RecSys}}.
\newblock


\bibitem[Shani et~al\mbox{.}(2005)]%
        {shani2005mdp}
\bibfield{author}{\bibinfo{person}{Guy Shani}, \bibinfo{person}{David Heckerman}, \bibinfo{person}{Ronen~I Brafman}, {and} \bibinfo{person}{Craig Boutilier}.} \bibinfo{year}{2005}\natexlab{}.
\newblock \showarticletitle{An MDP-based recommender system.}
\newblock \bibinfo{journal}{\emph{Journal of Machine Learning Research}} (\bibinfo{year}{2005}).
\newblock


\bibitem[Shao et~al\mbox{.}(2009)]%
        {shao2009music}
\bibfield{author}{\bibinfo{person}{Bo Shao}, \bibinfo{person}{Dingding Wang}, \bibinfo{person}{Tao Li}, {and} \bibinfo{person}{Mitsunori Ogihara}.} \bibinfo{year}{2009}\natexlab{}.
\newblock \showarticletitle{Music recommendation based on acoustic features and user access patterns}.
\newblock \bibinfo{journal}{\emph{IEEE Transactions on Audio, Speech, and Language Processing}} (\bibinfo{year}{2009}).
\newblock


\bibitem[Shaw et~al\mbox{.}(2018)]%
        {shaw2018self}
\bibfield{author}{\bibinfo{person}{Peter Shaw}, \bibinfo{person}{Jakob Uszkoreit}, {and} \bibinfo{person}{Ashish Vaswani}.} \bibinfo{year}{2018}\natexlab{}.
\newblock \showarticletitle{Self-Attention with Relative Position Representations}. In \bibinfo{booktitle}{\emph{Proc. of NAACL}}. \bibinfo{pages}{464--468}.
\newblock


\bibitem[Shen et~al\mbox{.}(2021)]%
        {shen2021multi}
\bibfield{author}{\bibinfo{person}{Qi Shen}, \bibinfo{person}{Lingfei Wu}, \bibinfo{person}{Yitong Pang}, \bibinfo{person}{Yiming Zhang}, \bibinfo{person}{Zhihua Wei}, \bibinfo{person}{Fangli Xu}, {and} \bibinfo{person}{Bo Long}.} \bibinfo{year}{2021}\natexlab{}.
\newblock \showarticletitle{Multi-behavior Graph Contextual Aware Network for Session-based Recommendation}.
\newblock \bibinfo{journal}{\emph{arXiv preprint arXiv:2109.11903}} (\bibinfo{year}{2021}).
\newblock


\bibitem[Shen et~al\mbox{.}(2023)]%
        {shen2021temporal}
\bibfield{author}{\bibinfo{person}{Qi Shen}, \bibinfo{person}{Shixuan Zhu}, \bibinfo{person}{Yitong Pang}, \bibinfo{person}{Yiming Zhang}, {and} \bibinfo{person}{Zhihua Wei}.} \bibinfo{year}{2023}\natexlab{}.
\newblock \showarticletitle{Temporal aware multi-interest graph neural network for session-based recommendation}. In \bibinfo{booktitle}{\emph{Proc. of ACML}}.
\newblock


\bibitem[Shen and Jin(2012)]%
        {shen2012learning}
\bibfield{author}{\bibinfo{person}{Yelong Shen} {and} \bibinfo{person}{Ruoming Jin}.} \bibinfo{year}{2012}\natexlab{}.
\newblock \showarticletitle{Learning personal+ social latent factor model for social recommendation}. In \bibinfo{booktitle}{\emph{Proc. of KDD}}.
\newblock


\bibitem[Shih and Chi(2018)]%
        {shih2018automatic}
\bibfield{author}{\bibinfo{person}{Shun-Yao Shih} {and} \bibinfo{person}{Heng-Yu Chi}.} \bibinfo{year}{2018}\natexlab{}.
\newblock \showarticletitle{Automatic, personalized, and flexible playlist generation using reinforcement learning}.
\newblock \bibinfo{journal}{\emph{arXiv preprint arXiv:1809.04214}} (\bibinfo{year}{2018}).
\newblock


\bibitem[Singha~Roy et~al\mbox{.}(2023)]%
        {singha2023scalable}
\bibfield{author}{\bibinfo{person}{Aayush Singha~Roy}, \bibinfo{person}{Edoardo D'Amico}, \bibinfo{person}{Elias Tragos}, \bibinfo{person}{Aonghus Lawlor}, {and} \bibinfo{person}{Neil Hurley}.} \bibinfo{year}{2023}\natexlab{}.
\newblock \showarticletitle{Scalable Deep Q-Learning for Session-Based Slate Recommendation}. In \bibinfo{booktitle}{\emph{Proc. of RecSys}}.
\newblock


\bibitem[Song et~al\mbox{.}(2019a)]%
        {song2019session_1}
\bibfield{author}{\bibinfo{person}{Bo Song}, \bibinfo{person}{Yi Cao}, \bibinfo{person}{Weifeng Zhang}, {and} \bibinfo{person}{Congfu Xu}.} \bibinfo{year}{2019}\natexlab{a}.
\newblock \showarticletitle{Session-based recommendation with hierarchical memory networks}. In \bibinfo{booktitle}{\emph{Proc. of CIKM}}.
\newblock


\bibitem[Song et~al\mbox{.}(2019b)]%
        {song2019session}
\bibfield{author}{\bibinfo{person}{Weiping Song}, \bibinfo{person}{Zhiping Xiao}, \bibinfo{person}{Yifan Wang}, \bibinfo{person}{Laurent Charlin}, \bibinfo{person}{Ming Zhang}, {and} \bibinfo{person}{Jian Tang}.} \bibinfo{year}{2019}\natexlab{b}.
\newblock \showarticletitle{Session-based social recommendation via dynamic graph attention networks}. In \bibinfo{booktitle}{\emph{Proc. of WSDM}}.
\newblock


\bibitem[Su et~al\mbox{.}(2024)]%
        {su2024roformer}
\bibfield{author}{\bibinfo{person}{Jianlin Su}, \bibinfo{person}{Murtadha Ahmed}, \bibinfo{person}{Yu Lu}, \bibinfo{person}{Shengfeng Pan}, \bibinfo{person}{Wen Bo}, {and} \bibinfo{person}{Yunfeng Liu}.} \bibinfo{year}{2024}\natexlab{}.
\newblock \showarticletitle{Roformer: Enhanced transformer with rotary position embedding}.
\newblock \bibinfo{journal}{\emph{Neurocomputing}}  \bibinfo{volume}{568} (\bibinfo{year}{2024}), \bibinfo{pages}{127063}.
\newblock


\bibitem[Su et~al\mbox{.}(2023)]%
        {su2023enhancing}
\bibfield{author}{\bibinfo{person}{Jiajie Su}, \bibinfo{person}{Chaochao Chen}, \bibinfo{person}{Weiming Liu}, \bibinfo{person}{Fei Wu}, \bibinfo{person}{Xiaolin Zheng}, {and} \bibinfo{person}{Haoming Lyu}.} \bibinfo{year}{2023}\natexlab{}.
\newblock \showarticletitle{Enhancing Hierarchy-Aware Graph Networks with Deep Dual Clustering for Session-based Recommendation}. In \bibinfo{booktitle}{\emph{Proc. of WWW}}.
\newblock


\bibitem[Su and Khoshgoftaar(2009)]%
        {su2009survey}
\bibfield{author}{\bibinfo{person}{Xiaoyuan Su} {and} \bibinfo{person}{Taghi~M Khoshgoftaar}.} \bibinfo{year}{2009}\natexlab{}.
\newblock \showarticletitle{A survey of collaborative filtering techniques}.
\newblock \bibinfo{journal}{\emph{Advances in artificial intelligence}} (\bibinfo{year}{2009}), \bibinfo{pages}{421425}.
\newblock


\bibitem[Sun et~al\mbox{.}(2018)]%
        {sun2018gaussian}
\bibfield{author}{\bibinfo{person}{Chi Sun}, \bibinfo{person}{Hang Yan}, \bibinfo{person}{Xipeng Qiu}, {and} \bibinfo{person}{Xuanjing Huang}.} \bibinfo{year}{2018}\natexlab{}.
\newblock \showarticletitle{Gaussian word embedding with a wasserstein distance loss}.
\newblock \bibinfo{journal}{\emph{arXiv preprint arXiv:1808.07016}} (\bibinfo{year}{2018}).
\newblock


\bibitem[Sun et~al\mbox{.}(2019)]%
        {sun2019bert4rec}
\bibfield{author}{\bibinfo{person}{Fei Sun}, \bibinfo{person}{Jun Liu}, \bibinfo{person}{Jian Wu}, \bibinfo{person}{Changhua Pei}, \bibinfo{person}{Xiao Lin}, \bibinfo{person}{Wenwu Ou}, {and} \bibinfo{person}{Peng Jiang}.} \bibinfo{year}{2019}\natexlab{}.
\newblock \showarticletitle{BERT4Rec: Sequential recommendation with bidirectional encoder representations from transformer}. In \bibinfo{booktitle}{\emph{Proc. of CIKM}}.
\newblock


\bibitem[Tan et~al\mbox{.}(2016)]%
        {tan2016improved}
\bibfield{author}{\bibinfo{person}{Yong~Kiam Tan}, \bibinfo{person}{Xinxing Xu}, {and} \bibinfo{person}{Yong Liu}.} \bibinfo{year}{2016}\natexlab{}.
\newblock \showarticletitle{Improved recurrent neural networks for session-based recommendations}. In \bibinfo{booktitle}{\emph{Proceedings of the 1st workshop on deep learning for recommender systems}}.
\newblock


\bibitem[Tian et~al\mbox{.}(2022)]%
        {tian2022multi}
\bibfield{author}{\bibinfo{person}{Yu Tian}, \bibinfo{person}{Jianxin Chang}, \bibinfo{person}{Yanan Niu}, \bibinfo{person}{Yang Song}, {and} \bibinfo{person}{Chenliang Li}.} \bibinfo{year}{2022}\natexlab{}.
\newblock \showarticletitle{When multi-level meets multi-interest: A multi-grained neural model for sequential recommendation}. In \bibinfo{booktitle}{\emph{Proc. of SIGIR}}. \bibinfo{pages}{1632--1641}.
\newblock


\bibitem[Touvron et~al\mbox{.}(2023)]%
        {touvron2023llama}
\bibfield{author}{\bibinfo{person}{Hugo Touvron}, \bibinfo{person}{Thibaut Lavril}, \bibinfo{person}{Gautier Izacard}, \bibinfo{person}{Xavier Martinet}, \bibinfo{person}{Marie-Anne Lachaux}, \bibinfo{person}{Timoth{\'e}e Lacroix}, \bibinfo{person}{Baptiste Rozi{\`e}re}, \bibinfo{person}{Naman Goyal}, \bibinfo{person}{Eric Hambro}, \bibinfo{person}{Faisal Azhar}, {et~al\mbox{.}}} \bibinfo{year}{2023}\natexlab{}.
\newblock \showarticletitle{Llama: Open and efficient foundation language models}.
\newblock \bibinfo{journal}{\emph{arXiv preprint arXiv:2302.13971}} (\bibinfo{year}{2023}).
\newblock


\bibitem[Turgut et~al\mbox{.}(2023)]%
        {turgut2023prod2vec}
\bibfield{author}{\bibinfo{person}{Hacer Turgut}, \bibinfo{person}{Tan~Doruk Yetki}, \bibinfo{person}{{\"O}m{\"u}r Bali}, {and} \bibinfo{person}{Tayfun~Arda Y{\"u}cel}.} \bibinfo{year}{2023}\natexlab{}.
\newblock \showarticletitle{Prod2Vec-Var: A Session Based Recommendation System with Enhanced Diversity}. In \bibinfo{booktitle}{\emph{Proc. of CIKM}}. \bibinfo{pages}{5253--5254}.
\newblock


\bibitem[Vargas and Castells(2011)]%
        {vargas2011rank}
\bibfield{author}{\bibinfo{person}{Sa{\'u}l Vargas} {and} \bibinfo{person}{Pablo Castells}.} \bibinfo{year}{2011}\natexlab{}.
\newblock \showarticletitle{Rank and relevance in novelty and diversity metrics for recommender systems}. In \bibinfo{booktitle}{\emph{Proceedings of the fifth ACM conference on Recommender systems}}. \bibinfo{pages}{109--116}.
\newblock


\bibitem[Vaswani et~al\mbox{.}(2017)]%
        {vaswani2017attention}
\bibfield{author}{\bibinfo{person}{Ashish Vaswani}, \bibinfo{person}{Noam Shazeer}, \bibinfo{person}{Niki Parmar}, \bibinfo{person}{Jakob Uszkoreit}, \bibinfo{person}{Llion Jones}, \bibinfo{person}{Aidan~N Gomez}, \bibinfo{person}{{\L}ukasz Kaiser}, {and} \bibinfo{person}{Illia Polosukhin}.} \bibinfo{year}{2017}\natexlab{}.
\newblock \showarticletitle{Attention is all you need}.
\newblock \bibinfo{journal}{\emph{Proc. of NeurIPS}} (\bibinfo{year}{2017}).
\newblock


\bibitem[Velickovic et~al\mbox{.}(2017)]%
        {velickovic2017graph}
\bibfield{author}{\bibinfo{person}{Petar Velickovic}, \bibinfo{person}{Guillem Cucurull}, \bibinfo{person}{Arantxa Casanova}, \bibinfo{person}{Adriana Romero}, \bibinfo{person}{Pietro Lio}, {and} \bibinfo{person}{Yoshua Bengio}.} \bibinfo{year}{2017}\natexlab{}.
\newblock \showarticletitle{Graph attention networks}.
\newblock \bibinfo{journal}{\emph{stat}} (\bibinfo{year}{2017}).
\newblock


\bibitem[Vilnis and McCallum(2014)]%
        {vilnis2014word}
\bibfield{author}{\bibinfo{person}{Luke Vilnis} {and} \bibinfo{person}{Andrew McCallum}.} \bibinfo{year}{2014}\natexlab{}.
\newblock \showarticletitle{Word representations via gaussian embedding}.
\newblock \bibinfo{journal}{\emph{arXiv preprint arXiv:1412.6623}} (\bibinfo{year}{2014}).
\newblock


\bibitem[Wang et~al\mbox{.}(2006)]%
        {wang2006unifying}
\bibfield{author}{\bibinfo{person}{Jun Wang}, \bibinfo{person}{Arjen~P De~Vries}, {and} \bibinfo{person}{Marcel~JT Reinders}.} \bibinfo{year}{2006}\natexlab{}.
\newblock \showarticletitle{Unifying user-based and item-based collaborative filtering approaches by similarity fusion}. In \bibinfo{booktitle}{\emph{Proc. of SIGIR}}.
\newblock


\bibitem[Wang et~al\mbox{.}(2021b)]%
        {wang2021session}
\bibfield{author}{\bibinfo{person}{Jianling Wang}, \bibinfo{person}{Kaize Ding}, \bibinfo{person}{Ziwei Zhu}, {and} \bibinfo{person}{James Caverlee}.} \bibinfo{year}{2021}\natexlab{b}.
\newblock \showarticletitle{Session-based recommendation with hypergraph attention networks}. In \bibinfo{booktitle}{\emph{Proc. of SDM}}.
\newblock


\bibitem[Wang and Lim(2023)]%
        {wang2023zero}
\bibfield{author}{\bibinfo{person}{Lei Wang} {and} \bibinfo{person}{Ee-Peng Lim}.} \bibinfo{year}{2023}\natexlab{}.
\newblock \showarticletitle{Zero-Shot Next-Item Recommendation using Large Pretrained Language Models}.
\newblock \bibinfo{journal}{\emph{arXiv preprint arXiv:2304.03153}} (\bibinfo{year}{2023}).
\newblock


\bibitem[Wang et~al\mbox{.}(2022a)]%
        {wang2022self}
\bibfield{author}{\bibinfo{person}{Liuyin Wang}, \bibinfo{person}{Xianghong Xu}, \bibinfo{person}{Kai Ouyang}, \bibinfo{person}{Huanzhong Duan}, \bibinfo{person}{Yanxiong Lu}, {and} \bibinfo{person}{Hai-Tao Zheng}.} \bibinfo{year}{2022}\natexlab{a}.
\newblock \showarticletitle{Self-Supervised Dual-Channel Attentive Network for Session-based Social Recommendation}. In \bibinfo{booktitle}{\emph{2022 IEEE 38th International Conference on Data Engineering (ICDE)}}.
\newblock


\bibitem[Wang et~al\mbox{.}(2019b)]%
        {wang2019collaborative}
\bibfield{author}{\bibinfo{person}{Meirui Wang}, \bibinfo{person}{Pengjie Ren}, \bibinfo{person}{Lei Mei}, \bibinfo{person}{Zhumin Chen}, \bibinfo{person}{Jun Ma}, {and} \bibinfo{person}{Maarten de Rijke}.} \bibinfo{year}{2019}\natexlab{b}.
\newblock \showarticletitle{A collaborative session-based recommendation approach with parallel memory modules}. In \bibinfo{booktitle}{\emph{Proc. of SIGIR}}.
\newblock


\bibitem[Wang et~al\mbox{.}(2020a)]%
        {wang2020kerl}
\bibfield{author}{\bibinfo{person}{Pengfei Wang}, \bibinfo{person}{Yu Fan}, \bibinfo{person}{Long Xia}, \bibinfo{person}{Wayne~Xin Zhao}, \bibinfo{person}{ShaoZhang Niu}, {and} \bibinfo{person}{Jimmy Huang}.} \bibinfo{year}{2020}\natexlab{a}.
\newblock \showarticletitle{KERL: A knowledge-guided reinforcement learning model for sequential recommendation}. In \bibinfo{booktitle}{\emph{Proc. of SIGIR}}.
\newblock


\bibitem[Wang et~al\mbox{.}(2021a)]%
        {wang2021survey}
\bibfield{author}{\bibinfo{person}{Shoujin Wang}, \bibinfo{person}{Longbing Cao}, \bibinfo{person}{Yan Wang}, \bibinfo{person}{Quan~Z Sheng}, \bibinfo{person}{Mehmet~A Orgun}, {and} \bibinfo{person}{Defu Lian}.} \bibinfo{year}{2021}\natexlab{a}.
\newblock \showarticletitle{A survey on session-based recommender systems}.
\newblock \bibinfo{journal}{\emph{ACM Computing Surveys (CSUR)}} (\bibinfo{year}{2021}).
\newblock


\bibitem[Wang et~al\mbox{.}(2017)]%
        {wang2017perceiving}
\bibfield{author}{\bibinfo{person}{Shoujin Wang}, \bibinfo{person}{Liang Hu}, {and} \bibinfo{person}{Longbing Cao}.} \bibinfo{year}{2017}\natexlab{}.
\newblock \showarticletitle{Perceiving the next choice with comprehensive transaction embeddings for online recommendation}. In \bibinfo{booktitle}{\emph{Proc. of ECML}}.
\newblock


\bibitem[Wang et~al\mbox{.}(2018b)]%
        {wang2018attention}
\bibfield{author}{\bibinfo{person}{Shoujin Wang}, \bibinfo{person}{Liang Hu}, \bibinfo{person}{Longbing Cao}, \bibinfo{person}{Xiaoshui Huang}, \bibinfo{person}{Defu Lian}, {and} \bibinfo{person}{Wei Liu}.} \bibinfo{year}{2018}\natexlab{b}.
\newblock \showarticletitle{Attention-based transactional context embedding for next-item recommendation}. In \bibinfo{booktitle}{\emph{Proc. of AAAI}}.
\newblock


\bibitem[Wang et~al\mbox{.}(2022b)]%
        {wang2022sequential}
\bibfield{author}{\bibinfo{person}{Shoujin Wang}, \bibinfo{person}{Qi Zhang}, \bibinfo{person}{Liang Hu}, \bibinfo{person}{Xiuzhen Zhang}, \bibinfo{person}{Yan Wang}, {and} \bibinfo{person}{Charu Aggarwal}.} \bibinfo{year}{2022}\natexlab{b}.
\newblock \showarticletitle{Sequential/session-based recommendations: Challenges, approaches, applications and opportunities}. In \bibinfo{booktitle}{\emph{Proc. of SIGIR}}. \bibinfo{pages}{3425--3428}.
\newblock


\bibitem[Wang and Isola(2020)]%
        {2020arXiv200510242W}
\bibfield{author}{\bibinfo{person}{Tongzhou Wang} {and} \bibinfo{person}{Phillip Isola}.} \bibinfo{year}{2020}\natexlab{}.
\newblock \showarticletitle{Understanding contrastive representation learning through alignment and uniformity on the hypersphere}. In \bibinfo{booktitle}{\emph{Proc. of ICML}}. \bibinfo{pages}{9929--9939}.
\newblock


\bibitem[Wang et~al\mbox{.}(2021c)]%
        {wang2021deconfounded}
\bibfield{author}{\bibinfo{person}{Wenjie Wang}, \bibinfo{person}{Fuli Feng}, \bibinfo{person}{Xiangnan He}, \bibinfo{person}{Xiang Wang}, {and} \bibinfo{person}{Tat-Seng Chua}.} \bibinfo{year}{2021}\natexlab{c}.
\newblock \showarticletitle{Deconfounded recommendation for alleviating bias amplification}. In \bibinfo{booktitle}{\emph{Proc. of KDD}}.
\newblock


\bibitem[Wang et~al\mbox{.}(2023)]%
        {wang2023diffusion}
\bibfield{author}{\bibinfo{person}{Wenjie Wang}, \bibinfo{person}{Yiyan Xu}, \bibinfo{person}{Fuli Feng}, \bibinfo{person}{Xinyu Lin}, \bibinfo{person}{Xiangnan He}, {and} \bibinfo{person}{Tat-Seng Chua}.} \bibinfo{year}{2023}\natexlab{}.
\newblock \showarticletitle{Diffusion recommender model}. In \bibinfo{booktitle}{\emph{Proc. of SIGIR}}. \bibinfo{pages}{832--841}.
\newblock


\bibitem[Wang et~al\mbox{.}(2018c)]%
        {wang2018streaming}
\bibfield{author}{\bibinfo{person}{Weiqing Wang}, \bibinfo{person}{Hongzhi Yin}, \bibinfo{person}{Zi Huang}, \bibinfo{person}{Qinyong Wang}, \bibinfo{person}{Xingzhong Du}, {and} \bibinfo{person}{Quoc Viet~Hung Nguyen}.} \bibinfo{year}{2018}\natexlab{c}.
\newblock \showarticletitle{Streaming ranking based recommender systems}. In \bibinfo{booktitle}{\emph{Proc. of SIGIR}}.
\newblock


\bibitem[Wang et~al\mbox{.}(2020c)]%
        {wang2020beyond}
\bibfield{author}{\bibinfo{person}{Wen Wang}, \bibinfo{person}{Wei Zhang}, \bibinfo{person}{Shukai Liu}, \bibinfo{person}{Qi Liu}, \bibinfo{person}{Bo Zhang}, \bibinfo{person}{Leyu Lin}, {and} \bibinfo{person}{Hongyuan Zha}.} \bibinfo{year}{2020}\natexlab{c}.
\newblock \showarticletitle{Beyond clicks: Modeling multi-relational item graph for session-based target behavior prediction}. In \bibinfo{booktitle}{\emph{Proc. of WWW}}.
\newblock


\bibitem[Wang et~al\mbox{.}(2019a)]%
        {wang2019heterogeneous}
\bibfield{author}{\bibinfo{person}{Xiao Wang}, \bibinfo{person}{Houye Ji}, \bibinfo{person}{Chuan Shi}, \bibinfo{person}{Bai Wang}, \bibinfo{person}{Yanfang Ye}, \bibinfo{person}{Peng Cui}, {and} \bibinfo{person}{Philip~S Yu}.} \bibinfo{year}{2019}\natexlab{a}.
\newblock \showarticletitle{Heterogeneous graph attention network}. In \bibinfo{booktitle}{\emph{Proc. of WWW}}.
\newblock


\bibitem[Wang et~al\mbox{.}(2024)]%
        {wang2024diff}
\bibfield{author}{\bibinfo{person}{Yuhao Wang}, \bibinfo{person}{Ziru Liu}, \bibinfo{person}{Yichao Wang}, \bibinfo{person}{Xiangyu Zhao}, \bibinfo{person}{Bo Chen}, \bibinfo{person}{Huifeng Guo}, {and} \bibinfo{person}{Ruiming Tang}.} \bibinfo{year}{2024}\natexlab{}.
\newblock \showarticletitle{Diff-MSR: A Diffusion Model Enhanced Paradigm for Cold-Start Multi-Scenario Recommendation}. In \bibinfo{booktitle}{\emph{Proc. of WSDM}}. \bibinfo{pages}{779--787}.
\newblock


\bibitem[Wang et~al\mbox{.}(2018a)]%
        {wang2018variational}
\bibfield{author}{\bibinfo{person}{Zhitao Wang}, \bibinfo{person}{Chengyao Chen}, \bibinfo{person}{Ke Zhang}, \bibinfo{person}{Yu Lei}, {and} \bibinfo{person}{Wenjie Li}.} \bibinfo{year}{2018}\natexlab{a}.
\newblock \showarticletitle{Variational recurrent model for session-based recommendation}. In \bibinfo{booktitle}{\emph{Proc. of CIKM}}.
\newblock


\bibitem[Wang et~al\mbox{.}(2020b)]%
        {wang2020global}
\bibfield{author}{\bibinfo{person}{Ziyang Wang}, \bibinfo{person}{Wei Wei}, \bibinfo{person}{Gao Cong}, \bibinfo{person}{Xiao-Li Li}, \bibinfo{person}{Xian-Ling Mao}, {and} \bibinfo{person}{Minghui Qiu}.} \bibinfo{year}{2020}\natexlab{b}.
\newblock \showarticletitle{Global context enhanced graph neural networks for session-based recommendation}. In \bibinfo{booktitle}{\emph{Proc. of SIGIR}}.
\newblock


\bibitem[Wei et~al\mbox{.}(2022)]%
        {wei2022gsl4rec}
\bibfield{author}{\bibinfo{person}{Chunyu Wei}, \bibinfo{person}{Bing Bai}, \bibinfo{person}{Kun Bai}, {and} \bibinfo{person}{Fei Wang}.} \bibinfo{year}{2022}\natexlab{}.
\newblock \showarticletitle{Gsl4rec: Session-based recommendations with collective graph structure learning and next interaction prediction}. In \bibinfo{booktitle}{\emph{Proc. of WWW}}.
\newblock


\bibitem[Wei et~al\mbox{.}(2021)]%
        {wei2021model}
\bibfield{author}{\bibinfo{person}{Tianxin Wei}, \bibinfo{person}{Fuli Feng}, \bibinfo{person}{Jiawei Chen}, \bibinfo{person}{Ziwei Wu}, \bibinfo{person}{Jinfeng Yi}, {and} \bibinfo{person}{Xiangnan He}.} \bibinfo{year}{2021}\natexlab{}.
\newblock \showarticletitle{Model-agnostic counterfactual reasoning for eliminating popularity bias in recommender system}. In \bibinfo{booktitle}{\emph{Proc. of KDD}}.
\newblock


\bibitem[Wilm et~al\mbox{.}(2023)]%
        {wilm2023scaling}
\bibfield{author}{\bibinfo{person}{Timo Wilm}, \bibinfo{person}{Philipp Normann}, \bibinfo{person}{Sophie Baumeister}, {and} \bibinfo{person}{Paul-Vincent Kobow}.} \bibinfo{year}{2023}\natexlab{}.
\newblock \showarticletitle{Scaling Session-Based Transformer Recommendations using Optimized Negative Sampling and Loss Functions}. In \bibinfo{booktitle}{\emph{Proc. of RecSys}}.
\newblock


\bibitem[Wu and Yan(2017)]%
        {wu2017session}
\bibfield{author}{\bibinfo{person}{Chen Wu} {and} \bibinfo{person}{Ming Yan}.} \bibinfo{year}{2017}\natexlab{}.
\newblock \showarticletitle{Session-aware information embedding for e-commerce product recommendation}. In \bibinfo{booktitle}{\emph{Proceedings of the 2017 ACM on conference on information and knowledge management}}.
\newblock


\bibitem[Wu et~al\mbox{.}(2023a)]%
        {wu2023generic}
\bibfield{author}{\bibinfo{person}{Huizi Wu}, \bibinfo{person}{Hui Fang}, \bibinfo{person}{Zhu Sun}, \bibinfo{person}{Cong Geng}, \bibinfo{person}{Xinyu Kong}, {and} \bibinfo{person}{Yew-Soon Ong}.} \bibinfo{year}{2023}\natexlab{a}.
\newblock \showarticletitle{A generic reinforced explainable framework with knowledge graph for session-based recommendation}. In \bibinfo{booktitle}{\emph{2023 IEEE 39th International Conference on Data Engineering (ICDE)}}.
\newblock


\bibitem[Wu et~al\mbox{.}(2023b)]%
        {wu2023causality}
\bibfield{author}{\bibinfo{person}{Huizi Wu}, \bibinfo{person}{Cong Geng}, {and} \bibinfo{person}{Hui Fang}.} \bibinfo{year}{2023}\natexlab{b}.
\newblock \showarticletitle{Causality and correlation graph modeling for effective and explainable session-based recommendation}.
\newblock \bibinfo{journal}{\emph{ACM Transactions on the Web}} (\bibinfo{year}{2023}), \bibinfo{pages}{1--25}.
\newblock


\bibitem[Wu et~al\mbox{.}(2022)]%
        {wu2020graph}
\bibfield{author}{\bibinfo{person}{Shiwen Wu}, \bibinfo{person}{Fei Sun}, \bibinfo{person}{Wentao Zhang}, \bibinfo{person}{Xu Xie}, {and} \bibinfo{person}{Bin Cui}.} \bibinfo{year}{2022}\natexlab{}.
\newblock \showarticletitle{Graph neural networks in recommender systems: a survey}.
\newblock \bibinfo{journal}{\emph{Comput. Surveys}} (\bibinfo{year}{2022}), \bibinfo{pages}{1--37}.
\newblock


\bibitem[Wu et~al\mbox{.}(2019)]%
        {wu2019session}
\bibfield{author}{\bibinfo{person}{Shu Wu}, \bibinfo{person}{Yuyuan Tang}, \bibinfo{person}{Yanqiao Zhu}, \bibinfo{person}{Liang Wang}, \bibinfo{person}{Xing Xie}, {and} \bibinfo{person}{Tieniu Tan}.} \bibinfo{year}{2019}\natexlab{}.
\newblock \showarticletitle{Session-based recommendation with graph neural networks}. In \bibinfo{booktitle}{\emph{Proc. of AAAI}}.
\newblock


\bibitem[Wu et~al\mbox{.}(2013)]%
        {wu2013personalized}
\bibfield{author}{\bibinfo{person}{Xiang Wu}, \bibinfo{person}{Qi Liu}, \bibinfo{person}{Enhong Chen}, \bibinfo{person}{Liang He}, \bibinfo{person}{Jingsong Lv}, \bibinfo{person}{Can Cao}, {and} \bibinfo{person}{Guoping Hu}.} \bibinfo{year}{2013}\natexlab{}.
\newblock \showarticletitle{Personalized next-song recommendation in online karaokes}. In \bibinfo{booktitle}{\emph{Proc. of RecSys}}.
\newblock


\bibitem[Wu et~al\mbox{.}(2020)]%
        {wu2020connecting}
\bibfield{author}{\bibinfo{person}{Zonghan Wu}, \bibinfo{person}{Shirui Pan}, \bibinfo{person}{Guodong Long}, \bibinfo{person}{Jing Jiang}, \bibinfo{person}{Xiaojun Chang}, {and} \bibinfo{person}{Chengqi Zhang}.} \bibinfo{year}{2020}\natexlab{}.
\newblock \showarticletitle{Connecting the dots: Multivariate time series forecasting with graph neural networks}. In \bibinfo{booktitle}{\emph{Proc. of KDD}}.
\newblock


\bibitem[Xia et~al\mbox{.}(2021a)]%
        {xia2021self_1}
\bibfield{author}{\bibinfo{person}{Xin Xia}, \bibinfo{person}{Hongzhi Yin}, \bibinfo{person}{Junliang Yu}, \bibinfo{person}{Yingxia Shao}, {and} \bibinfo{person}{Lizhen Cui}.} \bibinfo{year}{2021}\natexlab{a}.
\newblock \showarticletitle{Self-Supervised Graph Co-Training for Session-based Recommendation}. In \bibinfo{booktitle}{\emph{Proc. of CIKM}}.
\newblock


\bibitem[Xia et~al\mbox{.}(2021b)]%
        {xia2021self}
\bibfield{author}{\bibinfo{person}{Xin Xia}, \bibinfo{person}{Hongzhi Yin}, \bibinfo{person}{Junliang Yu}, \bibinfo{person}{Qinyong Wang}, \bibinfo{person}{Lizhen Cui}, {and} \bibinfo{person}{Xiangliang Zhang}.} \bibinfo{year}{2021}\natexlab{b}.
\newblock \showarticletitle{Self-supervised hypergraph convolutional networks for session-based recommendation}. In \bibinfo{booktitle}{\emph{Proc. of AAAI}}.
\newblock


\bibitem[Xia et~al\mbox{.}(2023a)]%
        {xia2023efficient}
\bibfield{author}{\bibinfo{person}{Xin Xia}, \bibinfo{person}{Junliang Yu}, \bibinfo{person}{Qinyong Wang}, \bibinfo{person}{Chaoqun Yang}, \bibinfo{person}{Nguyen Quoc~Viet Hung}, {and} \bibinfo{person}{Hongzhi Yin}.} \bibinfo{year}{2023}\natexlab{a}.
\newblock \showarticletitle{Efficient on-device session-based recommendation}.
\newblock \bibinfo{journal}{\emph{ACM Transactions on Information Systems}} (\bibinfo{year}{2023}), \bibinfo{pages}{1--24}.
\newblock


\bibitem[Xia et~al\mbox{.}(2023b)]%
        {xia2023towards}
\bibfield{author}{\bibinfo{person}{Xin Xia}, \bibinfo{person}{Junliang Yu}, \bibinfo{person}{Guandong Xu}, {and} \bibinfo{person}{Hongzhi Yin}.} \bibinfo{year}{2023}\natexlab{b}.
\newblock \showarticletitle{Towards communication-efficient model updating for on-device session-based recommendation}. In \bibinfo{booktitle}{\emph{Proc. of CIKM}}. \bibinfo{pages}{2795--2804}.
\newblock


\bibitem[Xian et~al\mbox{.}(2019)]%
        {xian2019reinforcement}
\bibfield{author}{\bibinfo{person}{Yikun Xian}, \bibinfo{person}{Zuohui Fu}, \bibinfo{person}{S. Muthukrishnan}, \bibinfo{person}{Gerard de Melo}, {and} \bibinfo{person}{Yongfeng Zhang}.} \bibinfo{year}{2019}\natexlab{}.
\newblock \showarticletitle{Reinforcement Knowledge Graph Reasoning for Explainable Recommendation}. In \bibinfo{booktitle}{\emph{Proc. of SIGIR}}. \bibinfo{pages}{285--294}.
\newblock


\bibitem[Xie et~al\mbox{.}(2022)]%
        {xie2022decoupled}
\bibfield{author}{\bibinfo{person}{Yueqi Xie}, \bibinfo{person}{Peilin Zhou}, {and} \bibinfo{person}{Sunghun Kim}.} \bibinfo{year}{2022}\natexlab{}.
\newblock \showarticletitle{Decoupled side information fusion for sequential recommendation}. In \bibinfo{booktitle}{\emph{Proc. of SIGIR}}. \bibinfo{pages}{1611--1621}.
\newblock


\bibitem[Xin et~al\mbox{.}(2024)]%
        {xin2024effectiveness}
\bibfield{author}{\bibinfo{person}{Xin Xin}, \bibinfo{person}{Liu Yang}, \bibinfo{person}{Ziqi Zhao}, \bibinfo{person}{Pengjie Ren}, \bibinfo{person}{Zhumin Chen}, \bibinfo{person}{Jun Ma}, {and} \bibinfo{person}{Zhaochun Ren}.} \bibinfo{year}{2024}\natexlab{}.
\newblock \showarticletitle{On the effectiveness of unlearning in session-based recommendation}. In \bibinfo{booktitle}{\emph{Proc. of WSDM}}. \bibinfo{pages}{855--863}.
\newblock


\bibitem[Xu et~al\mbox{.}(2019)]%
        {xu2019graph}
\bibfield{author}{\bibinfo{person}{Chengfeng Xu}, \bibinfo{person}{Pengpeng Zhao}, \bibinfo{person}{Yanchi Liu}, \bibinfo{person}{Victor~S Sheng}, \bibinfo{person}{Jiajie Xu}, \bibinfo{person}{Fuzhen Zhuang}, \bibinfo{person}{Junhua Fang}, {and} \bibinfo{person}{Xiaofang Zhou}.} \bibinfo{year}{2019}\natexlab{}.
\newblock \showarticletitle{Graph Contextualized Self-Attention Network for Session-based Recommendation.}. In \bibinfo{booktitle}{\emph{Proc. of IJCAI}}.
\newblock


\bibitem[Xu et~al\mbox{.}(2023)]%
        {xu2021deconfounded}
\bibfield{author}{\bibinfo{person}{Shuyuan Xu}, \bibinfo{person}{Juntao Tan}, \bibinfo{person}{Shelby Heinecke}, \bibinfo{person}{Vena~Jia Li}, {and} \bibinfo{person}{Yongfeng Zhang}.} \bibinfo{year}{2023}\natexlab{}.
\newblock \showarticletitle{Deconfounded causal collaborative filtering}.
\newblock \bibinfo{journal}{\emph{ACM Transactions on Recommender Systems}} (\bibinfo{year}{2023}), \bibinfo{pages}{1--25}.
\newblock


\bibitem[Xu et~al\mbox{.}(2022)]%
        {xu2022modeling}
\bibfield{author}{\bibinfo{person}{Xianghong Xu}, \bibinfo{person}{Kai Ouyang}, \bibinfo{person}{Liuyin Wang}, \bibinfo{person}{Jiaxin Zou}, \bibinfo{person}{Yanxiong Lu}, \bibinfo{person}{Hai-Tao Zheng}, {and} \bibinfo{person}{Hong-Gee Kim}.} \bibinfo{year}{2022}\natexlab{}.
\newblock \showarticletitle{Modeling Latent Autocorrelation for Session-based Recommendation}. In \bibinfo{booktitle}{\emph{Proc. of CIKM}}.
\newblock


\bibitem[Yang et~al\mbox{.}(2021b)]%
        {yang2021explanation}
\bibfield{author}{\bibinfo{person}{Aobo Yang}, \bibinfo{person}{Nan Wang}, \bibinfo{person}{Hongbo Deng}, {and} \bibinfo{person}{Hongning Wang}.} \bibinfo{year}{2021}\natexlab{b}.
\newblock \showarticletitle{Explanation as a Defense of Recommendation}. In \bibinfo{booktitle}{\emph{Proc. of WSDM}}. \bibinfo{pages}{1029--1037}.
\newblock


\bibitem[Yang et~al\mbox{.}(2014)]%
        {yang2014continuous}
\bibfield{author}{\bibinfo{person}{Chong Yang}, \bibinfo{person}{Xiaohui Yu}, {and} \bibinfo{person}{Yang Liu}.} \bibinfo{year}{2014}\natexlab{}.
\newblock \showarticletitle{Continuous KNN join processing for real-time recommendation}. In \bibinfo{booktitle}{\emph{Proc. of ICDM}}.
\newblock


\bibitem[Yang et~al\mbox{.}(2023a)]%
        {yang2023loam}
\bibfield{author}{\bibinfo{person}{Heeyoon Yang}, \bibinfo{person}{YunSeok Choi}, \bibinfo{person}{Gahyung Kim}, {and} \bibinfo{person}{Jee-Hyong Lee}.} \bibinfo{year}{2023}\natexlab{a}.
\newblock \showarticletitle{LOAM: Improving Long-tail Session-based Recommendation via Niche Walk Augmentation and Tail Session Mixup}. In \bibinfo{booktitle}{\emph{Proc. of SIGIR}}.
\newblock


\bibitem[Yang et~al\mbox{.}(2023b)]%
        {yang2023diffusion}
\bibfield{author}{\bibinfo{person}{Ling Yang}, \bibinfo{person}{Zhilong Zhang}, \bibinfo{person}{Yang Song}, \bibinfo{person}{Shenda Hong}, \bibinfo{person}{Runsheng Xu}, \bibinfo{person}{Yue Zhao}, \bibinfo{person}{Wentao Zhang}, \bibinfo{person}{Bin Cui}, {and} \bibinfo{person}{Ming-Hsuan Yang}.} \bibinfo{year}{2023}\natexlab{b}.
\newblock \showarticletitle{Diffusion models: A comprehensive survey of methods and applications}.
\newblock \bibinfo{journal}{\emph{Comput. Surveys}} (\bibinfo{year}{2023}), \bibinfo{pages}{1--39}.
\newblock


\bibitem[Yang et~al\mbox{.}(2021a)]%
        {yang2021top}
\bibfield{author}{\bibinfo{person}{Mengyue Yang}, \bibinfo{person}{Quanyu Dai}, \bibinfo{person}{Zhenhua Dong}, \bibinfo{person}{Xu Chen}, \bibinfo{person}{Xiuqiang He}, {and} \bibinfo{person}{Jun Wang}.} \bibinfo{year}{2021}\natexlab{a}.
\newblock \showarticletitle{Top-N Recommendation with Counterfactual User Preference Simulation}. In \bibinfo{booktitle}{\emph{Proc. of CIKM}}.
\newblock


\bibitem[Yang et~al\mbox{.}(2023c)]%
        {yang2023multiple}
\bibfield{author}{\bibinfo{person}{Yaming Yang}, \bibinfo{person}{Jieyu Zhang}, \bibinfo{person}{Yujing Wang}, \bibinfo{person}{Zheng Miao}, {and} \bibinfo{person}{Yunhai Tong}.} \bibinfo{year}{2023}\natexlab{c}.
\newblock \showarticletitle{Multiple Connectivity Views for Session-based Recommendation}. In \bibinfo{booktitle}{\emph{Proc. of RecSys}}.
\newblock


\bibitem[Yap et~al\mbox{.}(2012)]%
        {yap2012effective}
\bibfield{author}{\bibinfo{person}{Ghim-Eng Yap}, \bibinfo{person}{Xiao-Li Li}, {and} \bibinfo{person}{Philip~S Yu}.} \bibinfo{year}{2012}\natexlab{}.
\newblock \showarticletitle{Effective next-items recommendation via personalized sequential pattern mining}. In \bibinfo{booktitle}{\emph{Proc. of DASFAA}}.
\newblock


\bibitem[Ye et~al\mbox{.}(2020)]%
        {ye2020cross}
\bibfield{author}{\bibinfo{person}{Rui Ye}, \bibinfo{person}{Qing Zhang}, {and} \bibinfo{person}{Hengliang Luo}.} \bibinfo{year}{2020}\natexlab{}.
\newblock \showarticletitle{Cross-Session Aware Temporal Convolutional Network for Session-based Recommendation}. In \bibinfo{booktitle}{\emph{Proc. of ICDM}}.
\newblock


\bibitem[Yin et~al\mbox{.}(2023)]%
        {yin2023understanding}
\bibfield{author}{\bibinfo{person}{Qing Yin}, \bibinfo{person}{Hui Fang}, \bibinfo{person}{Zhu Sun}, {and} \bibinfo{person}{Yew-Soon Ong}.} \bibinfo{year}{2023}\natexlab{}.
\newblock \showarticletitle{Understanding diversity in session-based recommendation}.
\newblock \bibinfo{journal}{\emph{ACM Transactions on Information Systems}} (\bibinfo{year}{2023}), \bibinfo{pages}{1--34}.
\newblock


\bibitem[Ying et~al\mbox{.}(2018)]%
        {ying2018graph}
\bibfield{author}{\bibinfo{person}{Rex Ying}, \bibinfo{person}{Ruining He}, \bibinfo{person}{Kaifeng Chen}, \bibinfo{person}{Pong Eksombatchai}, \bibinfo{person}{William~L Hamilton}, {and} \bibinfo{person}{Jure Leskovec}.} \bibinfo{year}{2018}\natexlab{}.
\newblock \showarticletitle{Graph convolutional neural networks for web-scale recommender systems}. In \bibinfo{booktitle}{\emph{Proc. of KDD}}.
\newblock


\bibitem[Yu et~al\mbox{.}(2023)]%
        {yu2023causality}
\bibfield{author}{\bibinfo{person}{Dianer Yu}, \bibinfo{person}{Qian Li}, \bibinfo{person}{Hongzhi Yin}, {and} \bibinfo{person}{Guandong Xu}.} \bibinfo{year}{2023}\natexlab{}.
\newblock \showarticletitle{Causality-guided graph learning for session-based recommendation}. In \bibinfo{booktitle}{\emph{Proc. of CIKM}}. \bibinfo{pages}{3083--3093}.
\newblock


\bibitem[Yu et~al\mbox{.}(2020)]%
        {yu2020tagnn}
\bibfield{author}{\bibinfo{person}{Feng Yu}, \bibinfo{person}{Yanqiao Zhu}, \bibinfo{person}{Qiang Liu}, \bibinfo{person}{Shu Wu}, \bibinfo{person}{Liang Wang}, {and} \bibinfo{person}{Tieniu Tan}.} \bibinfo{year}{2020}\natexlab{}.
\newblock \showarticletitle{TAGNN: Target attentive graph neural networks for session-based recommendation}. In \bibinfo{booktitle}{\emph{Proc. of SIGIR}}.
\newblock


\bibitem[Yuan et~al\mbox{.}(2020)]%
        {yuan2020future}
\bibfield{author}{\bibinfo{person}{Fajie Yuan}, \bibinfo{person}{Xiangnan He}, \bibinfo{person}{Haochuan Jiang}, \bibinfo{person}{Guibing Guo}, \bibinfo{person}{Jian Xiong}, \bibinfo{person}{Zhezhao Xu}, {and} \bibinfo{person}{Yilin Xiong}.} \bibinfo{year}{2020}\natexlab{}.
\newblock \showarticletitle{Future data helps training: Modeling future contexts for session-based recommendation}. In \bibinfo{booktitle}{\emph{Proc. of WWW}}.
\newblock


\bibitem[Yuan et~al\mbox{.}(2022)]%
        {yuan2022micro}
\bibfield{author}{\bibinfo{person}{Jiahao Yuan}, \bibinfo{person}{Wendi Ji}, \bibinfo{person}{Dell Zhang}, \bibinfo{person}{Jinwei Pan}, {and} \bibinfo{person}{Xiaoling Wang}.} \bibinfo{year}{2022}\natexlab{}.
\newblock \showarticletitle{Micro-behavior encoding for session-based recommendation}. In \bibinfo{booktitle}{\emph{2022 IEEE 38th International Conference on Data Engineering (ICDE)}}.
\newblock


\bibitem[Yuan et~al\mbox{.}(2021)]%
        {yuan2021dual}
\bibfield{author}{\bibinfo{person}{Jiahao Yuan}, \bibinfo{person}{Zihan Song}, \bibinfo{person}{Mingyou Sun}, \bibinfo{person}{Xiaoling Wang}, {and} \bibinfo{person}{Wayne~Xin Zhao}.} \bibinfo{year}{2021}\natexlab{}.
\newblock \showarticletitle{Dual Sparse Attention Network For Session-based Recommendation}. In \bibinfo{booktitle}{\emph{Proc. of AAAI}}.
\newblock


\bibitem[Zangerle et~al\mbox{.}(2014)]%
        {zangerle2014nowplaying}
\bibfield{author}{\bibinfo{person}{Eva Zangerle}, \bibinfo{person}{Martin Pichl}, \bibinfo{person}{Wolfgang Gassler}, {and} \bibinfo{person}{G{\"u}nther Specht}.} \bibinfo{year}{2014}\natexlab{}.
\newblock \showarticletitle{\# nowplaying music dataset: Extracting listening behavior from twitter}. In \bibinfo{booktitle}{\emph{Proceedings of the first international workshop on internet-scale multimedia management}}.
\newblock


\bibitem[Zeng et~al\mbox{.}(2024)]%
        {zeng2024federated}
\bibfield{author}{\bibinfo{person}{Huimin Zeng}, \bibinfo{person}{Zhenrui Yue}, \bibinfo{person}{Qian Jiang}, {and} \bibinfo{person}{Dong Wang}.} \bibinfo{year}{2024}\natexlab{}.
\newblock \showarticletitle{Federated recommendation via hybrid retrieval augmented generation}.
\newblock \bibinfo{journal}{\emph{arXiv preprint arXiv:2403.04256}} (\bibinfo{year}{2024}).
\newblock


\bibitem[Zhang et~al\mbox{.}(2020b)]%
        {zhang2020personalized}
\bibfield{author}{\bibinfo{person}{Mengqi Zhang}, \bibinfo{person}{Shu Wu}, \bibinfo{person}{Meng Gao}, \bibinfo{person}{Xin Jiang}, \bibinfo{person}{Ke Xu}, {and} \bibinfo{person}{Liang Wang}.} \bibinfo{year}{2020}\natexlab{b}.
\newblock \showarticletitle{Personalized graph neural networks with attention mechanism for session-aware recommendation}.
\newblock \bibinfo{journal}{\emph{IEEE Transactions on Knowledge and Data Engineering}} (\bibinfo{year}{2020}).
\newblock


\bibitem[Zhang et~al\mbox{.}(2023)]%
        {zhang2023efficiently}
\bibfield{author}{\bibinfo{person}{Peiyan Zhang}, \bibinfo{person}{Jiayan Guo}, \bibinfo{person}{Chaozhuo Li}, \bibinfo{person}{Yueqi Xie}, \bibinfo{person}{Jae~Boum Kim}, \bibinfo{person}{Yan Zhang}, \bibinfo{person}{Xing Xie}, \bibinfo{person}{Haohan Wang}, {and} \bibinfo{person}{Sunghun Kim}.} \bibinfo{year}{2023}\natexlab{}.
\newblock \showarticletitle{Efficiently leveraging multi-level user intent for session-based recommendation via atten-mixer network}. In \bibinfo{booktitle}{\emph{Proc. of WSDM}}.
\newblock


\bibitem[Zhang et~al\mbox{.}(2021b)]%
        {zhang2021knowledge}
\bibfield{author}{\bibinfo{person}{Rongzhi Zhang}, \bibinfo{person}{Yulong Gu}, \bibinfo{person}{Xiaoyu Shen}, {and} \bibinfo{person}{Hui Su}.} \bibinfo{year}{2021}\natexlab{b}.
\newblock \showarticletitle{Knowledge-enhanced Session-based Recommendation with Temporal Transformer}.
\newblock \bibinfo{journal}{\emph{arXiv preprint arXiv:2112.08745}} (\bibinfo{year}{2021}).
\newblock


\bibitem[Zhang et~al\mbox{.}(2019)]%
        {zhang2019deep}
\bibfield{author}{\bibinfo{person}{Shuai Zhang}, \bibinfo{person}{Lina Yao}, \bibinfo{person}{Aixin Sun}, {and} \bibinfo{person}{Yi Tay}.} \bibinfo{year}{2019}\natexlab{}.
\newblock \showarticletitle{Deep learning based recommender system: A survey and new perspectives}.
\newblock \bibinfo{journal}{\emph{ACM Computing Surveys (CSUR)}} (\bibinfo{year}{2019}).
\newblock


\bibitem[Zhang et~al\mbox{.}(2013)]%
        {zhang2013combining}
\bibfield{author}{\bibinfo{person}{Wei Zhang}, \bibinfo{person}{Jianyong Wang}, {and} \bibinfo{person}{Wei Feng}.} \bibinfo{year}{2013}\natexlab{}.
\newblock \showarticletitle{Combining latent factor model with location features for event-based group recommendation}. In \bibinfo{booktitle}{\emph{Proc. of KDD}}.
\newblock


\bibitem[Zhang et~al\mbox{.}(2022)]%
        {zhang2022price}
\bibfield{author}{\bibinfo{person}{Xiaokun Zhang}, \bibinfo{person}{Bo Xu}, \bibinfo{person}{Liang Yang}, \bibinfo{person}{Chenliang Li}, \bibinfo{person}{Fenglong Ma}, \bibinfo{person}{Haifeng Liu}, {and} \bibinfo{person}{Hongfei Lin}.} \bibinfo{year}{2022}\natexlab{}.
\newblock \showarticletitle{Price does matter! modeling price and interest preferences in session-based recommendation}. In \bibinfo{booktitle}{\emph{Proc. of SIGIR}}. \bibinfo{pages}{1684--1693}.
\newblock


\bibitem[Zhang et~al\mbox{.}(2020a)]%
        {zhang2020explainable}
\bibfield{author}{\bibinfo{person}{Yongfeng Zhang}, \bibinfo{person}{Xu Chen}, {et~al\mbox{.}}} \bibinfo{year}{2020}\natexlab{a}.
\newblock \showarticletitle{Explainable recommendation: A survey and new perspectives}.
\newblock \bibinfo{journal}{\emph{Foundations and Trends{\textregistered} in Information Retrieval}} (\bibinfo{year}{2020}), \bibinfo{pages}{1--101}.
\newblock


\bibitem[Zhang et~al\mbox{.}(2021a)]%
        {zhang2021causal}
\bibfield{author}{\bibinfo{person}{Yang Zhang}, \bibinfo{person}{Fuli Feng}, \bibinfo{person}{Xiangnan He}, \bibinfo{person}{Tianxin Wei}, \bibinfo{person}{Chonggang Song}, \bibinfo{person}{Guohui Ling}, {and} \bibinfo{person}{Yongdong Zhang}.} \bibinfo{year}{2021}\natexlab{a}.
\newblock \showarticletitle{Causal intervention for leveraging popularity bias in recommendation}. In \bibinfo{booktitle}{\emph{Proc. of SIGIR}}.
\newblock


\bibitem[Zhang et~al\mbox{.}(2014)]%
        {zhang2014explicit}
\bibfield{author}{\bibinfo{person}{Yongfeng Zhang}, \bibinfo{person}{Guokun Lai}, \bibinfo{person}{Min Zhang}, \bibinfo{person}{Yi Zhang}, \bibinfo{person}{Yiqun Liu}, {and} \bibinfo{person}{Shaoping Ma}.} \bibinfo{year}{2014}\natexlab{}.
\newblock \showarticletitle{Explicit factor models for explainable recommendation based on phrase-level sentiment analysis}. In \bibinfo{booktitle}{\emph{Proc. of SIGIR}}. \bibinfo{pages}{83--92}.
\newblock


\bibitem[Zhang and Yang(2018)]%
        {zhang2018overview}
\bibfield{author}{\bibinfo{person}{Yu Zhang} {and} \bibinfo{person}{Qiang Yang}.} \bibinfo{year}{2018}\natexlab{}.
\newblock \showarticletitle{An overview of multi-task learning}.
\newblock \bibinfo{journal}{\emph{National Science Review}} (\bibinfo{year}{2018}).
\newblock


\bibitem[Zhang and Nasraoui(2007)]%
        {zhang2007efficient}
\bibfield{author}{\bibinfo{person}{Zhiyong Zhang} {and} \bibinfo{person}{Olfa Nasraoui}.} \bibinfo{year}{2007}\natexlab{}.
\newblock \showarticletitle{Efficient hybrid Web recommendations based on Markov clickstream models and implicit search}. In \bibinfo{booktitle}{\emph{IEEE/WIC/ACM International Conference on Web Intelligence (WI'07)}}.
\newblock


\bibitem[Zhang and Wang(2021)]%
        {zhang2021graph}
\bibfield{author}{\bibinfo{person}{Zizhuo Zhang} {and} \bibinfo{person}{Bang Wang}.} \bibinfo{year}{2021}\natexlab{}.
\newblock \showarticletitle{Graph Neighborhood Routing and Random Walk for Session-based Recommendation}. In \bibinfo{booktitle}{\emph{Proc. of ICDM}}.
\newblock


\bibitem[Zhang and Wang(2023)]%
        {zhang2022graph}
\bibfield{author}{\bibinfo{person}{Zizhuo Zhang} {and} \bibinfo{person}{Bang Wang}.} \bibinfo{year}{2023}\natexlab{}.
\newblock \showarticletitle{Graph spring network and informative anchor selection for session-based recommendation}.
\newblock \bibinfo{journal}{\emph{Neural Networks}} (\bibinfo{year}{2023}), \bibinfo{pages}{43--56}.
\newblock


\bibitem[Zheng et~al\mbox{.}(2021)]%
        {zheng2021disentangling}
\bibfield{author}{\bibinfo{person}{Yu Zheng}, \bibinfo{person}{Chen Gao}, \bibinfo{person}{Xiang Li}, \bibinfo{person}{Xiangnan He}, \bibinfo{person}{Yong Li}, {and} \bibinfo{person}{Depeng Jin}.} \bibinfo{year}{2021}\natexlab{}.
\newblock \showarticletitle{Disentangling user interest and conformity for recommendation with causal embedding}. In \bibinfo{booktitle}{\emph{Proc. of WWW}}.
\newblock


\bibitem[Zheng et~al\mbox{.}(2020)]%
        {zheng2020dgtn}
\bibfield{author}{\bibinfo{person}{Yujia Zheng}, \bibinfo{person}{Siyi Liu}, \bibinfo{person}{Zekun Li}, {and} \bibinfo{person}{Shu Wu}.} \bibinfo{year}{2020}\natexlab{}.
\newblock \showarticletitle{Dgtn: Dual-channel graph transition network for session-based recommendation}. In \bibinfo{booktitle}{\emph{Proc. of ICDM}}.
\newblock


\bibitem[Zhou et~al\mbox{.}(2019)]%
        {zhou2019variational}
\bibfield{author}{\bibinfo{person}{Fan Zhou}, \bibinfo{person}{Zijing Wen}, \bibinfo{person}{Kunpeng Zhang}, \bibinfo{person}{Goce Trajcevski}, {and} \bibinfo{person}{Ting Zhong}.} \bibinfo{year}{2019}\natexlab{}.
\newblock \showarticletitle{Variational session-based recommendation using normalizing flows}. In \bibinfo{booktitle}{\emph{Proc. of WWW}}.
\newblock


\bibitem[Zhou et~al\mbox{.}(2021)]%
        {zhou2021temporal}
\bibfield{author}{\bibinfo{person}{Huachi Zhou}, \bibinfo{person}{Qiaoyu Tan}, \bibinfo{person}{Xiao Huang}, \bibinfo{person}{Kaixiong Zhou}, {and} \bibinfo{person}{Xiaoling Wang}.} \bibinfo{year}{2021}\natexlab{}.
\newblock \showarticletitle{Temporal Augmented Graph Neural Networks for Session-Based Recommendations}. In \bibinfo{booktitle}{\emph{Proc. of SIGIR}}.
\newblock


\bibitem[Zhu et~al\mbox{.}(2022)]%
        {zhu2022transition}
\bibfield{author}{\bibinfo{person}{Guanghui Zhu}, \bibinfo{person}{Haojun Hou}, \bibinfo{person}{Jingfan Chen}, \bibinfo{person}{Chunfeng Yuan}, {and} \bibinfo{person}{Yihua Huang}.} \bibinfo{year}{2022}\natexlab{}.
\newblock \showarticletitle{Transition Relation Aware Self-Attention for Session-based Recommendation}.
\newblock \bibinfo{journal}{\emph{arXiv preprint arXiv:2203.06407}} (\bibinfo{year}{2022}).
\newblock


\bibitem[Zhu et~al\mbox{.}(2023)]%
        {zhu2023membership}
\bibfield{author}{\bibinfo{person}{Zhihao Zhu}, \bibinfo{person}{Chenwang Wu}, \bibinfo{person}{Rui Fan}, \bibinfo{person}{Defu Lian}, {and} \bibinfo{person}{Enhong Chen}.} \bibinfo{year}{2023}\natexlab{}.
\newblock \showarticletitle{Membership inference attacks against sequential recommender systems}. In \bibinfo{booktitle}{\emph{Proc. of WWW}}. \bibinfo{pages}{1208--1219}.
\newblock


\bibitem[Zimdars et~al\mbox{.}(2001)]%
        {zimdars2001using}
\bibfield{author}{\bibinfo{person}{Andrew Zimdars}, \bibinfo{person}{David~Maxwell Chickering}, {and} \bibinfo{person}{Christopher Meek}.} \bibinfo{year}{2001}\natexlab{}.
\newblock \showarticletitle{Using temporal data for making recommendations}. In \bibinfo{booktitle}{\emph{Proc. of UAI}}. \bibinfo{pages}{580--588}.
\newblock


\end{thebibliography}










\end{document}